\documentclass[aps,prd,longbibliography,notitlepage,twocolumn,showpacs,amsmath,amssymb,superscriptaddress,nofootinbib,floatfix,10pt,preprintnumbers]{revtex4-1}

\usepackage{graphicx}
\usepackage{dcolumn}
\usepackage{bm}
\usepackage{epsf}
\usepackage{rotating}
\usepackage{epsfig,graphics,rotate,color}
\usepackage{wrapfig}
\usepackage{amssymb}
\usepackage{graphicx}
\usepackage{amsmath}
\usepackage{gensymb}
\usepackage{amsfonts}
\usepackage{braket}
\usepackage[utf8]{inputenc}
\usepackage[dvipsnames]{xcolor}
\usepackage[colorlinks=true,citecolor=blue,linkcolor=blue]{hyperref}
\usepackage{mathbbol}
\usepackage{placeins} 
\usepackage{tabularx}
\usepackage{algorithm,algorithmic}
\usepackage{cancel}
\newcommand{\vect}[1]{\boldsymbol{\mathbf{#1}}}

\begin{document}

\preprint{IPPP/18/60}

\title{Statistical challenges in the search for dark matter}
\author{Sara Algeri}%
\affiliation{Department of Mathematics, Imperial College London, SW72AZ, United Kingdom}
\author{Melissa van Beekveld}%
\affiliation{Theoretical High Energy Physics, IMAPP, Faculty of Science, Mailbox 79, Radboud University, The Netherlands}
\author{Nassim Bozorgnia}
\affiliation{Institute for Particle Physics Phenomenology, Department of Physics,\\
Durham University, Durham, DH1 3LE, United Kingdom}%
\author{Alyson Brooks}
\affiliation{Department of Physics \& Astronomy,
Rutgers University, 136 Frelinghuysen Road, Piscataway, NJ 08854 U.S.A.}%
\author{J. Alberto Casas}%
\affiliation{Instituto de F\'{\i}sica Te\'orica, IFT-UAM/CSIC, Universidad Aut\'onoma de Madrid, 28049 Madrid, Spain}%
\author{Jessi Cisewski-Kehe}
\affiliation{Department of Statistics and Data Science, \\ Yale University, New Haven, CT 06520 U.S.A.}
\author{Francis-Yan Cyr-Racine}
\affiliation{Department of Physics, Harvard University, Cambridge, MA 02138 U.S.A.}
\author{Thomas D. P. Edwards} \thanks{Editors}
\affiliation{Gravitation Astroparticle Physics Amsterdam (GRAPPA), Institute of Physics,
University of Amsterdam, 1090 GL Amsterdam, The Netherlands}
\author{Fabio Iocco}
\affiliation{ICTP South American Institute for Fundamental Research, and Instituto de Fısica Teorica - Universidade Estadual Paulista (UNESP), Rua Dr. Bento Teobaldo Ferraz 271, 01140-070 Sao Paulo, SP Brazil}
\author{Bradley J. Kavanagh} \thanks{Editors}
\affiliation{Gravitation Astroparticle Physics Amsterdam (GRAPPA), Institute of Physics,
University of Amsterdam, 1090 GL Amsterdam, The Netherlands}
\author{Judita Mamu\v{z}i\'{c}}%
\affiliation{Instituto de Física Corpuscular (IFIC) / Consejo Superior de Investigaciones Científicas (CSIC) - Universidad de Valencia (UV), Spain}
\author{Siddharth Mishra-Sharma}
\affiliation{Department of Physics, Princeton University, Princeton, NJ 08544 U.S.A.}
\author{Wolfgang Rau}
\affiliation{Arthur B. McDonald Canadian Astroparticle Physics Research Institute, Department of Physics, Engineering Physics and Astronomy, Queen's University, Kingston ON K7L 3N6, Canada}
\author{Roberto Ruiz de Austri}
\affiliation{Instituto de Física Corpuscular (IFIC) / Consejo Superior de Investigaciones Científicas (CSIC) - Universidad de Valencia (UV), Spain}
\author{Benjamin R. Safdi}
\affiliation{Leinweber Center for Theoretical Physics, Department of Physics,
University of Michigan, Ann Arbor, MI 48109 U.S.A.}
\author{Pat Scott} \thanks{Editors}
\affiliation{Department of Physics, Imperial College London, Blackett Laboratory, Prince Consort Road, London SW7 2AZ, United Kingdom}
\author{Tracy R. Slatyer}
\affiliation{Center for Theoretical Physics, Massachusetts Institute of Technology, Cambridge, MA 02139 U.S.A.}
\author{Yue-Lin Sming Tsai}
\affiliation{Institute of Physics, Academia Sinica, Taipei 11529, Taiwan}
\author{Aaron C. Vincent}
\thanks{Editors}
\affiliation{Arthur B. McDonald Canadian Astroparticle Physics Research Institute, Department of Physics, Engineering Physics and Astronomy, Queen's University, Kingston ON K7L 3N6, Canada}
\email{aaron.vincent@queensu.ca}
\affiliation{Visiting Fellow, Perimeter Institute for Theoretical Physics, 31 Caroline St. N., Waterloo, Ontario N2L 2Y5, Canada}
\author{Christoph Weniger}
\affiliation{Gravitation Astroparticle Physics Amsterdam (GRAPPA), Institute of Physics,
University of Amsterdam, 1090 GL Amsterdam, The Netherlands}
\author{Jennifer Rittenhouse West}
\affiliation{Department of Physics \& Astronmy,
University of California, Irvine, CA 92697 U.S.A. }
\author{Robert L. Wolpert}
\affiliation{Department of Statistical Science, Duke University, Durham, NC 27708 U.S.A}

\date{\today}

\begin{abstract}
The search for the particle nature of dark matter has given rise to a number of experimental, theoretical and statistical challenges. Here, we report on a number of these statistical challenges and new techniques to address them, as discussed in the \textit{DMStat} workshop held Feb 26 -- Mar 3 2018 at the Banff International Research Station for Mathematical Innovation and Discovery (BIRS) in Banff, Alberta.\footnote{\url{http://www.birs.ca/events/2018/5-day-workshops/18w5095}}
\end{abstract}

\maketitle

\tableofcontents
\section{Introduction}
The nature of dark matter (DM) is one of the most pressing puzzles in modern particle physics and astronomy. Overwhelming evidence from galactic dynamics to cosmology tells us that $\sim 85$\% of the matter content of the Universe is in a very different form from the familiar ``baryonic'' matter described by the Standard Model (SM) of particle physics. Precision measurements of the cosmic microwave background (CMB) suggest the existence of a new particle that is cold (i.e.\ non-relativistic at sufficiently early cosmic times), dark (very weakly-interacting with quarks, electrons and photons), and behaved like matter (a pressureless fluid) in the Early Universe \cite{Bertone:2004pz}.

Evidence for dark matter comes to us entirely via its gravitational influence; however, there are many good reasons to believe in a particle physics portal to the dark sector. In fact, many theories of physics beyond the standard model (BSM) such as supersymmetry naturally predict a nonzero relic abundance of ``dark'' particles.

In the absence of a definitive non-gravitational signal of DM, the space of possible models of particle dark matter has also thrived. Theoretical motivations such as the ``WIMP miracle,'' the ``baryon disaster,'' the Peccei-Quinn solution to the strong CP problem, and the neutrino mass problem motivate such candidates as the WIMP, asymmetric dark matter, the axion or the sterile neutrino. The full list of DM candidates is as varied as it is extensive.

Following several decades of searches, it has become increasingly clear that discovery is less likely to happen via a single ``smoking gun" signal, but rather by scrutinizing data from many experiments in many disparate fields. The main searches for dark matter are broadly categorized into direct detection, indirect detection, and production at colliders.

The next decade will present us with major advances in experiments designed to search for dark matter, as well as experiments with a broader focus on searches for BSM physics. Even though current and past searches have thus far come up empty, the parameter space that has been explored pales in comparison with what will become available in years to come. This includes an unparalleled quantity of astrophysical data, from e.g.\ the Square Kilometre Array (SKA)  radio telescope \cite{2009IEEEP..97.1482D} that will map the distribution of matter in the dark ages before the formation of the first galaxies via the 21 cm spin transition line of the hydrogen atom \cite{2012RPPh...75h6901P}; gamma ray telescopes such as the Cherenkov Telescope Array (CTA) \cite{Acharya:2017ttl} that will yield important information about the highest energies in the universe; and the next generation of galaxy surveys (eBOSS \cite{Dawson:2015wdb}, DESI \cite{Aghamousa:2016zmz}) which will map the distribution of structure in the universe. Starting in 2022, the Large Synoptic Survey Telescope (LSST) \cite{Abell:2009aa} will survey the southern sky to unprecedented depth, allowing for the discovery of new ultra-faint dwarf galaxies \cite{Kim:2017iwr} and increasing the sample of known galaxy-scale strong gravitational lenses by a factor of 10 \cite{Oguri:2010ns}. Taken together, these new observations will dramatically improve our knowledge of dark matter structure on kiloparsec scales and below, hence stress-testing the standard cold dark matter paradigm in a new regime. Concurrently, space based missions such as Gaia \cite{2016A&A...595A...1G,2018arXiv180409365G} will map the distribution of dark matter in our own neighbourhood for the first time, with the promise of sub-milliarcsecond astrometry. The PINGU upgrade to the IceCube neutrino detector at the South Pole \cite{TheIceCube-Gen2:2016cap} will be able to detect light DM candidates, as we embark on the first decade of neutrino astronomy.

Meanwhile, DM-specific searches such as XENONnT \cite{2017APS..APR.J9003A}, LUX-ZEPLIN \cite{Akerib:2018lyp} and ADMX \cite{Du:2018uak} (along with DM-focused analyses of collider data) will provide the best sensitivity for testing a large variety of hypotheses regarding the particle properties of DM.

Reconciling the vast landscape of theoretical models and the many disparate data sets is not an easy task, and it inevitably leads to a number of statistical challenges on scales that range from the interpretation of a single experiment, all the way to combination of models and large datasets with one another.

Our aim in this short review is to outline major statistical challenges that came up over the course of the \textit{DMStat} workshop\footnote{Held Feb 26 -- Mar 3 2018 at the Banff International Research Station for Mathematical Innovation and Discovery (BIRS) in Banff, Alberta (\url{http://www.birs.ca/events/2018/5-day-workshops/18w5095}).}, along with proposed approaches and solutions including software developed by the community. We begin with a brief review of the problem of dark matter (Sec.~\ref{sec:darkmatter}), followed in Sec.~\ref{sec:searches} by a description of current strategies for experimental dark matter searches and some challenges those searches have encountered. We then outline and discuss several statistical approaches (Sec.~\ref{sec:StatisticalChallenges}), including some novel techniques, and present some simple examples in Sec~\ref{sec:examples}.

\subsection{Disambiguation}

Although statistics acts in some sense as a lingua franca across the sciences, there are certain ``regional dialects'' that should be noted: we have identified a few terms in particular that have very different meanings when used by the astroparticle physics community versus statisticians.
\begin{description}
\item[Model] In physics, the word ``model'' may be entirely synonymous with ``theory'', in the sense of referring to an entire physical theory that one may wish to test with statistics -- or it may refer to a specific realisation of a theory.  This can be a restriction of the theory to a particular subspace of its possible forms, or very often, a specific numerical choice for all values of the free parameters of a theory. In contrast, in the field of statistics, the term ``model'' refers to an incompletely-specified probability distribution; draws from this distribution are meant to replicate the processes that generated the observable data. Data are used to \textit{estimate} the missing components of the model, typically a vector of unknown parameters. Such estimation problems are fundamental in \textit{statistical inference}.
\item[Simulation] In statistics, this usually refers to the process of generating a large number of random realizations from a probability model by Monte Carlo methods, in order to characterize the distribution of a quantity of interest, e.g.\ an estimator or a test statistic. In physics, the word simulation will usually refer to modeling the outcome of a physical experiment based on the model parameters (for example, the angular power spectrum of the cosmic microwave background, given a set of cosmological parameters). A physicist's simulation may or may not be deterministic in nature.
\item[Coverage] Often (mis)used in particle physics as a loosely defined qualitative notion of the completeness with which a theoretical parameter space has been sampled.  In statistics, coverage has a very specific and well-defined meaning, referring to the fraction of repeated experiments in which the true value of a quantity actually appears inside a confidence interval/region.  A 90\% confidence interval/contour is said to undercover if the true value would actually appear inside the interval in less than 90\% of repeated experiments, or to overcover if the true value would appear within the interval in more than 90\% of repeats.  Overcoverage is inefficient but relatively benign, as it increases the probability of a type II error (failure to reject a false null hypothesis, a ``false negative" finding); undercoverage leads to an increase in the rate of type I error (rejection of a true null hypothesis, a ``false positive" finding), so is generally considered more serious.

\item[Machine learning] A field that grew out of computer science and the study of artificial
intelligence, but is concerned with many of the same challenges faced by statisticians. Indeed,
the line between these fields is quite blurry, although each uses its own terminology, e.g.,
\textit{supervised learning} problems in machine learning are closely aligned with \textit{regression} problems in statistics. Methods developed within machine learning tend to be more
focused on broad applicability with computational efficiency, while statisticians will often tailor their models more to a particular application, and place more emphasis on theoretical properties of methods.

\end{description}

\section{The dark matter problem}
\label{sec:darkmatter}

While the flat rotation curves of spiral galaxies \cite{Rubin:1980zd} are often held up as conclusive evidence for a missing matter component in the Universe, equally strong evidence for dark matter arises over a range of scales: the motion of nearby stars above and below the galactic plane \cite{Read:2014qva,Sivertsson:2017rkp}; the velocities of galaxies within clusters \cite{1933AcHPh...6..110Z}; gravitational lensing by galaxies and clusters \cite{Clowe:2006eq,Parker:2005fh,Hoekstra:2013via,Velander:2013jga}; the rapid formation time of galaxies \cite{Coil:2012vw,Vogelsberger:2014dza}, as well as the angular power spectrum of the cosmic microwave background \cite{Ade:2015xua}. Each of these points to a large nonbaryonic component of matter in the Universe -- about 85\% of the total matter, or $\sim$25\% of the total energy density. If this missing matter is in the form of particles, it should be (almost) electrically neutral \cite{McDermott:2010pa} and have only very weak interactions with ordinary matter \cite{Taoso:2007qk}.

The Standard Model of particle physics does not provide a suitable particle to play the role of dark matter. However, a number of BSM theories of physics provide compelling candidates for this nonbaryonic particle. In supersymmetric (SUSY) extensions of the Standard Model, for example, the lightest new SUSY particle is typically stable and can be produced with sufficient density in the early Universe to account for all of the dark matter \cite{Jungman:1995df}. Such particles fall into the class of `Weakly Interacting Massive Particles' (WIMPs) \cite{Roszkowski:2017nbc}. Another popular candidate arises in solutions of the strong CP problem, which hint at the existence of a new light pseudoscalar, the axion \cite{Peccei:1977hh,Peccei:2006as}. Other exotic candidates are also plausible: primordial black holes \cite{Carr:2016drx}, sterile neutrinos \cite{Merle:2017jfn}, particles in a strongly-interacting dark sector \cite{Hochberg:2014dra}, or candidates motivated by potential problems on small astrophysical scales such as self-interacting dark matter \cite{Spergel:1999mh,Tulin:2017ara}.

Each of these models and candidates comes with its own set of parameters: particle masses, interaction strengths, etc. While these parameters can in principle be constrained by experiment, there is often little theoretical guidance as to which values should be preferred and so choosing appropriate priors is often a challenge \cite{Allanach:2006jc,Berger:2008cq,Cabrera:2016wwr}. Each model also provides its own set of experimental signatures, requiring an ability to sort signal from background, as we discuss shortly.

In what follows, we will focus on the standard WIMP paradigm for DM, but of course many of the statistical challenges we discuss are relevant also for other candidates. Indeed, an interesting question arises if we remain agnostic about the nature of the dark matter particle: how do we compare models with very different parameter spaces and observational signatures? Most commonly used tools for model comparison require them to be \textit{nested}, so that a modification of the parameters of one model yields the second model. We must therefore be careful if we choose to compare fundamentally distinct models such as the WIMP and the axion.

\section{Contemporary challenges in Dark Matter}
\label{sec:searches}Here we summarise some of the key observational challenges in the search for DM. This includes searches for DM scattering of Standard Model particles (Sec.~\ref{sec:Direct}), searches for the products from DM annihilation (Sec.~\ref{sec:Indirect}) and searches for DM production in colliders (Sec.~\ref{sec:Colliders}). We also discuss the impact of DM on galaxy formation and the challenge of constraining its properties with astronomical observations (Sec.~\ref{sec:Simulations}). Our goal is not to provide a thorough review of these topics but to point out some of the statistical issues involved and to provide a backdrop for the more detailed discussion of statistical challenges in Sec.~\ref{sec:StatisticalChallenges}.

\subsection{Direct searches}
\label{sec:Direct}
{\it Direct searches} are experiments searching for evidence of interactions of individual dark matter particles in terrestrial detectors. This type of experiment can be carried out for most dark matter candidates (except for very light particles which have energies that are below the threshold of existing particle detectors, and extremely heavy particles where the number density becomes too small and interactions are too rare even in the case of high cross sections).

\subsubsection{WIMP Signatures}
Here we concentrate on experiments originally developed for the detection of WIMPs or WIMP-like particles. Due to simple scattering kinematics and the fact that particles we can hope to detect would be gravitationally bound to our galaxy (and thus must have a velocity of less than $\sim$600\,km/s \cite{Piffl:2013mla}), the energy transfer through interactions with electrons is at most in the eV range, while interactions with nuclei yield typical recoil energies in the keV range. Thus, most of these experiments concentrate on the identification of nuclear recoils (NR), while discarding electron recoils (ER). However, recent developments have yielded detectors with very low energy thresholds capable of detecting energies in the range that can be expected from electron interacting dark matter \cite{Essig:2011nj,Essig:2017kqs}.

Other signatures that have been proposed for the identification of dark matter interactions include modulations of the detected signal with time. Due to the motion of the earth about the Sun and the Sun about the galaxy, the relative velocity of the detector and the dark matter particles changes over the course of the year, leading to a weak modulation of the interaction rate \cite{Freese:2012xd}.

If the direction of the incoming dark matter particles can be identified in a detector, one would also expect a modulation over the course of the day, since the rotation of the earth about its axis leads to a change of the orientation of the detector relative to the flux of incoming dark matter particles \cite{Mayet:2016zxu}.

\subsubsection{Detection Channels}
There are three main ways of detecting particles:
\begin{enumerate}
\item Particle interactions may ionize the target, and the liberated charges can be collected; the amount of charge detected allows for an estimate of the deposited energy. However, the amount of charge produced is usually very different for different interactions (ER interactions are usually much more efficient in ionizing than NR interactions).
\item In some cases excited electrons de-excite via the emission of photons in a process called {\it scintillation}. The fraction of energy converted to scintillation light is usually small (few percent) and again the scintillation efficiency is usually  much higher for ER than for NR interactions.
\item Eventually, most of the energy of the interaction is converted into thermal energy. This provides an opportunity to measure the total energy transfer of the interaction independent of the interaction type.
\end{enumerate}
Experiments have been designed to take advantage of each of the detection channels, and in many cases two of the channels are combined, allowing for the discrimination between ER background events and NR dark matter candidate events.

Below we list a selection of current experiments, to give an idea of the range of techniques:
\begin{itemize}
\item \textit{Inorganic scintillating crystals} are used by the DAMA/LIBRA experiment \cite{Bernabei:2008yh,Bernabei:2013xsa,Bernabei:2018yyw} which relies on the  annual modulation discussed above for signal identification. Such a signal as indeed been observed  beyond any statistical doubt, but its interpretation in terms of dark matter interactions is inconsistent with the absence of a compatible signal in other experiments. SABRE \cite{Froborg:2016ova} and COSINE \cite{Adhikari:2017esn} are upcoming attempts to test the DAMA signal using the same technique.

\item \textit{Semiconductor detectors}  made out of Si (DAMIC \cite{2012PhLB..711..264B,deMelloNeto:2015mca}) and Ge (CoGeNT \cite{Aalseth:2012if}) benefit from low thresholds and very high intrinsic material purity. CoGeNT observed a statistically significant rise of the observed event rate towards low energy and hinted at a possible interpretation as a dark matter signal. However, a careful reanalysis with a more realistic background model was able to explain the observation with conventional interactions \cite{Aalseth:2014arxiv}. This highlights that statistical significance must be accompanied with a very good understanding of background and detector response in order to avoid misinterpretations.

\item \textit{Super-heated liquid detectors} were developed by COUPP \cite{Behnke:2012ys} and PICASSO \cite{Archambault:2012pm}, now combined to form the PICO collaboration \cite{Amole:2017dex}. The level of superheat in these detectors is adjusted such that they are insensitive to ER background and only NR interactions trigger a phase transition.

\item \textit{Cryogenic detectors} combine the detection of thermal energy with the measurement of a charge signal (SuperCDMS \cite{Agnese:2016cpb,Agnese:2017njq,Agnese:2018col,Agnese:2017jvy}, EDELWEISS \cite{Hehn:2014bya,Hehn:2016nll}) or scintillation light (CRESST \cite{Angloher:2015ewa,Petricca:2017zdp} and COSINUS \cite{Angloher:2016ooq}) for an effective reduction of ER background.

\item \textit{Gaseous detectors} are either designed to reach very low thresholds and thus access dark matter particles down into the sub-GeV mass range (NEWS-G \cite{Arnaud:2017bjh}) or aim at identifying the recoil direction (DRIFT \cite{Battat:2016xxe}, MIMAC \cite{Couturier:2016blf}, DMTPC \cite{Leyton:2016nit} and NEWAGE \cite{Nakamura:2015tna}).

\item \textit{Noble liquid detectors} often use dual-phase time projection chambers, taking advantage of the high scintillation light yield of Xe (LUX \cite{Akerib:2016vxi}, XENON1T \cite{Aprile:2017iyp,Aprile:2018dbl}, PANDA-X II \cite{Cui:2017nnn}) and Ar (DarkSide \cite{Agnes:2014bvk,Aalseth:2017fik}) and the fact that electrons liberated in an interaction can be drifted over long distances in the inert material. Comparison of the initial scintillation light and the secondary scintillation produced by the charges that are extracted from the liquid into the gas phase allow for an effective ER background discrimination. In argon, an excellent ER discrimination can also be achieved by just looking at the  pulse-shape of the scintillation light in a simple liquid detector (DEAP \cite{Amaudruz:2017ekt}).

\end{itemize}

Finally, upcoming generations of these detectors (e.g.\ LZ \cite{LZ:sensitivity}, XENONnT \cite{XENONnT:sensitivity}, DarkSide \cite{DS:sensitivity}, PICO-500L \cite{PICO500:sensitivity}, SuperCDMS SNOLAB \cite{SuperCDMS:sensitivity}) are aiming to push the sensitivity to the ultimate limit given by NR interactions from solar or atmospheric neutrinos.

\subsubsection{Statistical methods in direct detection}
A very good tool for the analysis of data from experiments with very low background is the Optimum Interval Method \cite{Yellin:2002xd,Yellin:2008da,Yellin:2011xf}. This gives the best sensitivity in the presence of an unknown background. Since no assumption is required about the origin or spectral shape of the background, this is also a very conservative method. Instead of spectral information, other parameters (timing, pulse shape parameters, etc.) could be used and it would be very useful to have an expansion of this method to a 2- or more-dimensional parameter space. This method can by construction only produce limits on the dark matter interaction rate and does not allow signal extraction.

Recently, many experiments have adopted an analysis method in which known backgrounds are explicitly taken into account. The spectral shapes of these backgrounds are determined and included in a Maximum Likelihood fit. This provides better sensitivity in the presence of a background, but requires that the background features be determined independently. It also allows the extraction of a dark matter signal. A problem arises if there are backgrounds whose distributions are not known, or not well known.

Combining results from different experiments or in some cases from different detectors in the same experiment may also cause a challenge. In particular if the performance and the backgrounds of the different detectors are different, it is non-trivial to find an unbiased method that extracts the best joint sensitivity (see Sec.~\ref{sec:combining} for a simple toy example).

\subsection{Indirect searches}
\label{sec:Indirect}
Indirect searches for DM mainly rely on a search for the high-energy products of DM self-annihilation into Standard Model particles (for a dedicated review, see Ref.~\cite{Gaskins:2016cha}). The WIMP hypothesis has been a compelling driver of these searches, since thermal production through annihilation in the early Universe implies ongoing (albeit suppressed) annihilation today. Nonetheless, WIMPs are not the only DM candidates that are expected to yield an indirect signal: asymmetric DM can annihilate with a relic symmetric DM component, and axions and sterile neutrinos can decay or oscillate to standard model particles.
Indirect searches make use of the large DM concentrations present in astronomical bodies including the Galactic centre, dwarf satellite galaxies, galaxy clusters, as well as the full isotropic background at high redshifts.

Signals of DM annihilation or decay can include:
\begin{enumerate}
\item Cosmic rays produced by nearby DM annihilation in the MW halo can be detected by space observatories including PAMELA \cite{PAMELA,PAMELA2} and AMS-02 \cite{Aguilar:2013qda}, or balloon experiments such as ATIC \cite{ATIC4TALK,ATIC2005}, HEAO \cite{Gruber:1999yr}, TRACER \cite{Boyle:2007wi}, and CREAM \cite{2011ApJ...728..122Y}. Cosmic rays mainly probe local (within $\sim$1 kpc) cosmic ray sources. Because they are composed of charged particles, their arrival directions do not point back towards their sources; rather, they diffuse through the turbulent magnetic structures of the interstellar medium (ISM).
\item Gamma rays, produced copiously by internal bremsstrahlung, decay of heavy unstable DM annihilation products, or interactions with the ISM, are searched for with space-borne experiments such as Fermi-LAT \cite{2009ApJ...697.1071A}, DAMPE \cite{TheDAMPE:2017dtc}, INTEGRAL/SPI \cite{Vedrenne:2003} and Chandra \cite{Weisskopf:2000tx} (among others). At very high energies (and thus very high DM mass), ground-based air Cherenkov telescopes such as MAGIC \cite{MAGIC}, VERITAS \cite{Weekes:2001pd}, HESS \cite{Hinton:2004eu} and in the future CTA \cite{Acharya:2017ttl} can constrain signals from high-mass DM.
\item Neutrinos from DM annihilation and decay can be searched for at neutrino telescopes such as SuperKamiokande \cite{1998PhRvL..81.1158F}, IceCube \cite{Ahrens:2003ix} and ANTARES \cite{Collaboration:2011nsa}. Because of the difficulty in detecting neutrinos, these bounds are fairly weak. However, neutrino telescopes are sensitive to DM which is captured in the Sun via elastic scattering. Since these particles sink to the Solar centre and annihilate, the neutrino signal (or lack thereof) from $>$ GeV dark matter becomes one of the cleanest (if model-dependent) indirect signals of DM \cite{Barger:2001ur}.
\item Finally, DM annihilation at high-redshift into photons and charged particles change the ionization floor during the post-recombination dark ages \cite{Padmanabhan:2005es}. This extra fraction of free electrons rescatters CMB photons, leading to a suppression in the angular power spectrum at high multipoles, akin to a ``blurring'' of the last scattering surface. Energy injection at lower-redshift (e.g., from DM decay or annihilation in clusters) leads to an increase in correlation on large scales. The polarization of the CMB signal is particularly sensitive to this effect, because Thomson scattering is polarized.
\end{enumerate}

A common issue that plagues indirect searches for DM is the simple fact that astrophysical backgrounds are quite poorly understood. A DM-like signal will inevitably come with a number of plausible astrophysical interpretations.

Given known backgrounds, indirect searches nevertheless provide strong and fairly model-independent constraints on new physics. CMB bounds from Planck \cite{Ade:2015xua}, gamma ray observations of the Milky Way's dwarf satellite galaxies (e.g.~\cite{Fermi-LAT:2016uux}), and low-energy ($\sim$10 GeV) positron observations by AMS-02 \cite{Bergstrom:2013jra} provide some of the strongest limits on WIMP dark matter. Solar neutrino observations provide the best limits on spin-dependent WIMP-nucleus scattering \cite{Choi:2015ara,Aartsen:2016exj}.

\subsection{Collider searches}
\label{sec:Colliders}
If non-gravitational interactions between DM and the Standard Model (SM) exist, particles of DM could be produced in proton-proton collisions at the Large Hadron Collider (LHC)~\cite{1748-0221-3-08-S08001}. The LHC operates at the highest center of mass energy and provides the highest luminosity in current high energy experiments. Therefore, sensitivity to very low production cross-sections of DM particles can be achieved at general purpose experiments like ATLAS~\cite{Aad:2008zzm} and CMS~\cite{Chatrchyan:2008aa}, while sensitivity to certain DM models can be also obtained at specialized experiments like LHCb~\cite{Alves:2008zz} and ALICE~\cite{Aamodt:2008zz}.
As DM particles are not expected to interact with the detector material, the typical signature will have missing transverse energy in the detector. The main backgrounds for the analyses come from limited detector resolution, neutrinos in the final states of SM processes, and non-collision background processes.

Several approaches in DM searches are used at LHC experiments~\cite{Abercrombie:2015wmb}.
The DM particles are not expected to leave a signal in interaction with the material of the detectors, but they can be observed if they are produced in association with a visible SM particle $X(=g,q,\gamma,Z,W,h)$. These are the so-called “mono-X” or $\cancel{\it{E}}_{T}$+$X$ searches, where $\cancel{\it{E}}_{T}$ is the missing transverse energy in the detector.
Another approach in DM searches is to use effective field theories (EFT). They rely on the assumption that production of DM occurs through a contact interaction, involving a quark-antiquark pair (or two gluons) and two DM particles. Kinematics of $\cancel{\it{E}}_{T}$+$X$ models can significantly differ from the contact interaction approach.
EFT assumes a heavy mediator in the interaction of DM and SM particles, but if the mediator is not heavy, models that explicitly include mediators need to be used. They provide an extension to the EFT approach, and use “simplified models”, constructed for specific particles and their interactions. Models using a mediator can predict significantly different signals, where decays back to SM particles are viable.
In this context, due to different kinematics, analyses can be optimized for two types of signatures. The first use the $\cancel{\it{E}}_{T}$+$X$ signature, and can be interpreted using both the EFT and simplified models, and the second use simplified models that probe DM -- SM couplings.
A number of additional physics scenarios account for DM. These include e.g.\ the two Higgs doublet model (2HDM)~\cite{Bauer:2017ota}, or various models in the framework of Supersymmetry (SUSY)~\cite{Golfand:1971iw,Volkov:1973ix,Wess:1974tw,Wess:1974jb,Ferrara:1974pu,Salam:1974ig}.

A few assumptions are made in DM searches at LHC experiments. To ensure that DM particles are produced in p-p collisions, it is assumed that interactions between SM and DM particles exist.
Most of the analyses assume DM to be a weakly interacting massive particle (WIMP), which is stable on collider time scales and does not interact with the detector material.
Typically, minimal flavor violation (MFV) is assumed \cite{DAmbrosio:2002vsn}, which results in the same flavor structure of couplings of DM to ordinary particles as in the SM.
Results of $\cancel{\it{E}}_{T}$+$X$ and simplified models that probe DM -- SM couplings are typically presented using vector and axial-vector mediators; fixed values of mediator couplings to quarks, leptons and DM; mediator width set using the minimal width formula; and the mediator and DM particle masses as free parameters \cite{Boveia:2016mrp}. Other physics scenarios have their results presented as a function of free parameters in the model under study, e.g.\ Higgs branching ratio or SUSY particle masses, for 2HDM and SUSY models respectively.

\subsubsection{Collider signatures}
A wide range of models are tested and dedicated DM searches are performed at ATLAS and CMS. These include the $\cancel{\it{E}}_{T}$ + X and searches using simplified models that probe DM - SM couplings. Since the interactions of DM with SM particles are not known, a number of additional scenarios are considered:
\begin{itemize}
\item In the $\cancel{\it{E}}_{T}$+$X$ search, DM is produced in association with the particle X coming from an Initial State Radiation (ISR) jet, photon, W boson or Z boson. The DM production cross-section scales with quark-X coupling, and the signal is expected as an excess in the tail of the $\cancel{\it{E}}_{T}$ distribution. The analysis typically has a requirement on $\cancel{\it{E}}_{T}$, and a selection for the particle X, and the interpretation is done for different mediator and DM particle mass. The highest cross-section is for gluon ISR, and the highest sensitivity in the mediator and DM particle mass can be achieved using the $\cancel{\it{E}}_{T}$+$\mathrm{jet}$ analysis, compared to $\cancel{\it{E}}_{T}$+$\gamma$, $\cancel{\it{E}}_{T}$+$Z$ and $\cancel{\it{E}}_{T}$+$Z$ searches.
\item In analyses that probe the DM - SM couplings, the mediator can decay back to SM particles. Then the signal appears as a localized excess in the invariant mass distribution of two fermions. Typical searches perform a scan on a di-fermion invariant mass distribution. The search for dijet resonance represents one of the most important analyses due to the high production cross-section and a number of approaches are used, while the dilepton resonance search is well motivated by the clean signature which provides strong constraints for small mediator-lepton couplings. The exclusion is done using different sets of assumptions on the DM, mediator and couplings. For an axial-vector mediator, dijet searches have smaller sensitivity for very low mediator masses, but very high exclusion for high mediator masses, for any mass of DM particles.
\item At LHC energies there is no top quark content in protons, and a mono-top final state is a clear signature of new physics, and represents an important scenario in DM searches.
\item DM can be produced in association with heavy flavor particles, and interesting searches are $\cancel{\it{E}}_{T}$+$t\bar{t}$ and $\cancel{\it{E}}_{T}$+$b$.
\item Since ISR of Higgs bosons is strongly suppressed, models where the Higgs couples to DM represent an interesting scenario. The typical signature has visible decays of the Higgs (e.g.\ $H\rightarrow bb$ or $H\rightarrow\gamma\gamma$) and $\cancel{\it{E}}_{T}$.
\item Invisible Higgs decay occurs for a model where DM couples to the Higgs boson, and the mass of the DM particle is smaller than half of the Higgs boson mass. This gives rise to an `invisible' branching fraction for the Higgs boson.
\item The two Higgs doublet model with a light pseudoscalar mediator which decays to DM produces an enhanced $\cancel{\it{E}}_{T}$+$X$ signature. Due to resonant production of heavy scalar and heavy pseudoscalar Higgses, enhancement occurs in the $\cancel{\it{E}}_{T}$+$Z$ and $\cancel{\it{E}}_{T}$+$h$ channels.
\item Supersymmetry (SUSY) predicts a DM candidate (e.g.\ neutralino, gravitino), and systematic searches are done for different productions of SUSY particles. A typical signature has a long chain of cascade decays of SUSY particles, with the lightest SUSY particle (LSP) at the end of the chain.
\end{itemize}

\subsubsection{Statistical approaches at the LHC}
In LHC searches analyses are optimized to maximize signal and reduce the background in the signal region (SR) selection~\cite{Baak:2014wma}. The SM background is estimated using data-driven techniques, from the control region (CR) selections, designed to be dominated by one type of SM background, and orthogonal to the SRs, where normalization factors for each background are estimated using data.
Statistical interpretation is performed using the frequentist approach, where a hypothesis is tested using statistics only. For an analysis with multiple bins in the discriminating variable distribution, the likelihood for the observed number of events is modeled by the Poisson distribution, which considers the expected and observed number of events in each bin, and nuisance parameters to account for uncertainties in each bin as:
\begin{equation}
L(\mu,\theta)=\prod_{i=1}^{N}P(\mu s_{i}(\theta) + b_{i}(\theta)) \times \prod_{j=1}^{M}P_{j}(\theta)\,,
\end{equation}
where $\mu$ is the signal strength, $s$ is the expected number of signal events, $b$ is the expected number of background events, $\theta$ are the nuisance parameters, N is the number of signal region bins, and M is the number of backgrounds considered.

In order to quantify a possible excess, a local $p$-value is calculated using the profiled log-likelihood ratio test statistic
\begin{equation}
q_{\mu} = -2 \ln \frac{L(\mu,\hat{\hat{\theta}})}{L(\hat{\mu},\hat{\theta})}\,,
\end{equation}
where $L(\mu,\hat{\hat{\theta}})$ maximizes the likelihood for a specific signal strength $\mu$, and $L(\hat{\mu},\hat{\theta})$ is the global maximum likelihood.
For the case of a statistical test for the discovery of a positive signal, a one-sided likelihood for the background hypothesis only ($\mu = 0$) is used:
\begin{equation}
q_0 = \left\{
  \begin{array}{cc}
    -2 \ln \lambda(0) & \hat{\mu} \geq 0\,, \\
    0                   & \hat{\mu} < 0\,,
  \end{array}\right.
\end{equation}
where $\lambda$ is the likelihood ratio \cite{Cowan:2010js}. The test statistic's distribution asymptotically follows a $\chi^{2}$ distribution~\cite{Wilks:1938dza}, and the $p$-value is calculated as an integral for values higher than the observed test statistic $q_{0,obs}$. The significance is calculated using the inverse Gaussian cumulative distribution function ($Z = \Phi^{-1}(1-p)$).
For a case of high significance, the global significance needs to be calculated, as a probability for finding such an excess from statistical fluctuations of the background when looking in a large number of SR bins. As the number of SR bins considered increases, the global significance becomes smaller.

If no significant excess is observed, upper limits on the visible production cross-section are set using the one-sided profile log-likelihood (with the signal strength $\mu$ as a free parameter) using the test statistic:
\begin{equation}
q_{\mu} = \left\{
  \begin{array}{cc}
    -2 \ln \lambda(\mu) & \hat{\mu} \leq \mu, \\
    0                   & \hat{\mu} > \mu.
  \end{array}\right.
\end{equation}
To set exclusion limits, the $CL_{S}$ technique~\cite{0954-3899-28-10-313} is used. It accounts for a small number of expected signal events compared to the number of expected background events. Exclusion limits are calculated using:
\begin{equation}
CL_{s} = \frac{p_{s+b}}{p_{b}}\,,
\end{equation}
where $p_{s+b} = P(q\geq q_{obs}|s+b)$ using $\mu$ = 1, and $p_{b} = P(q \geq q_{obs}|b)$ for $\mu$ = 0. The $p_{b}$ is a conditioning factor to account for the goodness of fit of the background-only hypothesis, designed to prevent downwards fluctuations in the background leading to the unreasonably strong exclusion of signal models. The exclusion limits are typically set at 95\% CL, for each DM model point, and exclusion regions are drawn for $CL_{s} \leq 0.05$.

With recent developments of machine learning (ML) techniques, a number of improvements are being developed in DM searches. Firstly, performance in object reconstruction can be improved using ML which allows for better signal separation. A number of applications are being implemented for e.g.\ lepton reconstruction or b-jet tagging. Secondly, signal separation in the analyses can be improved using ML techniques with a number of new methods being investigated for DM searches, e.g.\ Boosted Decision Trees~\cite{Hastie:2009fk}, Deep-learning Networks~\cite{0483bd9444a348c8b59d54a190839ec9}, Generative Adversarial Networks~\cite{DBLP:journals/corr/GoodfellowPMXWOCB14}, etc. Further details of ML techniques in DM searches can be found in Sec.~\ref{sec:ML}.

Collider searches represent an important avenue in DM searches, as they provide precise constraints on DM masses, for given assumptions on the mediator and couplings. The $\cancel{\it{E}}_{T}$+$jet$ and a dijet resonance searches are expected to have good sensitivity to many DM models using the full Run 3 integrated luminosity~\cite{Chala:2015ama}. If DM is not found, this would represent an important constraint. Current analyses use simplified models for optimization but more complex models of DM interactions need to be considered for the future. Additionally, initial assumptions on the DM particles need to be relaxed, e.g.\ additional scenarios with long-lived DM particles need to be considered to extend the reach of DM searches. Statistical interpretation represents a crucial aspect in quantifying the significance of a potential excess and for setting exclusion limits. In addition, better sensitivity can be achieved through improved object reconstruction and signal separation by using modern machine learning techniques.

\subsection{Gravitational probes and structure formation}
\label{sec:Simulations}

Predictions for the distribution of large scale structure from cold dark matter (CDM) cosmology are in excellent agreement with observations on large scales \cite{Hlozek:2011pc}. On smaller scales, it may be possible to probe the micro-physics of DM by studying the properties of galaxies and comparing with the results of numerical simulations of galaxy formation \cite{Brooks:2014qya}.

For around a decade, a number of discrepancies between the results of numerical simulations and observations of galaxies have been put forward as indicators of physics beyond the standard CDM paradigm. These ``small-scale" problems include the presence of bulge-less disk galaxies \cite{vandenBosch:2001bp}, the core-cusp problem \cite{deBlok:2009sp,Oh:2010mc}, the missing satellites problem \cite{Klypin:1999uc} and the Too-Big-To-Fail problem \cite{BoylanKolchin:2011de,BoylanKolchin:2011dk}. However, many of these issues were first observed in DM-only simulations. The inclusion of gas and stars (and associated feedback mechanisms) in more realistic hydrodynamical simulations has alleviated many of these ``small-scale'' tensions (see e.g.,~\cite{Brook:2010bn,Brook:2011nz,Pontzen:2011ty,Brooks2013,DiCintio:2014xia,BZ2014,Chan2015,Wetzel2016}).

The effects of baryonic feedback must be included in any realistic simulation of galaxy formation, even those which include non-standard dark matter models. These include Self-Interacting dark matter \cite{Vogelsberger2014,Elbert:2014bma,Fry:2015rta,Elbert:2016dbb} (originally invoked to solve the core/cusp problem \cite{spergel00,Loeb2011}) and warm dark matter \cite{Governato:2014gja,Herpich2014,Lovell:2016fec} (which may suppress structure on small scales). Simulations involving even more exotic models, such as ultra-light fuzzy dark matter \cite{Zhang:2016uiy,Zhang:2017chj}, could yield testable predictions but are still in their infancy. In all cases, the complicated (and sometimes poorly understood) sub-grid physics of baryonic feedback can make it difficult to derive strong constraints on the DM properties from galaxy simulations.

Perhaps the most promising probe will be in the properties of ultra faint dwarf galaxies, many more of which will be detected and studied in the LSST-era \cite{Tollerud:2008ze,Tollerud:2010bj,Walsh2009}. In such galaxies, the effects of baryonic feedback tend to be less pronounced, and thus small-scale problems (such as the presence of cored density profiles) in ultra faint dwarfs would be strongly suggestive of new physics beyond the CDM paradigm \cite{Governato:2012fa,Tollet:2015gqa}. Unfortunately, modeling uncertainties in dwarf galaxies means that detecting the presence of large cores may be difficult. In particular, different approaches, assumptions and priors tend to produce different estimates for the density profiles of dwarfs \cite{Walker:2011zu,Bonnivard:2015xpq}, even when the same data set is used. Reconciling these estimates in a statistically meaningful way represents a key challenge for using galaxy formation to probe the properties of DM (see Sec.~\ref{sec:nuisances}).

This challenge has motivated the development of new techniques to probe even smaller DM structures. Due to their small masses and shallow gravitational potentials, these structures are likely to be devoid of stars and gas, and are thus essentially dark \cite{Dooley:2016xkj,Kim:2017iwr}. Techniques to detect these small DM subhalos thus rely on the latter's gravitational influence on their surrounding. These include phase-space perturbations to local stellar streams \cite{Carlberg:2009ae,Carlberg:2012ur,Carlberg:2011xj,Carlberg:2013eya,Carlberg:2013gxa,2014ApJ...788..181N,2016ApJ...820...45C,Erkal:2014tda,Erkal:2015kqa,2016MNRAS.457.3817S,2016MNRAS.463..102E,2017MNRAS.466..628B,2017MNRAS.470...60E,2016PhRvL.116l1301B,Banik:2018pjp}, to the Milky Way disk \cite{2012ApJ...750L..41W,Feldmann:2013hqa}, or to halo stars \cite{Buschmann:2017ams} that could be detected with precise astrometric observations such as those enabled by the Gaia satellite \cite{2016A&A...595A...1G,2018arXiv180409365G}. Other techniques rely on the gravitational lensing signatures of these small, dark subhalos, both in our local neighborhood  \cite{Erickcek:2010fc,Li:2012qha,VanTilburg:2018ykj}, and at cosmological distances from our galaxy (see e.g.~Refs.~\cite{Mao:1998aa,Metcalf:ad,Dalal:2002aa,Koopmans:aa,Vegetti:2008aa,Vegetti:2009aa,Vegetti_2010_1,Vegetti_2010_2,Vegetti_2012,2014MNRAS.442.2017V,Hezaveh:2012ai,Hezaveh_2016_2,Hezaveh_2014,Fadely:2009aa,Daylan:2017kfh,Cyr-Racine:2018htu}). Although promising, these different methods of probing the small-scale DM structure all come with their own statistical challenges, including the delicate balance between allowing for enough model complexity while avoiding over-fitting the noise.

\section{Statistical challenges and approaches}
\label{sec:StatisticalChallenges}
The two important reasons for scientists to turn to statistics can be summarized as 1) discovery and 2) parameter estimation. In this review, we will focus mainly on the first question, that of discovery, since it is obviously the ultimate goal of dark matter searches. The question therefore becomes one of \textit{model comparison}, i.e., answering the question: given available data, does this theory of dark matter do better than the null hypothesis $H_0$?

The statistical approach to such a problem then depends on a number of criteria: whether $H_0$ is the presumed model, or if we are comparing two equally ``plausible'' alternatives (for example, the normal vs inverted neutrino mass hierarchies); whether $H_0$ is a special case of  -- nested within -- the alternative; whether parameters have unknown or meaningless values under $H_0$, and finally whether one adopts a Bayesian or frequentist framework.

The $p$-value is the most commonly used discovery criterion. In this framework, the conclusion of the statistical test for discovery is based on a test statistic $TS$ which has the property that larger values of $TS$ represent stronger evidence against $H_0$ and in favor of $H_1$. The $p$-value has the intuitive definition
\begin{equation}
p = Pr(TS(y) \geq TS(y_{obs}) | H_0),
\end{equation}
or: the probability that the test statistic $TS(y)$ is more extreme than the observed $TS(y_{obs})$ under the null hypothesis. The logarithm of the likelihood ratio
\begin{eqnarray}
\label{eqn:Lratio}
-2 \log \frac{\mathrm{max}_{\theta}P_{H_0}(y|\theta)}{\mathrm{max}_{\theta}P_{H_1}(y|\theta)}
\end{eqnarray}
 is a convenient test statistic, since Wilks' theorem tells us that for nested models, and under certain regularity conditions, its probability distribution follows a chi-squared distribution with degrees of freedom equal to the difference in the dimension of the parameter space under $H_0$ and $H_1$ \cite{wilks}.

More often than not, Wilks' theorem will not be applicable. This is usually due to comparison between non-nested models, or boundary issues where a given parameter has no meaning under $H_0$. In these cases, the ``bootstrap'' method of generating a PDF via Monte Carlo simulation must be used, which can be prohibitively expensive when searching for a 5-sigma effect.

The question that a phenomenologist seeks to answer when performing a statistical analysis is: how likely is the model to be true, given the data. However, $p$-values do not measure the relative likelihood of hypotheses, nor do they accurately reflect $P(\theta|y)$. Instead, they can be viewed as a measure of the ``false alarm'' rate: how often one should expect data this extreme if the null hypothesis is true. However, a large $p$-value does not validate the null hypothesis $H_0$, nor does a small $p$-value, in itself, suggest that $P(H_0|y)$ is small. In this sense, the $p$-value is anti-conservative: typically $p \ll P(H_0|y)$, meaning that interpreting the $p$-value as the probability that the null hypothesis is true may significantly overstate evidence for New Physics.

One way to help reduce this confusion is to report (in addition to the $p$-value) what is referred to as $P(H_0|y)_\mathrm{min}$. Here, $P(H_0|y)_\mathrm{min}$ is a lower bound on the probability of the null hypothesis given the data $P(H_0|y)$: the smallest possible value of $P(H_0|y)$ which can be obtained over a large class of priors. These two summary statistics provide different information about the problem at hand. In addition, reporting both numbers -- $p$ and $P(H_0|y)_\mathrm{min}$ -- highlights to the reader that they are \textit{not the same thing} and encourages a more careful interpretation of the results. We present a toy example of this in Sec.~\ref{sec:ReportBoth}.

On the practical side, it is not always trivial to find the maximum likelihood, especially when the parameter space is complicated and high-dimensional. A number of tools are available for efficiently exploring parameter spaces and calculating likelihoods, posterior probabilities and Bayesian evidences. These include: Markov Chain Monte Carlo samplers such as \texttt{CosmoMC} \cite{Lewis:2002ah} and \texttt{GreAT} \cite{Putze:2014aba}; ensemble samplers such as \texttt{emcee} \cite{2013PASP..125..306F}; nested samplers such as \texttt{MultiNest} \cite{Feroz:2007kg,Feroz:2008xx,Feroz:2013hea} and \texttt{POLYCHORD} \cite{2015MNRAS.453.4384H}; differential evolution samplers such as \texttt{Diver} \cite{Workgroup:2017htr}; and global optimizers such as AMPGO\footnote{See \url{http://infinity77.net/global_optimization/ampgo.html} for an implementation.} \cite{Lasdon2010}. A number of these were compared in Ref.~\cite{Workgroup:2017htr} but in general the best tool will depend (unfortunately) on the particular problem under investigation.\\

The rest of this section is split into two subsections. First we describe the details of a number of novel techniques that were primarily developed to overcome issues in modern DM data analysis. Second, we discuss recent progress in tackling some specific statistical problems in DM searches, as well as a number of opportunities and challenges that still remain in the field.

\subsection{Novel Techniques}
\subsubsection{Selecting between non-nested models}
\label{sec:mix}
How do we perform model selection between wildly different, non-nested models? For example, how do we choose which DM model (axions vs. WIMPs; scalar vs. fermion) is preferred by the data when the parameter spaces differ? In a frequentist framework, Wilks' theorem fails in this setting, and thus comparisons based on the $\chi^2$ approximation of the Likelihood Ratio Test (LRT) become meaningless. In a Bayesian framework, comparing evidences can be misleading, as the comparison between prior volumes becomes arbitrary. For example, the weight of a decade in axion masses versus a decade in WIMP masses are certainly not equivalent to one another. 

\begin{figure}[h]
\includegraphics[width=0.5\textwidth]{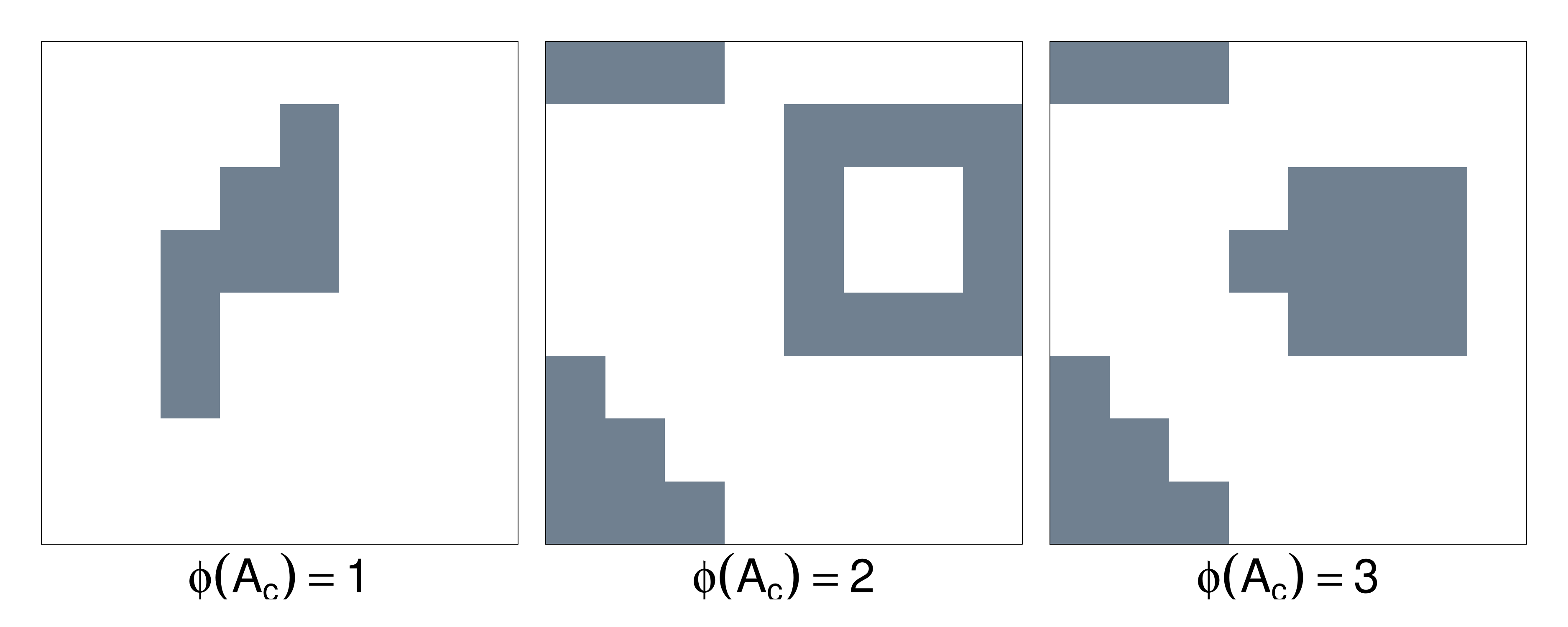}
\caption{From Ref.~\cite{algeri18}. Euler characteristic in two dimensions.\label{DTchar}}
\end{figure}
One proposed solution is to make the models nested by means of a comprehensive model which includes the models to be tested as special cases \citep{algeri16,algeri18}. For instance, let $f(y,\bm{\psi})$ and $g(y,\bm{\theta})$ be the models to be compared with vectors of parameters $\bm{\psi}$ and $\bm{\theta}$ respectively. Consider the mixture model
\begin{equation}
\label{mixture}
(1-\eta)f(y,\bm{\psi}) + \eta g(y,\bm{\theta})
\end{equation}
where $\eta \in [0,1]$. Despite the fact that $\eta$ does not have any physical interpretation (as the data are assumed to be generated by either $f$ or $g$), the asymptotic normality of its Maximum Likelihood Estimate (MLE) allows one to approximate the distribution of the LRT when testing
\begin{equation}
\label{test0}
H_0:\eta=0 \quad \text{vs}\quad \eta>0.
\end{equation}
Specifically, as described in Refs.~\cite{gv10,vg11}, in order to circumvent the problem of non-identifiability of $\bm{\theta}$, one can construct the profile LRT statistic for each value of $\bm{\theta}$ fixed, i.e.,
\begin{equation}
\label{LRT}
LRT(\bm{\theta})=-2\sum_{i=1}^n \log \frac{f(y_i,\hat{\bm{\psi}}_0)}{(1-\hat{\eta}_{\bm{\theta}})f(y_i,\bm{\psi}_{\bm{\theta}}) + \hat{\eta}_{\bm{\theta}} g(y_i,\bm{\theta})}\,,
\end{equation}
where $\hat{\bm{\psi}}_0$ is the MLE of $\bm{\psi}$ under $H_0$, $\hat{\eta}_{\bm{\theta}}$ and $\bm{\psi}_{\bm{\theta}}$ are the MLEs of $\eta$ and $\bm{\psi}$ under $H_1$ with $\bm{\theta}$ fixed. Letting $\bm{\theta}$ vary, $\{LRT(\bm{\theta})\}$ corresponds to a random field\footnote{A random field is a stochastic process where the index is multidimensional.} with index $\bm{\theta}$. A $p$-value for the test in \eqref{test0} is given by $P(\sup_{\bm{\theta}}\{LRT(\bm{\theta})\}>c)$, where $c$ is the maximum of \eqref{LRT} observed over a grid of values for $\bm{\theta}$.
Although the asymptotic distribution of $LRT(\bm{\theta})$ is known to be a 50:50 mixture of $\chi^2_1$ and zero \cite{chernoff}, we can only approximate the asymptotic distribution of the supremum of the random field $\{LRT(\bm{\theta})\}$.

One possible way to do so is to consider the so-called Euler characteristics of the set $\mathcal{A}_c$ of points $\mathcal{A}_c=\{\bm{\theta}\in \bm{\Theta}:LRT(\bm{\theta})>c\}$. We denote with $\phi(\mathcal{A}_c)$ the Euler characteristic of $\mathcal{A}_c$. In two dimensions, $\phi(\mathcal{A}_{c_k})$ corresponds to the numbers of connected components less the number of holes (see Figure \ref{DTchar}). In an arbitrary number of dimensions one can consider a quadrilateral mesh over $\mathcal{A}_c$. In this case, $\phi(\mathcal{A}_c)$ is  computed by adding and subtracting hypercubes of increasing dimensionality (i.e., number of points $-$ number of edges + number of squares $-$ number of cubes + number of 4-dimensional hypercubes etc).
The $p$-value of interest can then be computed as
\begin{equation}
\label{globalp}
P(\sup_{\bm{\theta}}\{LRT(\bm{\theta})\}>c)\approx E[\phi(\mathcal{A}_c)].
\end{equation}

As originally investigated by Ref.~\cite{vg11}, the advantage of referring to the expected Euler characteristic $E[\phi(\mathcal{A}_c)]$ in Eq.~\eqref{globalp} is that it can be estimated via a small Monte Carlo simulation of $\{LRT(\bm{\theta})\}$ under $H_0$ as described below.
Following Ref.~\cite{taylor2003}, we write $E[\phi(\mathcal{A}_c)]$ as
\begin{equation}
\label{lipscitz}
 E[\phi(\mathcal{A}_c)]=\sum^{D}_{d=0}\mathcal{L}_d(\bm{\Theta})\rho_d(c),
\end{equation}
where  $D$ is the dimensionality of $\bm{\theta}$, and the functionals  $\rho_d(c)$, namely the Euler characteristic densities, are  known in the statistical literature and only depend on the  marginal distribution of each component  of $\{LRT(\bm{\theta})\}$, i.e., the above-mentioned 50:50 mixture of $\chi^2_1$ and zero. For instance, if $D=2$, Eq.~\eqref{lipscitz} takes the form
\begin{equation*}
E[\phi(\mathcal{A}_c)]=\frac{c^{\frac{1}{2}}e^{-\frac{c}{2}}}{(2\pi)^{\frac{3}{2}}}\mathcal{L}_2(\bm{\Theta})+\frac{e^{-\frac{c}{2}}}{2\pi}\mathcal{L}_1(\bm{\Theta})+\frac{P(\chi^2_1>c)}{2}\mathcal{L}_0(\bm{\Theta}).
\end{equation*}
(see Ref.~\cite{algeri18} for more details.)
The functionals $\mathcal{L}_d(\bm{\Theta})$ in Eq.~\eqref{lipscitz}   are known as the  Lipschitz-Killing curvatures and their analytical expression for $d>0$ is typically hard to compute in practice. However, this problem can be overcome using the following steps:

\begin{itemize}
\item \textbf{Step 1:} Simulate $y_1,\dots,y_n$ from $f(y,\bm{\psi})$ 100--1000 times via Monte Carlo.
\item \textbf{Step 2:} For each Monte Carlo replicate in Step 1 compute \eqref{LRT} over a grid of values for $\bm{\theta}$.
\item \textbf{Step 3:} Select $c_1,\dots,c_D$ arbitrary small thresholds.
\item \textbf{Step 4:} For each $c_k$, $k=1,\dots,D$, in Step 1 compute $E[\phi(\mathcal{A}_{c_k})]$ over the Monte Carlo simulation obtained in Steps 1-2.
\item \textbf{Step 5:} Obtain the solutions
$\mathcal{L}^*_d(\bm{\Theta})$   of the system of $D$ linear equations
 \vspace{0.2cm}{\fontsize{3mm}{3mm}\selectfont{
\begin{equation*}
\begin{cases}
E[\phi(\mathcal{A}_{c_1})]-{\mathcal{L}}_0(\bm{\Theta})\rho_0(c_1)&=\quad \sum^{D}_{d=1}\mathcal{L}_d(\bm{\Theta})\rho_d(c_1)\\
E[\phi(\mathcal{A}_{c_2})]-{\mathcal{L}}_0(\bm{\Theta})\rho_0(c_2) &=\quad \sum^{D}_{d=1}\mathcal{L}_d(\bm{\Theta})\rho_d(c_2)\\
& \vdots \\
E[\phi(\mathcal{A}_{c_D})]-{\mathcal{L}}_0(\bm{\Theta})\rho_0(c_D)&=\quad  \sum^{D}_{d=1}\mathcal{L}_d(\bm{\Theta})\rho_d(c_D).\\
\end{cases}
\end{equation*}}}
\item \textbf{Step 6:} Compute $E[\phi(\mathcal{A}_c)]$, and consequently $P(\sup_{\bm{\theta}}\{LRT(\bm{\theta})\}>c)$, as
 \begin{equation*}
E[\phi(\mathcal{A}_c)]=\frac{{\mathcal{L}}_0(\bm{\Theta})P(\chi^2_1>c)}{2}+\sum^{D}_{d=1}{\mathcal{L}^*}_d(\bm{\Theta})\rho_d(c),
\end{equation*}
where ${\mathcal{L}}_0$ is the Euler characteristic of $\bm{\Theta}$, (e.g.\ it is one if $\bm{\Theta}$ is a disc, a square, a cube or it is zero if $\bm{\Theta}$ is a circle).
\end{itemize}
A more detailed discussion on the computation of the expected Euler characteristics $E[\phi(\mathcal{A}_{c_k})]$ is given in Ref.\cite{algeri18}. Applications to realistic simulated data from the Fermi LAT are discussed in Refs.~\cite{algeri16, algeri18}.

\subsubsection{Exploiting the count statistics of the signal}
\label{subsec:sig_statistics}
DM detectors are typically counting experiments, i.e.\ looking for signal events above a background.

In the case of DM indirect detection the observed signal is pixelized observed counts, \emph{e.g.}\ as seen in data from the \emph{Fermi}-LAT gamma-ray telescope. In this case the data would be a Poisson realization of the modeled dark matter signal, which could represent emission from large-scale structures such as the smooth Galactic halo, as well as point/extended structures such as dwarf spheroidal galaxies, extragalactic halos and Galactic subhalos.

The associated likelihood is then a product of the Poisson probabilities associated with the observed counts $n_i^{p}$ in each pixel of the region-of-interest:
\begin{equation}
\mathcal{L}(d | {\boldsymbol \theta}) = \prod_p \frac{\mu^{p}({\boldsymbol \theta})^{n^{p}} e^{-\mu^{p}({\boldsymbol \theta})}}{n^{p}!}\,,
\label{eq:pi}
\end{equation}
where $d$ denotes the data, ${\boldsymbol \theta}$ represents the set of model parameters (e.g.\ modeled backgrounds or signal) and $\mu^{p}({\boldsymbol \theta})$ is the number of expected counts in a given pixel and energy bin, usually characterized through spatial templates which model the emission associated with one or more physical process and/or source class.

A particular problem with indirect detection is distinguishing the signal events on top of a complex and uncertain background. One particular background of common interest comes from point sources in the signal region of interest, for example millisecond pulsars, which have non-power-law gamma-ray spectra similar to those expected from annihilation of weak-scale DM. Typically, known point sources are masked or individually modeled in an analysis, but this cannot be done when point sources cannot be detected individually. In this case, the collective emission of dim, sub-threshold point sources could be confused with a diffuse DM \emph{signal}.

In the presence of unresolved sources with unknown positions, detections of multiple photons from the same pixel no longer behave as independent events, as the detection of one photon increases the probability that a source is present in the pixel. The likelihood consequently deviates from the Poissonian form, and is instead characterized by non-Poissonian noise in the data. Following~\cite{Malyshev:2011zi}, the non-Poissonian likelihood can be conveniently cast in the language of probability generating functions, which for a discrete probability distribution with $p_k$ with $k=0,1,2,\ldots$ are defined as $P(t) \equiv \sum_{k=0}^{\infty} p_k t^k$ and allows us to recover the associated probabilities as $p_k = \frac{1}{k!} \left. \frac{d^k P(t)}{dt^k} \right|_{t=0}$. Exploiting the fact that the probability generating function for a sum of independent random variables is simply the product of the respective generating functions, the generating function for a smooth (Poissonian) template (associated with the likelihood in Eq.~\ref{eq:pi}) takes the form
\begin{equation}
P_{\rm P}(t; {\bm \theta}) = \prod_p \text{exp}\left[ \mu_p( {\bm \theta}) (t - 1) \right],
\end{equation}
while for a non-Poissonian template characterizing the distribution of an underlying unresolved point source population this takes the form
\begin{equation}
P_{\rm NP}(t; {\bm \theta}) = \prod_p \exp \left[ \sum_{m=1}^{\infty} x_{p,m}( {\bm \theta}) ( t^m - 1) \right] \,,
\end{equation}
where the $x_{p,m}$ have the interpretation of being the average number of point sources contributing $m$ photon counts within a pixel $p$. Further details on characterizing the non-Poissonian likelihood associated with a point source population, and numerical recipes, may be found in~\cite{Lee:2015fea,Lee:2014mza,Mishra-Sharma:2016gis}.

While DM emission from individual sources has been traditionally studied in the context of Poissonian template fitting~\cite{2010ApJ...717..825D,2012ApJ...761...91A,Hooper:2013rwa,Chang:2018bpt,Lisanti:2017qoz,Lisanti:2017qlb,Fermi-LAT:2016uux}, a comprehensive study taking into account all potential DM emitters (dwarf galaxies, sub- and above-threshold extragalactic halos, sub-threshold subhalos as well as the smooth Galactic halo) would require accurate modeling of the underlying sources and robust characterization of the \mbox{(non-)Poissonian} signal likelihood.

\subsubsection{Euclideanized signals}
\label{sec:euc}

Efficient forecasting of experimental sensitivities is key for developing the most relevant searches for dark matter particles. The sensitivity of future experiments can be quantified in various ways.  This includes the discovery reach, expected exclusion limits, and --- assuming a significant detection has been made --- the ability to discriminate various models and regions in the model parameter space.  Traditionally, the latter is done by defining a number of `benchmark points' in the model parameter space of interest, and studying with simulated mock data how well this scenario -- if realized in nature -- could be constrained with the experiment at hand.

The `Euclideanized signals' approach that was introduced in Ref.~\cite{Edwards:2018lsl} provides a way to study the model discrimination power of future instruments in a fundamentally \emph{benchmark-free} way.  This is achieved by efficient approximation methods for calculating the expected log-likelihood ratios, which in turn allow us to consider a very large number of reference points in the parameter space simultaneously.  Instead of considering, say, 10 benchmark points, one would consider thousands or millions of points, covering the entire parameter space of interest exhaustively.

Euclideanized signals are a mapping of a complicated model parameter space into a (typically high-dimensional) space where statistical distinctness corresponds to the Euclidean distance. Various clustering algorithms allow for the efficient pair-wise comparison and grouping of points according to their Euclidean distance, even for millions of points.  Once the mapping is done, it is then easy to study which parts of the parameter space are in principle distinguishable from other regions.

In Ref.~\cite{Edwards:2017kqw} a mapping of model parameters $\theta \mapsto x$ was defined, which allows one to approximate the log-likelihood ratio with Euclidean distances,
\begin{equation}
  \text{TS}(\vec \theta')_{\mathcal{D}(\vec\theta)} \equiv- 2 \ln
  \frac{\mathcal{L}(\mathcal{D}(\vec\theta)|\vec \theta')}{
  {\displaystyle \max_{\vec\theta''}}\mathcal{L}(\mathcal{D}(\vec\theta)|\vec\theta'')}
  \simeq \lVert \vec x(\vec\theta) - \vec x(\vec\theta') \rVert^2\,.
  \label{eqn:TSeuc}
\end{equation}
This mapping is valid for Poisson likelihoods (which trivially includes also Gaussian likelihoods), with arbitrary signal parameterization, and general background uncertainties modelled as Gaussian random fields. In Ref.~\cite{Edwards:2017kqw} it was shown with randomly generated signal and background models that the approximation technique yields estimates for the log-likelihood ratio that are correct to within 20\% (for up to $5\sigma$ distances).

The Euclideanized signal method makes two analyses computationally possible:
\begin{itemize}
\item \textit{Benchmark-free forecasting:} an exhaustive study of the model-discrimination power of experiments without resorting to a small number of benchmark scenarios,
\item \textit{Signal diversity:} estimating the number of discriminable signals that are predicted by a specific model.
\end{itemize}
We further discuss these in detail below focusing primarily on benchmark free forecasting.

\medskip

\textit{Benchmark-free model comparison ---}
Here we want to calculate, based on the collection of Euclideanized signals, whether two subsets of a global model can exist within a confidence region, with radius $r$, of each other. We define the radius as $r_{\alpha}(\mathcal{M}) = \sqrt{\chi^2_{k=d,\mathrm{ISF}}(1-\alpha)}$ where $\chi^2_{k=d,\mathrm{ISF}}$ is the inverse survival function of the Chi-squared distribution with $k=d$ degrees of freedom such that for $k=2$ and $\alpha = 0.046$ (corresponding to 95.45\% CL) $r_{0.046}(\mathcal{M}) = 2.486$. Here the two subsets of the global model correspond to the distinct models (A and B) we wish to compare and distinguish. Both model A and B are nested within (subsets of) the global model.

If within $r_{0.046}(\mathcal{M})$ of a parameter point from model A there exists a parameter point from B, then the point in A is not discriminable from B. On the other hand, if there exists no point from model B within $r_{0.046}(\mathcal{M})$ then the models can be in principle distinguished at 95\% CL. Here is a step by step guide to performing these calculations with reference to a direct detection (DD) example as presented in Ref.~\cite{Edwards:2018lsl}.

\begin{itemize}
\item \textbf{Step 1:} Sample the parameter space of $\mathcal{M}$, calculating signals for each point. For the problem at hand, we found that one obtains stable results if there are more than around 10 points within every $1\sigma\, (68\% \,\mathrm{CL})$ confidence contour. In the case of DD, the global model $\mathcal{M}$ may correspond to the non-relativistic effective field theory operators $\mathcal{O}_1$ and $\mathcal{O}_4$ (these are simply the usual spin independent and dependent interactions, respectively) \cite{Fitzpatrick:2012ix}. We then have a three-parameter model i.e.~the mass of the dark matter particle and the two individual DM-nucleon couplings to each operator. The sub-models A and B then correspond to two boundaries of $\mathcal{M}$ where one or the other of the DM-nucleon couplings is set to zero.

\item \textbf{Step 2:} Euclideanize the signals using experimental parameters such that each parameter point has an associated new vector $\vec{x}_i$. This step can be done using \texttt{swordfish} \cite{Edwards:2017kqw}.

\item \textbf{Step 3:} For each point $i$ in model A (i.e.~points with model parameters $\vec{\theta}_i$ corresponding to model A), find all points within $r_{0.046}(\mathcal{M})$. We denote the set of model parameters for these neighboring points as $\{ \vec{\theta}^{(i)}_j\}$.

The number of degrees of freedom used to calculate $r_{0.046}(\mathcal{M})$ for model comparison is equal to the difference in the dimensionality of the models of interest. If we consider comparison of the overall model $\mathcal{M}$, which has d degrees of freedom, with the sub-model A with $\mathrm{d'}$ then $\mathrm{k=d-d'}$. For the DD example, $\mathrm{d}=3$ and $\mathrm{d'}=2$ therefore $k=1$ and $r_{0.046}(\mathcal{M}) = 2.0$.

\item \textbf{Step 4:} Each point $i$ is then defined as discriminable or not according to the list of parameter points $\{ \vec{\theta}^{(i)}_j\}$. If $\{ \vec{\theta}^{(i)}_j\}$ contains a point from model B then $i$ is not discriminable and vice versa.
\end{itemize}

\begin{figure}
\includegraphics[width=0.5\textwidth]{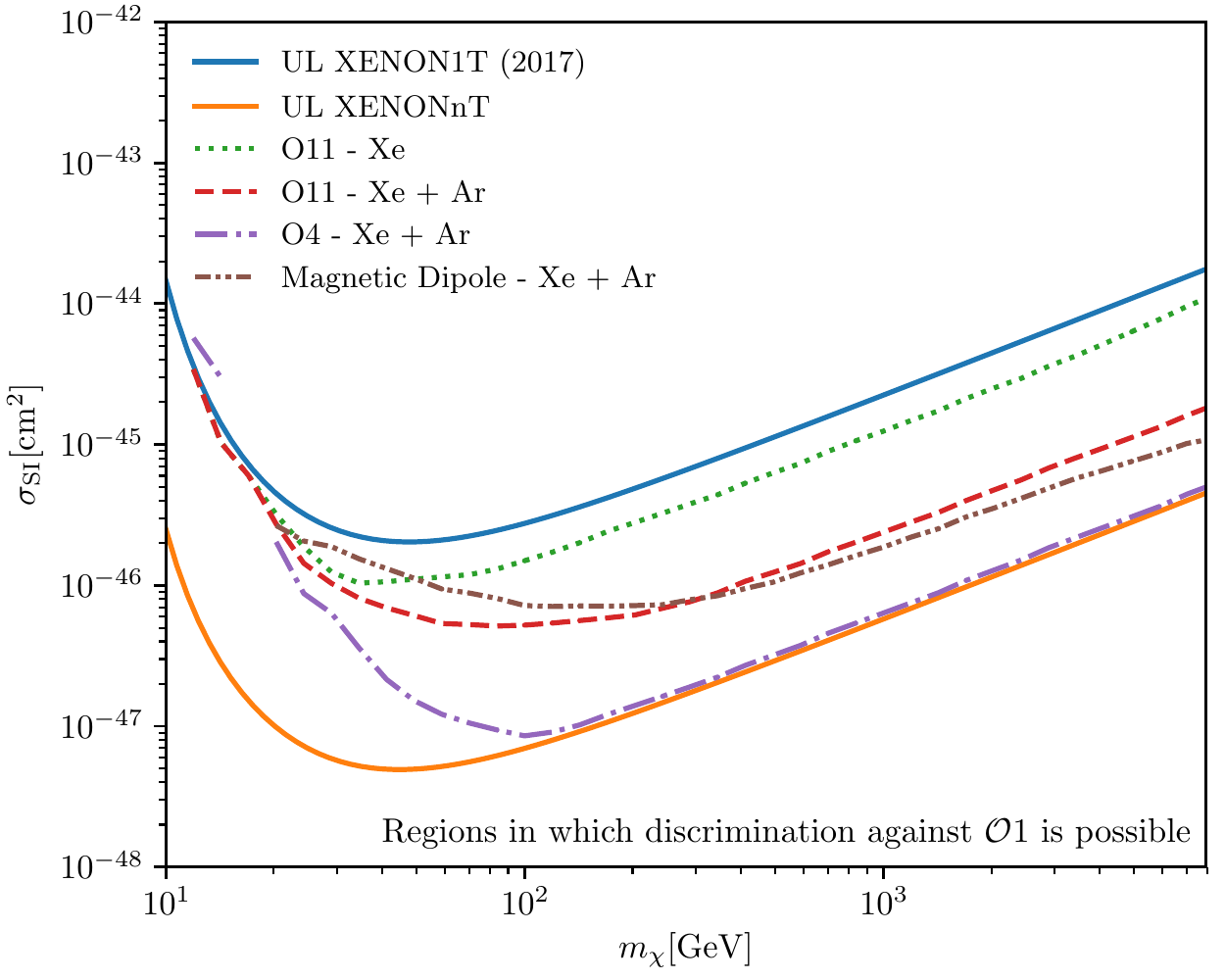}
\caption{Taken from Ref.~\cite{Edwards:2018lsl}.
\textit{Blue and Orange Lines:} 90\% confidence limits (CL) on the standard spin-independent cross section for XENON1T (2017) \cite{Aprile:2017iyp}, and a future experiment with 100 times the exposure. To the left/below each broken line, it is not possible to discriminate an $\mathcal{O}_1$-signal (with the indicated cross-section and DM mass) from the corresponding best-fit $\mathcal{O}_4$, $\mathcal{O}_{11}$ or magnetic dipole signal.  Above/right of each broken line, such a discrimination is possible with at least $2\sigma$ significance. All lines include DM halo uncertainties.}
\label{fig:discrim}
\end{figure}

In this way we are able to make benchmark-free statements about the discriminability of models such as those presented in Ref.~\cite{Edwards:2018lsl} and shown in Fig. \ref{fig:discrim}.

\medskip

\begin{figure}
\includegraphics[width=0.5\textwidth]{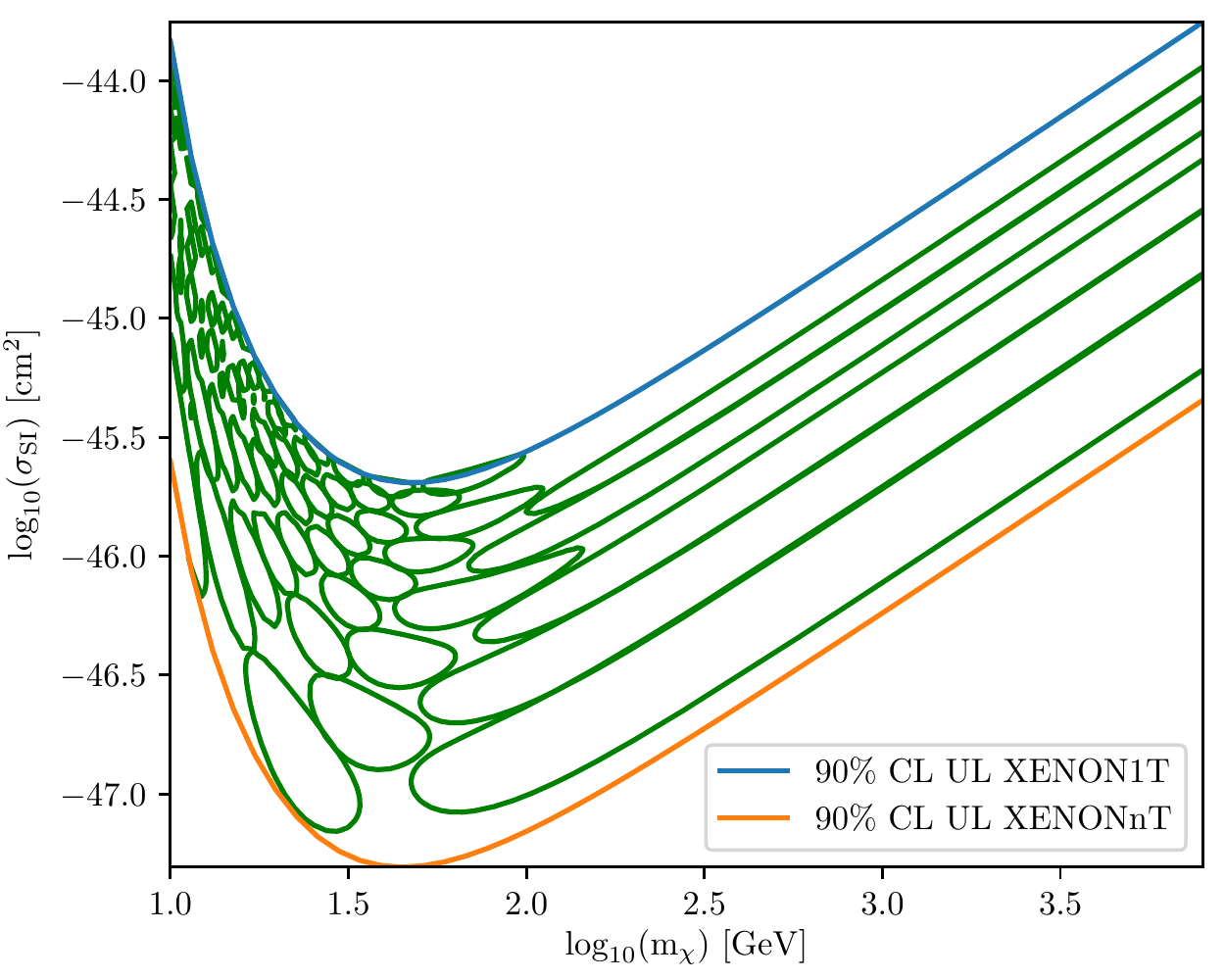}
\caption{Taken from Ref.~\cite{Edwards:2018lsl}.
\textit{Blue and Orange Lines:} 90\% confidence limits (CL) on the standard spin-independent cross section for XENON1T (2017) \cite{Aprile:2017iyp}, and a future experiment with 100 times the exposure.
\textit{Green Ellipses:} $68\%$ confidence contours for tightly packed set of points. Approximately describes the number of discriminable signals in the parameter space between blue and orange lines.}
\label{fig:Tiling}
\end{figure}

\textit{Signal Diversity ---} The number of discriminable models is approximately defined as the number of points one could fit in to a parameter space whilst maintaining $2\sigma$ discrimination between all points. We can visualize these regions by tightly packing confidence contours into the parameter space, as is shown in Fig.~\ref{fig:Tiling} for the case of a typical direct detection experiment. As can be seen in Fig.~\ref{fig:Tiling}, this number is $\mathcal{O}(10-100)$ but calculating $\nu^\alpha_{\mathcal{M},X}$ (defined below) in higher dimensional models with multiple experiments can be difficult. We therefore estimate the quantity using the following method:

\begin{itemize}
\item \textbf{Step 1:} Same as previous step 1
\item \textbf{Step 2:} Same as previous step 2
\item \textbf{Step 3:} Calculate the number of points $w_i$ within $r_{0.046}(\mathcal{M})/2$ of the point $\vec{x}_i$. The number of degrees of freedom is equal to the number of free parameters of the model. Again, for the DD example we have a three parameter model therefore $k=3$ and $r_{0.046}(\mathcal{M}) = 2.833$.
\item \textbf{Step 4:} The volume (here defined)  is then approximated by $\nu_{\mathcal{M}, X}^{0.046} = c_{ff}\sum_i w_i\;$ where $c_{ff}$ is a filling factor dependent on the dimensionality.
\end{itemize}

We can visualise the diversity and global degeneracy breaking abilities of experimental configurations (for nested models) using Infometric Venn diagrams as introduced in Ref.~\cite{Edwards:2018lsl}.

\subsubsection{ABC: when you can't actually afford a likelihood}
The increasing complexity of the models we use to describe physical processes have made the computation of a likelihood often difficult to handle or even impossible. Fortunately many methods have been developed to allow for the forward simulation of these complex situations, for example calculating the end state observables from an LHC event or the distribution of galaxies from a $\Lambda$CDM cosmological simulation.

Approximate Bayesian Computation (ABC) is a computationally-intensive framework for approximating a Bayesian posterior distribution when a likelihood function is not available or intractable \citep{beaumont2002approximate,csillery2010approximate}.  The basic ABC algorithm was introducted in \cite{tavare1997inferring, pritchard1999population}, but was also hinted at conceptually in \cite{rubin1984bayesianly}.  ABC has proved to be a useful tool for several problems in astronomy (e.g.\ \cite{cameron2012approximate, weyant2013likelihood,akeret2015approximate,ishida2015cosmoabc}).

The basic ABC algorithm proceeds simply, with only a few steps.  The overall idea is to use a forward model to generate a simulated dataset given draws from the prior distribution(s).  If that simulated dataset is ``close enough'' to the real observations, then the draws from the prior(s) that produced the simulated dataset are considered ``good'' draws, and those values are retained.  If not, the values are discarded.  This is repeated until enough draws from the prior are accepted, and those values are used to approximate the Bayesian posterior.

More precisely, the steps of the basic ABC algorithm proceed as follows given unknown parameter(s) $\theta$, prior(s) $p(\theta)$, observations $y_{\text{obs}}$, and forward model $F(y \mid \theta)$:\begin{itemize}
\item[(i)] sample $\theta_{\text{prop}} \sim p(\theta)$,
\item[(ii)] compute $y_{\text{prop}} \sim F(y \mid \theta_{\text{prop}})$,
\item[(iii)] if $y_{\text{prop}} = y_{\text{obs}}$, then keep $\theta_{\text{prop}}$, if not, discard $\theta_{\text{prop}}$,
\item[(iv)] repeat until desired number of values of $\theta_{\text{prop}}$ have been accepted.
\end{itemize}

Rather than waiting until $y_{\text{prop}}$ is equal to $y_{\text{obs}}$ exactly, lower-dimensional summary statistics are used.  For example, rather than comparing the full set of observations, one could compare only their sample means.  The summary statistics are crucial for good performance of the ABC algorithm.  The summary statistics will only be as good as the amount of information they contain about the data.  There are some methods for developing summary statistics (e.g.\ \cite{fearnhead2012constructing,blum2013comparative}), but physically-motivated summary statistics can also be effective.  The performance of the summary statistics and distance functions should be checked in a similar scenario where the true posterior is available (which will likely require a simplified model).

In order to define what is meant by ``close enough'', a tolerance (or tolerances) $\epsilon$ is set.  Then for distance function $\Delta$, $\theta_{\text{prop}}$ is accepted if $\Delta(y_{\text{obs}}, y_{\text{prop}}) \leq \epsilon$.  The desire is for $\epsilon$ to be small.

For observations $y_{\text{obs}}$ with summary statistic(s) $S(y_{\text{obs}})$, distance function $\Delta$, (small) tolerance $\epsilon$, and desired  particle sample size $N$,
an ABC posterior can be based on $\{\theta^{(1)}, \theta^{(2)}, \ldots, \theta^{(N)}\} = \{\theta^{(i)}\}_{i = 1}^N$
$\{\theta^{(i)}\}_{i = 1}^N$ from Algorithm~\ref{alg:abc}.

\begin{algorithm}[H]
\begin{algorithmic}[1]
\FOR{$i=1$ to $N$}
\WHILE{$\Delta \left(S(y_{\text{obs}}), S(y_{\text{prop}})\right) > \epsilon$}
\STATE{Propose $\theta_{\text{prop}}$ by drawing $\theta_{\text{prop}}$ from prior $p(\theta)$ }
\STATE{Generate $y_{\text{prop}}$ from forward process $ F(x \mid \theta_{\text{prop}})$}
\STATE{Calculate summary statistics $\{S(y_{\text{obs}}), S(\theta_{\text{prop}})\}$}
\ENDWHILE
\STATE{$\theta^{(i)} \leftarrow \theta_{\text{prop}}$}
\ENDFOR
\end{algorithmic}
\caption{Basic ABC Algorithm }\label{alg:abc}
\end{algorithm}

Selecting a small enough $\epsilon$ can be challenging because setting $\epsilon$ too low leads to a lower acceptance rate and hence (possibly significantly) more computational resources and time.  One popular option for getting around this issue is to use the ABC-Population Monte Carlo (ABC-PMC) algorithm proposed in \cite{beaumont2009adaptive}.  The general idea with ABC-PMC is to set up a sequential version of ABC that uses the accepted particles from the previous iteration as the proposal for the current iteration, rather than drawing from the prior distribution.  This provides improved proposals at each iteration, which is coupled with a shrinking tolerance at iteration $t$, $\epsilon_t$.  The important point is that in order to target the correct posterior distribution (assuming everything else is selected correctly), importance weights need to be computed at each iteration.  Details about the importance weights and the specific steps of the algorithm can be found in \cite{beaumont2009adaptive}. Ref.\ \cite{ishida2015cosmoabc} also has a nice overview of the ABC-PMC with an astronomy application and a Python implementation.  There are other sequential extensions of the ABC algorithm (e.g.\ \cite{bonassi2015sequential, del2012adaptive}).

\subsection{Progress and Challenges}

\subsubsection{Quantifying nuisances}
\label{sec:nuisances}
The signal expected in both direct and indirect searches depends sensitively on quantities of astrophysical nature: for indirect searches the DM distribution within the target (along the line of sight), and for direct searches, the amount of DM in the proximity of the Sun/Earth, as well as its velocity distribution (in principle also important for indirect searches, but generally ignored in simplified scenarios).
The lack of knowledge (i.e.\ uncertainties of statistical and/or systematic nature) on these quantities can be considered as ``nuisances'' for the interpretation of the signal in terms of DM particle properties (mass, self-annihilation, or scattering cross section). 
The local DM density can be extracted from local and global measurements~\cite{Pato:2015dua, Read:2014qva, Catena:2009mf}, while
recent state-of-the-art hydrodynamical simulations of galaxy formation provide  information on the local DM velocity distribution~\cite{Bozorgnia:2016ogo, Kelso:2016qqj, Sloane:2016kyi, Bozorgnia:2017brl}, as well as the DM density profile of Milky Way-like galaxies~\cite{Calore:2015oya}. Each of these astrophysical quantities are estimated with their associated uncertainties, yet little work has been done to properly quantify all known astrophysical uncertainties in the interpretation of DM signals. The impressive precision and refinement of both statistical tools, and searches at colliders, to which direct and indirect searches must be coupled, compel us to cope with these issues by addressing the state of our ignorance, and to properly treat the impact of astrophysical uncertainties within the budget of overall nuisances affecting the signal.

One example of quantifying nuisances is the case of how uncertainties on quantities in our own Galaxy, the Milky Way, affect particle DM constraints: how can we quantify the effect of astrophysical uncertainties in the determination of new particle physics parameters?

The reconstruction of the DM distribution in the entire Galaxy --when obtained through a method based on global properties such as the ``Rotation Curve''-- relies on a host of ancillary measurements and determinations, among which are those related to the motion of the Sun within our own Galaxy (the Local Standard of Rest, and the relative motion between the Sun and the Galactic Centre, in the following generically referred to as ``Galactic Parameters''), and the spatial distribution of the visible component of the Milky Way (stars and interstellar gas). The latter is affected by a statistical uncertainty related to the overall normalization of the stellar mass, and from a (currently) irreducible systematic on the determination of the shape of stellar morphological components (disc-s, and bulge).
The impact of both classes of uncertainties on the determination of the DM profile (expressed in terms of the local DM density, and the inner slope of a generalized NFW profile) has been assessed in recent studies \cite{Iocco:2015xga,Pato:2015dua}, with the conclusion that although none of them can hinder the certainty of the presence of a {\it dark} component of matter in the Galaxy -- even within the solar circle -- the effect of both classes of uncertainties is sizable in the actual determination of the local DM density and its distribution, especially towards the region of the Galactic bulge.
These analyses have compelled a study of how such uncertainties propagate in the determination of DM parameters, in the case of model-specific analysis: in Ref.~\cite{Benito:2016kyp} the authors considered the effect of Galactic uncertainties on two minimal extensions of the Standard Model, in the context of direct and indirect DM searches: the Singlet Scalar (SSDM) and the Inert Doublet (IDM) DM models. The phenomenologies of these two particular models have a simple dependence on a limited set of parameters, which makes them ideal cases for quantifying the effect of Galactic uncertainties on the determination of their parameters.

The following constraints from DM direct and indirect detection in the parameter space of the SSDM and IDM were taken into account: the 2015 LUX exclusion limit on the spin-independent elastic WIMP-nucleon cross section~\cite{Akerib:2015rjg}, and the Fermi-LAT limit on the averaged velocity annihilation cross section from the analysis of dwarf spheroidal galaxies in the Milky Way~\cite{TheFermi-LAT:2015kwa}. The authors also considered the parameter space favored by the DM interpretation of the Galactic Centre GeV excess~\cite{Calore:2014nla}. In particular, it was studied how the LUX exclusion limit and the region favored by the GeV excess vary in the available parameter space of the SSDM and IDM, for three different cases of variation of astrophysical uncertainties: a) the statistical uncertainty on one  ``reference" baryonic morphology, b) the variation of the Galactic parameters for the reference morphology, and c) the baryonic morphologies that maximize or minimize the local DM density.

It was found that for the SSDM case, the statistical uncertainty on the reference morphology has a very small effect on the LUX limit, while the uncertainties on the values of the Galactic parameters or the baryonic morphology have large effects on the LUX constraint. The largest variation in the exclusion limit is due to the variation in the Galactic parameters which leads to the largest variation in the local DM density (varying from $0.055 \pm 0.004$\,GeV\,cm$^{-3}$ to $1.762 \pm 0.017$\,GeV\,cm$^{-3}$; see Table 1 in Ref.~\cite{Benito:2016kyp}). For the region favored by the GeV excess, again the statistical uncertainty on the reference morphology has a minor effect. However, varying the Galactic parameters and considering different baryonic morphologies have large effects on the favored region and can relieve or worsen the tension with the constraints from dwarf spheroidals. The largest shift in the GeV excess region is due to the variation of the baryonic morphologies.

For the IDM case, it was found that the effects of Galactic uncertainties on the LUX exclusion limit are similar to those discussed for the SSDM. The variation of the local DM density causes the largest uncertainty in direct detection limits, and hence the variation in the Galactic parameters for the reference morphology has the largest effect in the IDM parameter space. The analysis also showed that the regions which can simultaneously explain the GeV excess and reproduce the measured DM relic abundance are small and in most cases in tension with the dwarf spheroidal constraints. Varying the Galactic parameters or baryonic morphology shifts the GeV excess region such that the DM relic abundance cannot be reproduced.

In summary, the statistical uncertainties in the observed Milky Way rotation curve and the normalization of the baryonic mass component do not affect the constraints on the parameters of new physics models, while the uncertainties in the Galactic parameters and the baryonic morphology can significantly impact the allowed model parameter space. Quantifying astrophysical uncertainties will be especially important in case a DM signal is discovered in future direct and indirect experiments, and is required to accurately determine  the particle physics nature of DM.

\subsubsection{Global fits: let's just do everything (and worry later about trying to afford it)}

The profusion of experiments hunting for DM in the last decade has left us with an enormous amount of complementary data on its possible identity.  Unfortunately, the heterogeneity of that data makes it difficult to apply to the DM problem in a cohesive, efficient and consistent way.  Different experiments are optimised to look for different types of dark matter, involve entirely different (but sometimes correlated) experimental methods and uncertainties, make different (but related) theoretical assumptions in the analysis of their data, and take different attitudes to sharing their data with the rest of the community.  Most experimental collaborations looking for DM have apparatuses that in principle are sensitive to many different variants, but only have the resources to analyse their own data, in the context of a small number of well-chosen theories.

Experimental constraints on theoretically well-justified theories help theorists to efficiently construct new models, based on some real picture of physical quantities such as masses or couplings.  The converse is also true: theoretical predictions for observable quantities in concrete and complete models help experimentalists to build powerful cuts, choose the right energy range to search, and estimate the exposure required for a statistically-significant measurement.  Ideally, this process should be a circular feedback loop.  This loop has become difficult to implement in modern times,
because of the growing number and complexity of DM models, and the great number of different experiments searching for them.

Global fits in particle and astroparticle physics are the means by which we can use the full treasure-trove of existing data to analyse a broad range of theories, and thereby complete the traditional scientific feedback loop.  Working with the experimental collaborations to define forms of their likelihood functions applicable to a broad range of DM theories (e.g.\ \cite{2010JCAP...01..031S,2012JCAP...11..057S,2016JCAP...04..022A,Collaboration:2242860}) and then combining them into composite likelihood analyses, teams of theorists, phenomenologists and experimentalists have successfully produced broad-ranging analyses of many popular theories for DM.  These include a number of different versions of supersymmetry \cite{2006JHEP...05..002R,2006PhRvD..73a5013A,2007JHEP...08..023A,2008JHEP...12..024T,2009PhRvD..80i5013L,2012PhRvD..85g5012F,2012JHEP...06..098B,2013PhRvD..87k5010K,2013PhRvD..88e5012F,2014PhRvD..89e5017H,2014JHEP...09..081S,2015EPJC...75..422D,2015EPJC...75..500B,2016JCAP...06..050C,2016EPJC...76...96B,GAMBIT_GUT,GAMBIT_MSSM7,2018EPJC...78..256B}, extra dimensions \cite{2011PhRvD..83c6008B}, Higgs portal and other minimal WIMP DM models \cite{2012JCAP...10..042C,2014JCAP...06..030A,2016JHEP...11..070B,2016PhRvD..94f5034M,GAMBIT_SS}, axions \cite{2017arXiv171011138H} and DM effective field theories \cite{2014JHEP...10..155M,2016JHEP...09..077L}.

Care needs to be taken when combining data from different experiments.  In particular, consistent theoretical calculations and assumptions must be applied to all predictions (and even experimental likelihoods) in a global fit, along with the same assumptions about Standard Model parameters, nuclear physics and astrophysical aspects like the density and velocity distributions of DM \cite{darkbit,2017CoPhC.213..252H,2018arXiv180400044A}.  Similarly, different theoretical calculations and experimental analyses must be invoked for different theories, alternative theoretical calculations should be considered for the same model in order to estimate theory errors accurately, sophisticated statistical sampling schemes must be employed for analysing high-dimensional parameter spaces \cite{2010JHEP...04..057A,2011JHEP...06..042F,scannerbit} and coverage properties need to be checked carefully \cite{2011JHEP...03..012B,2011JCAP...07..002A,2012PhRvD..86b3507S}.  Ideally, the global fitting framework itself should automatically ensure that all of these requirements are satisfied. This is something that has only recently become possible \cite{gambit}; future work is focussed on extending this rigorous consistency-checking and automation as far as the Lagrangian level.

Once a new DM model is proposed, an initial choice of priors on its parameters must be made, based on some theoretical constraints or preferences of the researcher. If the likelihoods are strong,
the final fitted result will not be dependent on the chosen prior. However, because only the \textit{Planck} \cite{Planck15cosmo} relic density measurement provides a clear ``signal'' amongst the different DM experiments, the likelihoods for DM searches are
very often too weak to dominate over the impact of the prior on the posterior. Such prior dependence is therefore difficult to avoid until more constraining experimental data become available. On the other hand, it is still not clear what kind of prior distribution -- if any -- is more objective and therefore preferable. At this stage, showing results based on both Bayesian and frequentist statistics is generally considered best practice in global fits, as they provide complementary information about the impacts of priors, fine-tuning and the quality of sampling and fit available in different parts of the parameter space.

A comprehensive global fit requires a lot of CPU time for many likelihoods, such as the simulation of signals at the LHC.  If good sampling is required of the entire likelihood surface, it can be extremely difficult to study collider signatures in models with more than a few degrees of freedom.
Nevertheless, considering DM models in neighbouring parameter regions, and the similarity of their respective likelihood computations, it might not be necessary to perform full signal simulations across entire parameter spaces.  Some simplifications can be used in the global analysis, based on simplified modelling of detector effects \cite{ColliderBit}, the similarity of kinematics, some mathematical tricks, or model configurations (e.g.\ \cite{Tsai:2012cs,2014JHEP...10..155M,Cui:2016ppb,2016PhRvD..94f5034M,Cheng:2016slx,2017CoPhC.213..252H,Liu:2017kmx}). Of course, such simplifications could introduce some systematic uncertainties or limitations on the applicability of the resulting constraints, which must be carefully checked before performing the fit.

Treating different \textit{theories} on the same footing, and comparing them both rigorously and quantitatively, was one of the major topics of discussion at this workshop.  Indeed, this is of particular difficulty and importance for global fits, given their comprehensiveness and the fact that they purport to provide a complete and accurate summary of the current status of the search for different theories of DM, and the identity of DM more broadly. Possible approaches include mixture models (Sec. \ref{sec:mix}), comparison of global $p$-values, the use of Bayes factors, or some extension of the Euclideanized signals approach (Sec.\ \ref{sec:euc}).  These have their own challenges: global $p$-values are notoriously difficult to obtain in complicated high-dimensional parameter spaces, Bayes factors come with attendant prior dependence -- which in general only gets worse for non-nested models -- and application of Euclideanized signals to model selection first requires that the method be developed further.

\subsubsection{Machine Learning in DM Physics}
\label{sec:ML}
Machine Learning (ML) techniques have already been widely adopted throughout the high energy, astro, astro-particle, and particle physics communities. We here briefly comment on some use cases for DM physics and provide useful references. Details are beyond the scope of this work, but we point the interested reader to \url{darkmachines.org} for a community effort to increase the use of ML in DM physics (an associated white paper is in preparation). Also, see Refs.~\cite{Albertsson:2018maf,Guest:2018yhq} for recent publications in high energy physics.

\textit{Direct Detection} --- The simplicity of direct detection experiments and their low background design has made the procedure of data analysis relatively simple. They have therefore been robust to the revolution of ML techniques.
Nevertheless some progress has been made in improving the efficiency of posterior sampling  when considering the large number nuisance parameters associated with galactic halo uncertainties \cite{arinaBayesianAnalysisMultiple2014}. Boosted decision trees have also be used to improve traditional cut and count analyses by optimally and automatically selecting the most promising signal events \cite{Arnaud:2017usi,Tan:2016zwf}.

\textit{Indirect Detection} --- Unlike DD, a large variety of ML techniques have been widely adopted throughout the indirect detection (ID) community. For concreteness we mention three major applications here. Firstly, Neural Networks (NNs) have been used to assess the probability of the galactic center gamma-ray excess \cite{Calore:2014nla} being produced by a population of unresolved point sources \cite{Caron:2017udl}. Unlike the characterization of the likelihood proposed in Sec.~\ref{subsec:sig_statistics}, the approach of Ref. \cite{Caron:2017udl} relies on many simulated realizations of a population of Millisecond Pulsars towards the galactic center. Secondly, the Fermi-LAT has provided the first view into the varied population of gamma ray point sources. Classification of these point sources has proven to be a complicated task typically involving dedicated follow-up studies from telescopes in other wavebands \cite{Schinzel:2017irp}. There are close to 1000 objects in the 3FGL source catalog which have yet to be associated to any source type. For DM searches these point sources are of great interest when looking for low mass sub-halos that would appear as point sources for Fermi-LAT resolution ($<0.15^{\circ}$ above $10\mathrm{\,GeV}$) \cite{Ackermann:2013yma}. Much progress has been made in classifying these unassociated sources using different methods such as random forests and logistic regression \cite{Parkinson:2016oab} with searches for novel source classes such as sub-halos also being performed in Ref.~\cite{Mirabal:2016huj}. Finally, lensing signatures from sub-halos on a variety of scales, as those mentioned in Sec.~\ref{sec:Simulations}, can be sensitive to DM physics. Progress has been made primarily in finding strong lens candidates from the large volume of incoming data, see Ref.~\cite{2017MNRAS.472.1129P} for an example using Convolutional Neural Networks.

\textit{Collider Searches} --- There is an ongoing and dedicated effort to improving the use of machine learning techniques throughout the collider physics community, see \url{https://iml.web.cern.ch} and \url{http://diana-hep.org} for details. Specifically for DM searches, see Ref.~\cite{Bertone:2017adx} where distributed Gaussian processes and NNs were used to increase the speed of likelihood evaluations to a computationally feasible rate for parameter inference.

\subsubsection{The statistical interpretation of fine-tuning}
\label{monsters_III}
A theoretical model presents fine-tuning (or, equivalently, the absence of naturalness) 
when the observable quantities depend critically on fine adjustments of the 
fundamental parameters. For example, in the minimal supersymmetric standard model (MSSM) the Higgs mass, $m_h$, is related to the initial soft mass $m_{H_u}$ and the $\mu-$parameter by $-m_h^2/2\simeq m_{H_u}^2+\mu^2$. Hence, if these initial parameters are ${\cal O}(1)$ TeV, the Higgs mass is fine-tuned by $\sim 1\%$.

Beyond amusing (and unlikely) coincidences, the presence of severe fine-tuning is a warning that the model is implausible in the way it is formulated. The detection of fine-tuning is always interesting because it is telling us something potentially highly non-trivial about the model.
There are three possible attitudes in the presence of fine-tuning:  {\em (i)} discard the model as implausible; {\em (ii)} complete the model, i.e.\ find a reason for the apparently improbable correlations; {\em (iii)} ignore the fine-tuning, i.e.\ assume a fortunate coincidence or, alternatively, hope that someone else will find a reason for the odd correlations, as in attitude {\em (ii)}. 

Fine-tuning is an important but, admittedly, slippery and debatable subject. There are two reasons for that. First, it is not easy to quantify the fine-tuning in a universal, model-independent way (ideally with a sound statistical meaning). Second, once the amount of fine-tuning has been established, it is a subjective matter how much fine-tuning one should accept; after all, coincidences happen. These difficulties (which may seem Bayesian, due to their implicit subjectivity) do not contradict the fact that fine-tuning (or naturalness) is a deep, relevant issue for the structure of a theory.

\begin{figure}
\includegraphics[width=0.45\textwidth]{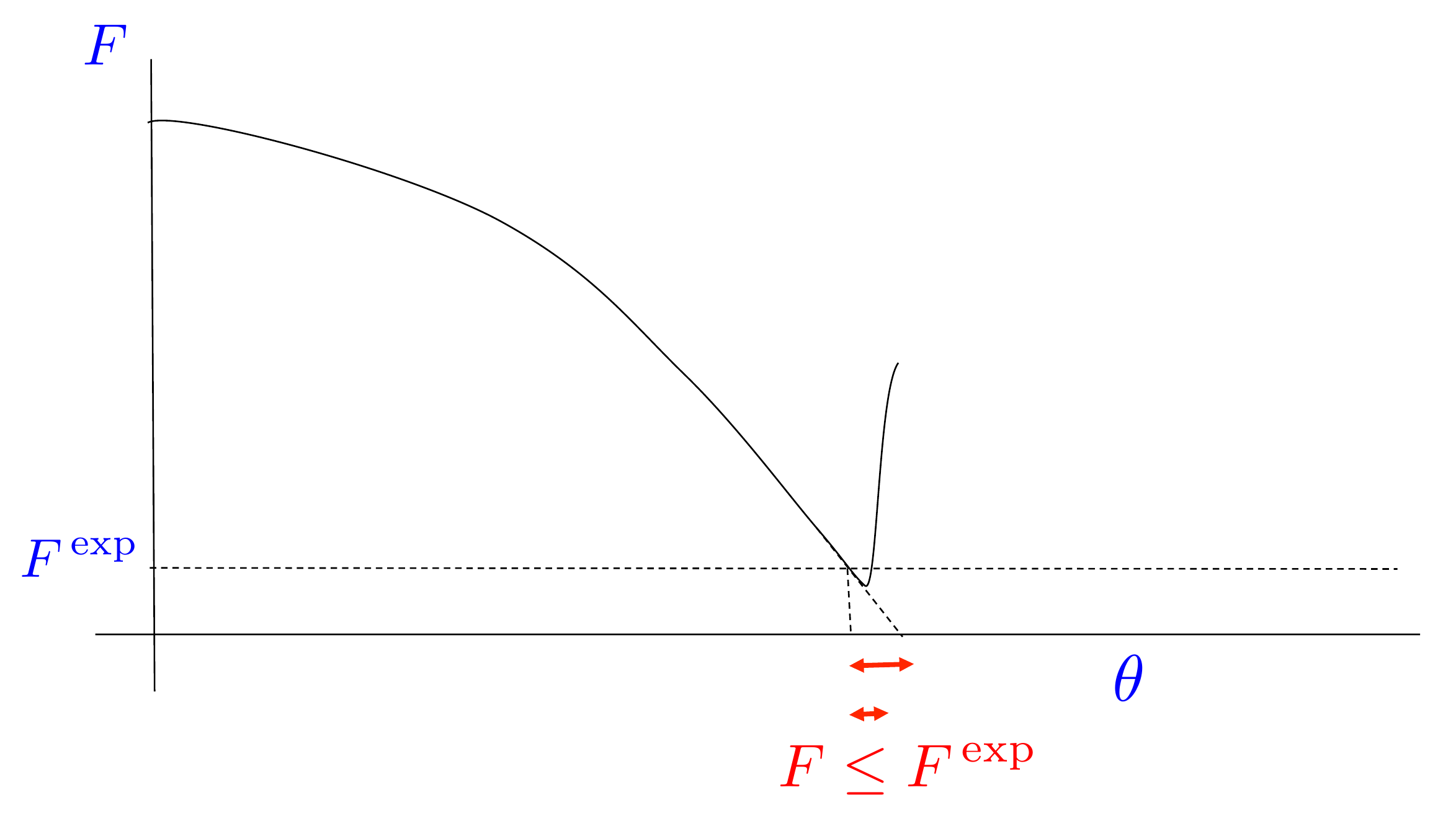}
\caption{An example of a fine-tuned quantity where the standard criterion ($\ref{BG}$) cannot be applied.}
\label{fig:FT}
\end{figure}

Consequently, the first and most important matter is how to quantify fine-tuning. Let $F(\theta_i)$ be the fine-tuned (observable) quantity, where $\theta_i$ are the fundamental (independent) parameters of the theory. Generically, this means that $F$ is very sensitive to small variations of one (or several) $\theta_i$. This has inspired the most popular (and perhaps standard) `measure' of fine-tuning \cite{BARBIERI198863}:
\begin{eqnarray}
\Delta \equiv \max |\Delta_{\theta_i}|, \ \ {\rm with}\ \ 
\Delta_{\theta_i}=\frac{\partial \log F}{\partial \log \theta_i}\ .
\label{BG}
\end{eqnarray}
It is understood that $\Delta\sim 10, 100, \cdots$ amounts to $\sim 10\%, 1 \%, \cdots$ fine-tuning.
While this seems reasonable, it would be nice to find a probabilistic interpretation of (\ref{BG}). Let us call $\theta$ and $\theta_0$ the parameter responsible for the tuning and the value that reproduces the experiment, $F(\theta_0)=F^{\rm exp}$.
Suppose, for the sake of argument, that $F$ has to be fine-tuned to a small value. Then $\theta_0$ should
lie at a small distance, $\delta\theta$, from the value that fully cancels $F$. Assuming that the natural range of $\theta$ is $\sim [0, \theta_0]$ with a flat prior, and that the expansion of $F(\theta)$ at first order captures its behavior in the neighborhood of interest, then it is straightforward that $\Delta^{-1}$ has the statistical meaning of a $p-$value \cite{Barbieri:1998uv, Casas:2014eca,Cabrera:2016wwr}:
\begin{eqnarray}
{P}(F\leq F^{\rm exp})=\frac{\delta \theta}{\theta_0}\simeq \Delta^{-1}\ .
\label{FTpvalue}
\end{eqnarray}
The previous assumptions are reasonable, but may be inappropriate in particular theoretical scenarios. Suppose for instance that the dependence of $F$ on $\theta$ is the one depicted in Fig. \ref{fig:FT}. Clearly, the standard criterion (\ref{BG}) underestimates the real fine-tuning, as it overestimates the actual interval of $\theta$ where $F\leq F^{\rm exp}$. This is not just an academic example. If the dark matter relic density is controlled by annihilation through some funnel (like Higgs or $Z$ funnels), the dependence of $\Omega_{\rm DM}$ on the DM-mass is exactly as in Fig. \ref{fig:FT}. In the borderline case, where the  previous interval tends to zero, the actual fine-tuning tends to infinity, whereas the standard criterion gives $\Delta\rightarrow 0$ ! The general lesson is that, before applying (\ref{BG}) blindly, one should check that the conditions for its validity are met. It is normally more sensible (and easier) to directly apply a $p-$value criterion to the parameter considered, instead of using the approximate expression (\ref{BG}). Examples of this, in the context of supersymmetric mechanisms for DM, can be found in \cite{Casas:2014eca}.
A more sophisticated way of giving an statistical meaning to fine-tuning, in a fully Bayesian spirit, would be the following. Pretend that the fine-tuned quantity, $F(\theta_i)$ has not been measured yet. Then, evaluate the Bayesian probability that $F\leq F^{\rm exp}$.  A similar-in-spirit procedure was applied in \cite{Cabrera:2012vu} to the electroweak fine-tuning of the MSSM.

\subsubsection{Global significance for overlapping signal regions}
When high local significance is observed in LHC searches for new physics, the probability of seeing such an excess only from statistical fluctuations of the background in \textit{any} of the analysis signal regions (SRs) needs to be quantified. This is described by the global significance, which takes into account trial factors, often called the ``look-elsewhere effect'' in physics. For a large number of signal regions, the probability of seeing a ``signal-like'' fluctuation in any one of them is higher. The probability $p_{0}$ of seeing one such high significance excess anywhere in a number of signal regions (assuming the background-only hypothesis $H_0$) is given as:
\begin{equation}
p_{0} = P(q_{x} \geq q_{x,obs}|H_{0})\,,
\end{equation}
for some test statistic $q_x$.
The global significance is obtained from the probability of observing a maximum local significance (across the $N_{\mathrm{SR}}$ signal regions considered) greater than the \textit{observed} maximum local significance. This is typically estimated using a number of pseudo-experiments $N_\mathrm{toy}$. A number of aspects in this calculation are described in~\cite{Lyons:1900zz,Demortier:2007zz,Gross:2010qma} for the case of non-overlapping SRs.
However, if SRs in the analysis have overlap in their discriminating variable selections, the correlations of SR selections need to be taken into account.

Analyses are often designed to use orthogonal selections in the parameter space of discriminating variables, to avoid considering correlations in the overlapping regions. However, there are cases of analyses that require overlapping selections. An example of such an analysis is the SUSY search for $\tilde{q}$ and $\tilde{g}$ production using a selection with two leptons, jets and missing transverse energy~\cite{Aaboud:2016ejt}. The analysis is optimized for $\tilde{q}$ and $\tilde{g}$ decays with jets and two leptons in the decay chain. The signal would produce an excess in the two-lepton invariant mass distribution $m_{ll}$. 
Depending on the mass differences of the SUSY particles, the $m_{ll}$ excess appears at different ranges of the distribution. In addition, different mass spectra produce jets of different transverse momentum, and different masses of the $\tilde{\chi}^{0}_{1}$ LSP give different sizes of missing energy in the signature, which are considered in the selection. Therefore, to account for a large number of SUSY models, SRs are designed with overlap in the selection. Consequently, when the global significance is calculated, the correlations of the overlap of SRs need to be taken into account.

\begin{figure}
\includegraphics[width=0.3\textwidth]{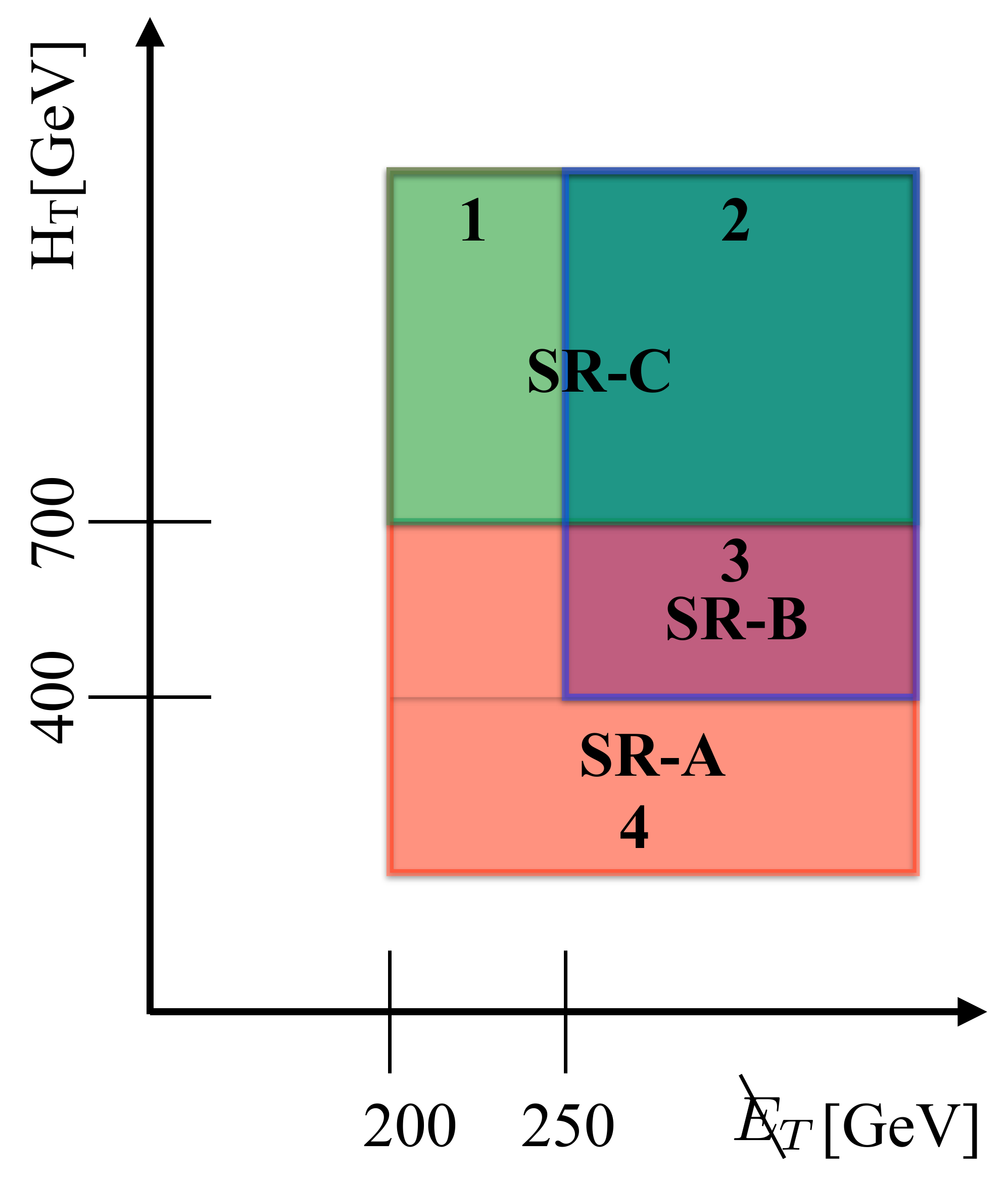}
\caption{Overlapping signal regions SR-A, SR-B and SR-C, in the SUSY analysis for $\tilde{q}$ and $\tilde{g}$ production using a selection with two leptons, jets and missing transverse energy~\cite{Aaboud:2016ejt}. Selections have overlap in the selection of missing transverse energy ($\cancel{\it{E}}_{T}$), scalar sum of transverse momentum of jets ($H_{T}$), and invariant mass of two leptons ($m_{ll}$). Global significance is calculated using orthogonal sub-regions 1,2,3 and 4.}
\label{fig:globalpval}
\end{figure}

Here, we briefly describe a novel technique to take overlapping SRs into account. It makes the assumption that systematic uncertainties for all background components are fully correlated across all SRs. First, the $N_{SR}$ considered signal regions are split into non-overlapping sub-regions. An example for the SUSY analysis with two leptons, jets and missing transverse energy is shown in Figure~\ref{fig:globalpval}. Next, for each non-overlapping sub-region, and for each background component, the yields of events are scaled in the following way:
\begin{itemize}
\item To account for systematic uncertainty, scaling is done by the random value obtained from a Gaussian with unit mean and width equal to the systematic uncertainty,
\item To account for the statistical uncertainty, scaling is done by the random value obtained from a Gaussian with unit mean and width equal to the statistical uncertainty,
\item To produce a pseudo-experiment, a random number from a Poisson distribution is drawn, with mean equal to the yield obtained in the previous steps.
\end{itemize}
Then the newly obtained yields from corresponding bins are summed into SRs. The $p$-value and corresponding significance is evaluated for each SR. The procedure is repeated for a large number of pseudo-experiments. 
The global $p$-value is then calculated as the fraction of pseudo-experiments in which the largest local significance is higher than the observed maximum local significance. This $p$-value is then converted into a one-sided significance.

The calculation of global significance represents a computationally demanding task. As a rule of thumb, a number of pseudo experiments $N_\mathrm{toy}$ is taken of the order of the inverse of the $p$-value, e.g. for a $p$-value of $10^{-4}$, $N_\mathrm{toy}$ is of order $10^{4}$. Calculations corresponding to 3-4 $\sigma$ significance are viable using hundreds of processing units on a modern computing cluster. However, the calculation of significance regions above 5$\sigma$ becomes computationally intractable. When the number of signal regions is $\mathcal{O}(10)$, the effect of the correction for the global significance becomes negligible at high significance. However, certain aspects of using asymptotic formulae need to be considered when the number of signal regions is large. A viable solution for large number of overlapping signal regions and high maximum local significance could be obtained by developing a method using the counting of up-crossings, as described in the global significance calculation using non-overlapping regions of Ref.~\cite{Gross:2010qma}.

\section{Examples and toy models}
\label{sec:examples}

In this section, we present a number of brief examples, based on statistical issues which arose during discussions at the workshop. While these toy examples represent simplified scenarios, it should be possible to extend them straightforwardly to more realistic applications in the field of DM searches.

\subsection{Parameter limits with non-compact support}
\label{sec:noncompact}
Consider the following toy problem: an experimental search for a new signal $s$ with an expected flux $\phi(E)$ as a function of energy $E$ proportional to:
\begin{equation}
\frac{d\phi_s}{dE} = \alpha^2 \frac{E^2}{(E^2-m^2)^2 + \Gamma^2}. \label{eq:newphys}
\end{equation}
This could parametrize some resonant scattering or annihilation process at $E_{res} = m$, with width $\Gamma$ and strength $\alpha$. This search could be performed in the presence of some background event rate:
\begin{equation}
\frac{d\phi_{bg}}{dE} = R_{BG} \left(\frac{E}{E_0}\right)^\gamma,
\end{equation}
governed by unknown nuisance parameters $R_{BG}$, $\gamma$. The total events observed in each energy bin $i$ with width $\Delta E_i$ of such a search would be:
\begin{equation}
N_i = T\Delta E_i\left(\frac{d\phi_s}{dE}(E_i)+\frac{d\phi_{bg}}{dE}(E_i) \right),
\end{equation}
where $T$ represents the \textit{exposure} of the experiment. One can then construct a likelihood $\log \mathcal{L} = \sum_i \log P(N_i)$ and proceed with the usual Bayesian analysis.

In the absence of a signal (i.e., the ``new physics'' rate is lower than the experiment's sensitivity can reach), we would like to place a limit on the (non-background) parameters governing the new physics we have been searching for \eqref{eq:newphys}, via the parameters $\alpha$ and $m$. Our theory does not specify a scale for these parameters, so the correct choice of prior that reflects our understanding of this theory would be log-uniform. Upon inspection of Eq.~\eqref{eq:newphys}, it is clear that in the absence of a clear signal, $\alpha$ can be arbitrarily small, and $m$ can be arbitrarily large. There is thus no well-motivated choice of a prior boundary.
\begin{figure}
\includegraphics[width=0.5\textwidth]{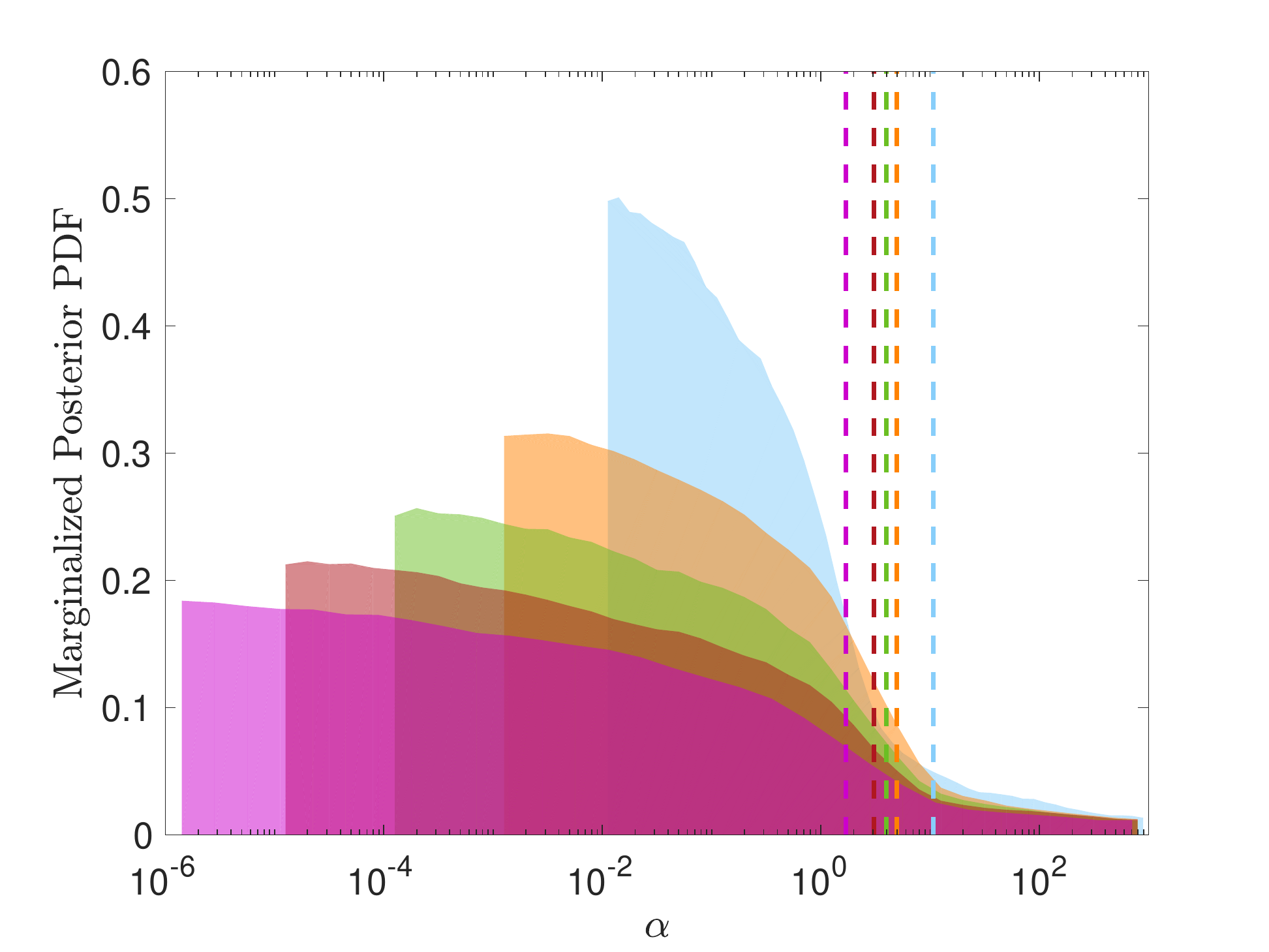}
\caption{Marginalized 1D posterior on the parameter $\alpha$ from \eqref{eq:newphys}, where the true value falls below the experimental sensitivity. Different colors correspond to lower boundaries on the prior from $\alpha_{min} = 10^{-6}$ to $10^{-2}$. Vertical dashed line show the corresponding 95\% CL limit inferred in each case.}
\label{fig:NClimit}
\end{figure}

This leads to the following conundrum: any change, e.g., in the lower prior boundary of $\alpha$ will affect the location of the 95\% credibility boundary, because the latter depends on the total posterior volume, even if the likelihood is completely flat down to $\alpha \rightarrow 10^{-\infty}$. The limit set on the theory in this way is therefore entirely dependent on the arbitrary size of the prior box. This is illustrated in Fig. \ref{fig:NClimit}, where shaded regions represent the marginalized posterior distribution in $\alpha$, and the vertical lines show the limit set in each case. This inherent fuzziness sends many dark matter phenomenologists (who are currently in the business of setting limits) running towards a more frequentist approach such as a profile likelihood where no such ambiguity exists.

\subsection{Combining two experiments}
\label{sec:combining}
Combining the evidence produced from $N$ similar experiments (i.e., ones which require only a few, commonly shared nuisance parameters $\vect{\eta}$) can be a relatively pain-free task with a simple product of likelihood functions
\begin{equation}
\mathcal{L}(d|\vect{\theta},\vect{\eta}) = \prod_{i=1}^N \mathcal{L}_i(d_i|\vect{\theta},\vect{\eta})\,.
\end{equation}
Here, $\vect{\theta}$ are the parameters of interest and $d_i$ are the observed data in each experiment.

Complications can arise however when these experiments have many distinct nuisance parameters which may or may not be governed by some common parameters. The overall likelihood function is then instead
\begin{equation}
\mathcal{L}(d|\vect{\theta},\vect{\eta}) = \prod_{i=1}^N \mathcal{L}_i(d_i|\vect{\theta}_i,\vect{\eta}_i)\,.
\end{equation}
In general, there can be a large number of nuisance parameters $\left\{\vect{\eta}_i\right\}$, which may not be independent and which therefore complicate the issue of specifying priors or joint likelihoods on these nuisance parameters. Furthermore, if we want to calculate (for example) posterior distributions for the parameters of interest, the required integrals over (correlated) $\left\{\vect{\eta}_i\right\}$ are high-dimensional and typically intractable.

We discuss here a Bayesian solution to this problem of combining \textit{dissimilar} experiments. More details can be found in \cite{Hakkila:2013cra,10.2307/4144397}. For clarity we discuss a simple setup, namely two toy single bin counting experiments (which we will refer to as DAMU and LAX) with uncertain background components. We assume that each experiment is contaminated with radioactive Unobtainium, governed by a contamination factor $c \in [0, 1]$. The number of expected background events is then $N_\mathrm{BG} = c R_\mathrm{Un}T$, where $R_\mathrm{Un}$ is the background rate expected from a pure Unobtainium source and $T$ is the exposure time (which we set to 1). From calibration using a purely Unobtainium source we can constrain the expected background rate to be
\begin{equation}
\label{eqn:unobevents}
P_U(R_\mathrm{Un}) = \frac{1}{\sqrt{2 \pi \sigma^2 R_\mathrm{avg}^2}} \exp \left( - \frac{(R_\mathrm{Un}/R_\mathrm{avg} - 1)^2}{2 \sigma^2}\right) \,,
\end{equation}
where $R_\mathrm{avg} = 1000$ events per unit time and $\sigma = 10\%$. Unfortunately our purification procedure is not perfect but the contamination in a given experiment can be constrained to be below $c_\mathrm{max}$. Below this value we assume a uniform prior on $c$, therefore the probability for getting a number of background events is described by
\begin{equation}
P_\mathrm{BG}(N_\mathrm{BG}) = \int_0^{c_\mathrm{max}} P_U\left(\frac{N_\mathrm{BG}}{c}\right)\,\frac{\mathrm{d}c}{c}\, .
\end{equation}
Finally, the signal is simply given by a number of events $N_\mathrm{sig} = \mu$ on which we want to place an informative limit by combining evidence from DAMU and LAX.

Our two toy experiments are now governed by different nuisance parameters $N_\mathrm{BG}^{\mathrm{(LAX)}}$ and $N_\mathrm{BG}^{\mathrm{(DAMU)}}$.

We can deal with this in a Bayesian manner by first identifying the set of parameters common to both experiments - in this case $\mu$ and $R_\mathrm{Un}$. The different contamination factors $c^{\mathrm{(LAX)}}$ and $c^{\mathrm{(DAMU)}}$ are now independent. We can then calculate the adjusted likelihood functions for each experiment $j$:
\begin{equation}
\mathcal{L}_j(\mu, R_\mathrm{Un}) = \int \mathrm{d}N_\mathrm{BG}^{(j)}\,\mathcal{L}(N_\mathrm{obs}^{(j)}|\mu, N_\mathrm{BG}^{(j)})\, P(N_\mathrm{BG}^{(j)}|R_\mathrm{Un})\, ,
\end{equation}
where $P(N_\mathrm{BG}^{(j)}|R_\mathrm{Un})$ simply corresponds to the flat prior on the contamination $c^{(j)}$ up to $c_\mathrm{max}^{(j)}$.
We can then calculate the marginal likelihood, incorporating our prior on $R_\mathrm{Un}$, $P(R_\mathrm{Un}|\mu) = P_U(R_\mathrm{Un})$,
\begin{equation}
\mathcal{L}(\mu) = \int \mathrm{d}R_\mathrm{Un}\, \left\{\prod_j \mathcal{L}_j(\mu, R_\mathrm{Un})\right\} P(R_\mathrm{Un}|\mu)\,.
\end{equation}

\begin{figure}[t!]
\includegraphics[width=0.5\textwidth]{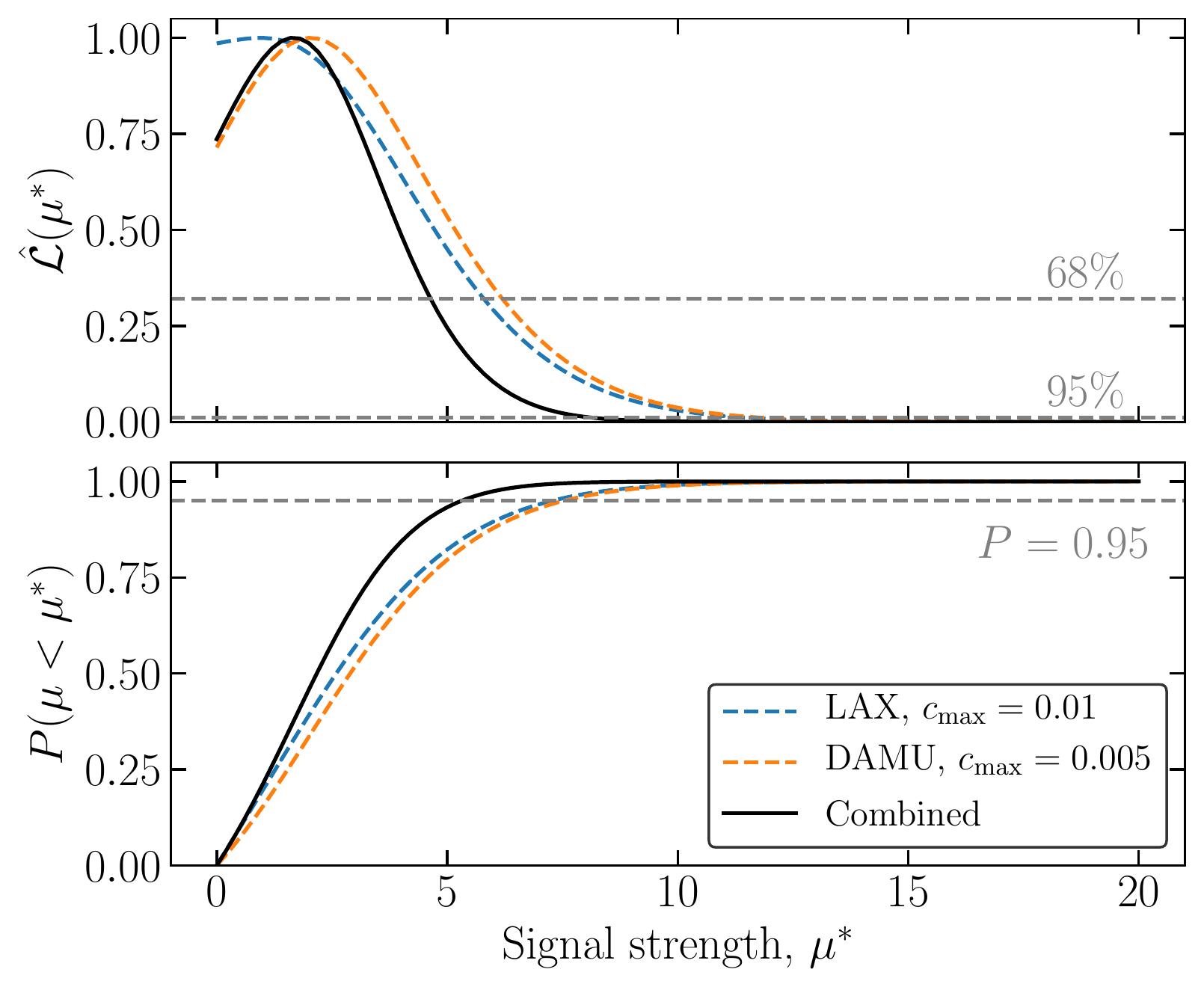}
\caption{Marginal likelihood ratio (top) and cumulative posterior distribution (bottom) for the LAX and DAMU toy experiments separately and combined. The combined curves correctly account for the correlation between nuisance parameters in the two experiments and can be used to easily set limits on the signal strength $\mu$.}
\label{fig:Combining}
\end{figure}

This likelihood function can now be used as usual, for exploring the parameter space, calculating posteriors and setting limits on the signal strength for the combined experiments. In Fig.~\ref{fig:Combining}, we show the marginal likelihood ratio $\hat{\mathcal{L}} = \mathcal{L}/\mathcal{L}_\mathrm{max}$ (top panel) and cumulative posterior distribution (bottom) for LAX and DAMU separately and combined. We assume that both experiments see a total of 4 events, but that the contamination in LAX ($c_\mathrm{max} = 0.01$) is more poorly constrained than in DAMU ($c_\mathrm{max} = 0.005$). The posterior is calculated assuming a flat prior on $\mu$.

As one can see, this parameterization of the problem requires the segmentation of relevant nuisance parameters to different experiments, i.e., $c^{\mathrm{(LAX)}}$ and $c^{\mathrm{(DAMU)}}$ even though they have a common baseline uncertainty set by the probability of events from pure Unobtainium  (Eq.~\ref{eqn:unobevents}). In this way a hierarchical structure can be created for convenient computation of the adjusted likelihood functions per experiment. For more complex situations, this method can make an intractable likelihood calculation possible by reducing the dimensionality of the required integrals to variables associated with each experiment individually. \\

\subsection{Presenting the $p$-value and the probability of the null}
\label{sec:ReportBoth}

As discussed in Sec.~\ref{sec:StatisticalChallenges}, it is important to emphasize the distinction between the $p$-value and the probability of the null hypothesis, given the observed data, $\rm{Pr}(H_0| {\bm y})$. These two numbers quantify different things. Taking the concrete example of looking for a bump-like signal on top of a smooth background (as in the case of the discovery of the Higgs Boson), the $p$-value gives a measure of the false alarm rate -- how often we expect the smooth background to fluctuate upwards to \textit{look} like a signal. Instead, $\rm{Pr}(H_0| {\bm y})$ gives the probability that the background-only hypothesis is correct (given the observed data), compared with the hypothesis that there is a signal. The probability of the null hypothesis will depend on the our choice of priors (for both the background and the signal). However, by considering a wide range of priors, we can determine a lower bound on $\rm{Pr}(H_0| {\bm y})$ and therefore the lowest possible probability that the background-only hypothesis is true.

We demonstrate this idea with a simple toy example of a single bin counting experiment. We assume that the expected number of background counts is known, $N_\mathrm{BG} = 49$. Given a number of observed counts $N_\mathrm{obs}$, we would like to quantify the level of agreement with the null hypothesis $H_0$ (that only background events contribute to the rate) and the alternative $H_1$ (that there is some non-zero signal).

The $p$-value is obtained as the probability of observing data as extreme or more extreme than what is observed, assuming $H_0$. Thus, we have:
\begin{align}
p = \begin{cases}
\sum_{k = N_\mathrm{obs}}^\infty P(k|N_\mathrm{BG}) &\quad \text{ for } N_\mathrm{obs} > N_\mathrm{BG}\,,\\
\frac{1}{2} &\quad \text{ for } N_\mathrm{obs} \leq N_\mathrm{BG}\,.
\end{cases}
\end{align}
Here, $P(k|N_\mathrm{BG})$ is the Poisson probability of observing $k$ events when $N_\mathrm{BG}$ events are expected. We set the $p$-value to $\frac{1}{2}$ for $N_\mathrm{obs} \leq N_\mathrm{BG}$ because an under-fluctuation does not correspond to data which is incompatible with $H_0$ in favor of $H_1$.

The probability of the null hypothesis is obtained using Bayes' theorem \cite{Bayes:1764vd}:
\begin{align}
P(H_0 | N_\mathrm{obs}) = \frac{P(N_\mathrm{obs} | H_0) P(H_0)}{P(N_\mathrm{obs} | H_0) P(H_0) + P(N_\mathrm{obs} | H_1) P(H_1)}\,.
\end{align}
Here, $P(H_0)$ and $P(H_1)$ are the prior probabilities that $H_0$ and $H_1$ respectively are true. In the absence of another well-motivated choice, we will assume $P(H_0) = P(H_1) = \frac{1}{2}$. As before, $P(N_\mathrm{obs} | H_0)$ is simply the Poisson probability of observing $N_\mathrm{obs}$ events, given $N_\mathrm{BG}$ expected background events. Instead, $P(N_\mathrm{obs} | H_1)$ is the probability of observing $N_\mathrm{obs}$ events, integrated over all possible numbers of signal events $N_\mathrm{sig}$:
\begin{align}
P(N_\mathrm{obs}|H_1) = \int P(N_\mathrm{obs}|N_\mathrm{BG} + N_\mathrm{sig}) P_\beta(N_\mathrm{sig})\,\mathrm{d}N_\mathrm{sig}\,.
\end{align}
Here, we write the prior on $N_\mathrm{sig}$ as $P_\beta(N_\mathrm{sig})$, which we parametrize by $\beta$. For concreteness, we will assume an exponential prior on the number of signal events:
\begin{align}
\label{eq:priorNsig}
P_\beta(N_\mathrm{sig}) = \beta \exp \left( - \beta N_\mathrm{sig}\right)\,.
\end{align}

\begin{figure}[t!]
\includegraphics[width=0.5\textwidth]{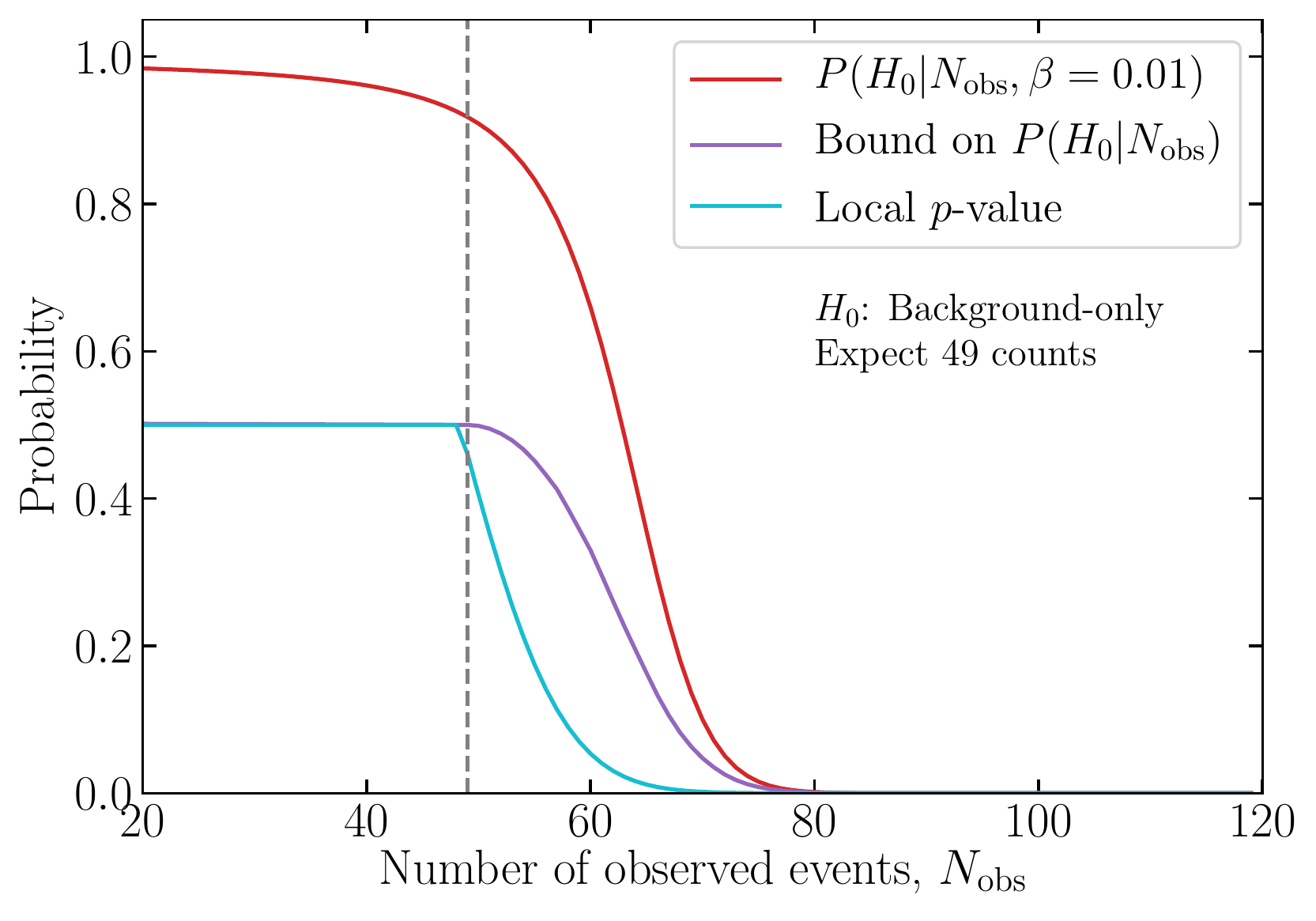}
\caption{Compatibility of an observed number of events  $N_\mathrm{obs}$ with the `background-only' null hypothesis $H_0$. The $p$-value quantifies the false alarm rate while $P(H_0|N_\mathrm{obs})$ quantifies the evidence in favor of $H_0$, as compared with the `signal + background' hypothesis $H_1$. The red line corresponds to the probability of the null hypothesis assuming a particular choice of prior on the number of signal events (see Eq.~\ref{eq:priorNsig}). The purple line shows the lower bound on the probability of the null, over a wide range of priors.}
\label{fig:ReportBoth}
\end{figure}

In Fig.~\ref{fig:ReportBoth}, we show the $p$-value as a function of the number of observed counts in this toy example. We also show $P(H_0|N_\mathrm{obs})$ for one specific prior on the number of signal counts, set by $\beta = 0.01$. Finally, we show the \textit{lower bound} on $P(H_0|N_\mathrm{obs})$ obtained by minimizing over a large class of priors (i.e.~by minimizing over $\beta$). This concrete example highlights what was discussed in Sec.~\ref{sec:StatisticalChallenges}: that the $p$-value is anti-conservative. In this case, when the $p$-value drops to 5\%, the probability of the null hypothesis is always larger than 30\%, even in the case of rather extreme priors. By presenting both the $p$-value and the bound on the probability of the null hypothesis, we give a more detailed picture of the evidence at hand as well as reminding the reader that these two things represent different information about the data.

\section{Conclusions}
The search for new physics has become a complex task, with many moving parts. By now, it seems likely that an eventual detection and identification of dark matter through its particle interactions will take place through careful statistical analysis. We have outlined a number of experimental lines of attack in the hunt for particle DM, along with their specific statistical challenges. \textbf{Direct detection} experiments search for very rare events by creating a signal region that is as background-free as possible. \textbf{Indirect detection} relies on the vast scale of the cosmos, with the drawback of a large and not-so-well understood background from standard model astrophysical processes. Finally, \textbf{collider} searches benefit from a very well-understood background, but the tremendous amount of data lead to a challenge of analysis and trials factors. 

We have outlined a number of statistical challenges that arise in these areas, and some techniques to help overcome them. The detection of new physics amounts to  a challenge of model selection, which comes with a number of pitfalls: the construction of a good test statistic, likelihood function or detection criterion; the choice of physical model parameters and their priors; and the efficient exploration of that parameter space. We have discussed Bayesian methods as well as the dangers of $p$-values and their misinterpretation. We have also presented a number of novel techniques for dealing with complicated and computationally expensive parameter spaces, such signal euclideanization and likelihood-free approaches (i.e.\ ABC). Combining experiments and approaches poses additional challenges, as nuisance parameters may or may not overlap, and systematic uncertainties may have varying effects. Approaches such as hierarchical modeling and global fitting help tackle these daunting issues with finite effort. We have also provided a small number of examples or challenges that help illustrate some of the issues of statistics that astroparticle physicists are confronting.

A key goal of the \textit{DMStat} workshop was to single out specific problems in the search for dark matter which present a statistical challenge. As we have seen, such problems are not hard to find. Thankfully, a wealth of advanced statistical tools have made confronting these challenges feasible. Indeed, we have been able to report significant progress in addressing a number of DM-specific problems, as well as highlighting a number of future challenges and avenues for further study. We are confident that the ongoing cooperation between statistics and DM physics will continue to yield progress and, perhaps some day soon, a discovery.

\acknowledgments
We are grateful to BIRS for hosting the \textit{DMStat} workshop. TE, CW and BJK acknowledge funding from the Netherlands Organization for Scientific Research (NWO) through the VIDI research program ``Probing the Genesis of Dark Matter" (680-47-532). F-YCR acknowledges the support of the National Aeronautical and Space Administration (NASA) ATP grant NNX16AI12G at Harvard University. PS is supported by STFC (ST/K00414X/1, ST/N000838/1, ST/P000762/1). TRS acknowledges support by the Office of High Energy Physics of the U.S. Department of Energy under grant Contract Numbers DE-SC0012567 and DE-SC0013999. ACV is supported by the Canada First Research Excellence Framework (CFREF). FI acknowledges support from the Simons Foundation and FAPESP process 2014/11070-2. JAC acknowledges support from MINECO, Spain, under contract FPA2016-78022-P and Centro de excelencia Severo Ochoa Program under grant SEV-2016-0597. JM acknowledges support by MINECO via the grant FPA2015-65652-C4-1-R and by the Severo Ochoa Excellence Centre Project SEV-2014-0398. SA acknowledges support from the Swedish Research Council through a grant with PI Jan Conrad.

\bibliographystyle{JHEP_pat}
\bibliography{main}

\providecommand{\href}[2]{#2}\begingroup\raggedright\begin{thebibliography}{100}

\bibitem{Bertone:2004pz}
G.~Bertone, D.~Hooper, and J.~Silk, {\it {Particle dark matter: Evidence,
  candidates and constraints}},  {\em Phys. Rept.} {\bf 405} (2005) 279--390,
  [\href{http://arxiv.org/abs/hep-ph/0404175}{{\tt hep-ph/0404175}}].

\bibitem{2009IEEEP..97.1482D}
P.~E. {Dewdney}, P.~J. {Hall}, R.~T. {Schilizzi}, and T.~J.~L.~W. {Lazio}, {\it
  {The Square Kilometre Array}},  {\em IEEE Proceedings} {\bf 97} (2009)
  1482--1496.

\bibitem{2012RPPh...75h6901P}
J.~R. {Pritchard} and A.~{Loeb}, {\it {21 cm cosmology in the 21st century}},
  {\em Reports on Progress in Physics} {\bf 75} (2012) 086901,
  [\href{http://arxiv.org/abs/1109.6012}{{\tt arXiv:1109.6012}}].

\bibitem{Acharya:2017ttl}
Cherenkov Telescope Array Consortium: B.~S. Acharya {\em et.~al.}, {\it
  {Science with the Cherenkov Telescope Array}},
  \href{http://arxiv.org/abs/1709.07997}{{\tt arXiv:1709.07997}}.

\bibitem{Dawson:2015wdb}
K.~S. Dawson {\em et.~al.}, {\it {The SDSS-IV extended Baryon Oscillation
  Spectroscopic Survey: Overview and Early Data}},  {\em Astron. J.} {\bf 151}
  (2016) 44, [\href{http://arxiv.org/abs/1508.04473}{{\tt arXiv:1508.04473}}].

\bibitem{Aghamousa:2016zmz}
DESI: A.~Aghamousa {\em et.~al.}, {\it {The DESI Experiment Part I:
  Science,Targeting, and Survey Design}},
  \href{http://arxiv.org/abs/1611.00036}{{\tt arXiv:1611.00036}}.

\bibitem{Abell:2009aa}
LSST Science, LSST Project: P.~A. Abell {\em et.~al.}, {\it {LSST Science Book,
  Version 2.0}},  \href{http://arxiv.org/abs/0912.0201}{{\tt arXiv:0912.0201}}.

\bibitem{Kim:2017iwr}
S.~Y. Kim, A.~H.~G. Peter, and J.~R. Hargis, {\it {There is No Missing
  Satellites Problem}},  \href{http://arxiv.org/abs/1711.06267}{{\tt
  arXiv:1711.06267}}.

\bibitem{Oguri:2010ns}
M.~Oguri and P.~J. Marshall, {\it {Gravitationally lensed quasars and
  supernovae in future wide-field optical imaging surveys}},  {\em Mon. Not.
  Roy. Astron. Soc.} {\bf 405} (2010) 2579--2593,
  [\href{http://arxiv.org/abs/1001.2037}{{\tt arXiv:1001.2037}}].

\bibitem{2016A&A...595A...1G}
{Gaia Collaboration}, T.~{Prusti}, {\em et.~al.}, {\it {The Gaia mission}},
  {\em \aap} {\bf 595} (2016) A1, [\href{http://arxiv.org/abs/1609.04153}{{\tt
  arXiv:1609.04153}}].

\bibitem{2018arXiv180409365G}
{Gaia Collaboration}, A.~G.~A. {Brown}, {\em et.~al.}, {\it {Gaia Data Release
  2. Summary of the contents and survey properties}},  {\em ArXiv e-prints}
  (2018) [\href{http://arxiv.org/abs/1804.09365}{{\tt arXiv:1804.09365}}].

\bibitem{TheIceCube-Gen2:2016cap}
IceCube: M.~G. Aartsen {\em et.~al.}, {\it {PINGU: A Vision for Neutrino and
  Particle Physics at the South Pole}},  {\em J. Phys.} {\bf G44} (2017)
  054006, [\href{http://arxiv.org/abs/1607.02671}{{\tt arXiv:1607.02671}}].

\bibitem{2017APS..APR.J9003A}
E.~{Aprile} and {Xenon Collaboration}, {\it {The XENONnT Dark Matter
  Experiment}},  in {\em APS April Meeting Abstracts} (2017) J9.003.

\bibitem{Akerib:2018lyp}
LUX-ZEPLIN: D.~S. Akerib {\em et.~al.}, {\it {Projected WIMP Sensitivity of the
  LUX-ZEPLIN (LZ) Dark Matter Experiment}},
  \href{http://arxiv.org/abs/1802.06039}{{\tt arXiv:1802.06039}}.

\bibitem{Du:2018uak}
ADMX: N.~Du {\em et.~al.}, {\it {A Search for Invisible Axion Dark Matter with
  the Axion Dark Matter Experiment}},  {\em Phys. Rev. Lett.} {\bf 120} (2018)
  151301, [\href{http://arxiv.org/abs/1804.05750}{{\tt arXiv:1804.05750}}].

\bibitem{Rubin:1980zd}
V.~C. Rubin, N.~Thonnard, and W.~K. Ford, Jr., {\it {Rotational properties of
  21 SC galaxies with a large range of luminosities and radii, from NGC 4605 /R
  = 4kpc/ to UGC 2885 /R = 122 kpc/}},  {\em Astrophys. J.} {\bf 238} (1980)
  471.

\bibitem{Read:2014qva}
J.~I. Read, {\it {The Local Dark Matter Density}},  {\em J. Phys.} {\bf G41}
  (2014) 063101, [\href{http://arxiv.org/abs/1404.1938}{{\tt
  arXiv:1404.1938}}].

\bibitem{Sivertsson:2017rkp}
S.~Sivertsson, H.~Silverwood, J.~I. Read, G.~Bertone, and P.~Steger, {\it {The
  Local Dark Matter Density from SDSS-SEGUE G-dwarfs}},  {\em Mon. Not. Roy.
  Astron. Soc.} (2017) [\href{http://arxiv.org/abs/1708.07836}{{\tt
  arXiv:1708.07836}}].

\bibitem{1933AcHPh...6..110Z}
F.~{Zwicky}, {\it {Die Rotverschiebung von extragalaktischen Nebeln}},  {\em
  Helvetica Physica Acta} {\bf 6} (1933) 110--127.

\bibitem{Clowe:2006eq}
D.~Clowe, M.~Bradac, {\em et.~al.}, {\it {A direct empirical proof of the
  existence of dark matter}},  {\em Astrophys. J.} {\bf 648} (2006) L109--L113,
  [\href{http://arxiv.org/abs/astro-ph/0608407}{{\tt astro-ph/0608407}}].

\bibitem{Parker:2005fh}
L.~C. Parker, M.~J. Hudson, and R.~G. Carlberg, {\it {Mass-to-light ratios of
  galaxy groups from weak lensing}},  {\em Astrophys. J.} {\bf 634} (2005)
  806--812, [\href{http://arxiv.org/abs/astro-ph/0508328}{{\tt
  astro-ph/0508328}}].

\bibitem{Hoekstra:2013via}
H.~Hoekstra, M.~Bartelmann, {\em et.~al.}, {\it {Masses of galaxy clusters from
  gravitational lensing}},  {\em Space Sci. Rev.} {\bf 177} (2013) 75--118,
  [\href{http://arxiv.org/abs/1303.3274}{{\tt arXiv:1303.3274}}].

\bibitem{Velander:2013jga}
M.~Velander {\em et.~al.}, {\it {CFHTLenS: The relation between galaxy dark
  matter haloes and baryons from weak gravitational lensing}},  {\em Mon. Not.
  Roy. Astron. Soc.} {\bf 437} (2014) 2111--2136,
  [\href{http://arxiv.org/abs/1304.4265}{{\tt arXiv:1304.4265}}].

\bibitem{Coil:2012vw}
A.~L. Coil, {\it {Large Scale Structure of the Universe}},
  \href{http://arxiv.org/abs/1202.6633}{{\tt arXiv:1202.6633}}.

\bibitem{Vogelsberger:2014dza}
M.~Vogelsberger, S.~Genel, {\em et.~al.}, {\it {Introducing the Illustris
  Project: Simulating the coevolution of dark and visible matter in the
  Universe}},  {\em Mon. Not. Roy. Astron. Soc.} {\bf 444} (2014) 1518--1547,
  [\href{http://arxiv.org/abs/1405.2921}{{\tt arXiv:1405.2921}}].

\bibitem{Ade:2015xua}
Planck: P.~A.~R. Ade {\em et.~al.}, {\it {Planck 2015 results. XIII.
  Cosmological parameters}},  {\em Astron. Astrophys.} {\bf 594} (2016) A13,
  [\href{http://arxiv.org/abs/1502.01589}{{\tt arXiv:1502.01589}}].

\bibitem{McDermott:2010pa}
S.~D. McDermott, H.-B. Yu, and K.~M. Zurek, {\it {Turning off the Lights: How
  Dark is Dark Matter?}},  {\em Phys. Rev.} {\bf D83} (2011) 063509,
  [\href{http://arxiv.org/abs/1011.2907}{{\tt arXiv:1011.2907}}].

\bibitem{Taoso:2007qk}
M.~Taoso, G.~Bertone, and A.~Masiero, {\it {Dark Matter Candidates: A Ten-Point
  Test}},  {\em JCAP} {\bf 0803} (2008) 022,
  [\href{http://arxiv.org/abs/0711.4996}{{\tt arXiv:0711.4996}}].

\bibitem{Jungman:1995df}
G.~Jungman, M.~Kamionkowski, and K.~Griest, {\it {Supersymmetric dark matter}},
   {\em Phys. Rept.} {\bf 267} (1996) 195--373,
  [\href{http://arxiv.org/abs/hep-ph/9506380}{{\tt hep-ph/9506380}}].

\bibitem{Roszkowski:2017nbc}
L.~Roszkowski, E.~M. Sessolo, and S.~Trojanowski, {\it {WIMP dark matter
  candidates and searches - current issues and future prospects}},  {\em Rept.
  Prog. Phys.} {\bf 81} (2018) 066201,
  [\href{http://arxiv.org/abs/1707.06277}{{\tt arXiv:1707.06277}}].

\bibitem{Peccei:1977hh}
R.~D. Peccei and H.~R. Quinn, {\it {CP Conservation in the Presence of
  Instantons}},  {\em Phys. Rev. Lett.} {\bf 38} (1977) 1440--1443.
  [,328(1977)].

\bibitem{Peccei:2006as}
R.~D. Peccei, {\it {The Strong CP problem and axions}},  {\em Lect. Notes
  Phys.} {\bf 741} (2008) 3--17,
  [\href{http://arxiv.org/abs/hep-ph/0607268}{{\tt hep-ph/0607268}}].
  [,3(2006)].

\bibitem{Carr:2016drx}
B.~Carr, F.~Kuhnel, and M.~Sandstad, {\it {Primordial Black Holes as Dark
  Matter}},  {\em Phys. Rev.} {\bf D94} (2016) 083504,
  [\href{http://arxiv.org/abs/1607.06077}{{\tt arXiv:1607.06077}}].

\bibitem{Merle:2017jfn}
A.~Merle, {\it {keV sterile neutrino Dark Matter}},  {\em PoS} {\bf NOW2016}
  (2017) 082, [\href{http://arxiv.org/abs/1702.08430}{{\tt arXiv:1702.08430}}].

\bibitem{Hochberg:2014dra}
Y.~Hochberg, E.~Kuflik, T.~Volansky, and J.~G. Wacker, {\it {Mechanism for
  Thermal Relic Dark Matter of Strongly Interacting Massive Particles}},  {\em
  Phys. Rev. Lett.} {\bf 113} (2014) 171301,
  [\href{http://arxiv.org/abs/1402.5143}{{\tt arXiv:1402.5143}}].

\bibitem{Spergel:1999mh}
D.~N. Spergel and P.~J. Steinhardt, {\it {Observational evidence for
  selfinteracting cold dark matter}},  {\em Phys. Rev. Lett.} {\bf 84} (2000)
  3760--3763, [\href{http://arxiv.org/abs/astro-ph/9909386}{{\tt
  astro-ph/9909386}}].

\bibitem{Tulin:2017ara}
S.~Tulin and H.-B. Yu, {\it {Dark Matter Self-interactions and Small Scale
  Structure}},  {\em Phys. Rept.} {\bf 730} (2018) 1--57,
  [\href{http://arxiv.org/abs/1705.02358}{{\tt arXiv:1705.02358}}].

\bibitem{Allanach:2006jc}
B.~C. Allanach, {\it {Naturalness priors and fits to the constrained minimal
  supersymmetric standard model}},  {\em Phys. Lett.} {\bf B635} (2006)
  123--130, [\href{http://arxiv.org/abs/hep-ph/0601089}{{\tt hep-ph/0601089}}].

\bibitem{Berger:2008cq}
C.~F. Berger, J.~S. Gainer, J.~L. Hewett, and T.~G. Rizzo, {\it {Supersymmetry
  Without Prejudice}},  {\em JHEP} {\bf 02} (2009) 023,
  [\href{http://arxiv.org/abs/0812.0980}{{\tt arXiv:0812.0980}}].

\bibitem{Cabrera:2016wwr}
M.~E. Cabrera, J.~A. Casas, A.~Delgado, S.~Robles, and R.~Ruiz~de Austri, {\it
  {Naturalness of MSSM dark matter}},  {\em JHEP} {\bf 08} (2016) 058,
  [\href{http://arxiv.org/abs/1604.02102}{{\tt arXiv:1604.02102}}].

\bibitem{Piffl:2013mla}
T.~Piffl {\em et.~al.}, {\it {The RAVE survey: the Galactic escape speed and
  the mass of the Milky Way}},  {\em Astron. Astrophys.} {\bf 562} (2014) A91,
  [\href{http://arxiv.org/abs/1309.4293}{{\tt arXiv:1309.4293}}].

\bibitem{Essig:2011nj}
R.~Essig, J.~Mardon, and T.~Volansky, {\it {Direct Detection of Sub-GeV Dark
  Matter}},  {\em Phys. Rev.} {\bf D85} (2012) 076007,
  [\href{http://arxiv.org/abs/1108.5383}{{\tt arXiv:1108.5383}}].

\bibitem{Essig:2017kqs}
R.~Essig, T.~Volansky, and T.-T. Yu, {\it {New Constraints and Prospects for
  sub-GeV Dark Matter Scattering off Electrons in Xenon}},  {\em Phys. Rev.}
  {\bf D96} (2017) 043017, [\href{http://arxiv.org/abs/1703.00910}{{\tt
  arXiv:1703.00910}}].

\bibitem{Freese:2012xd}
K.~Freese, M.~Lisanti, and C.~Savage, {\it {Colloquium: Annual modulation of
  dark matter}},  {\em Rev. Mod. Phys.} {\bf 85} (2013) 1561--1581,
  [\href{http://arxiv.org/abs/1209.3339}{{\tt arXiv:1209.3339}}].

\bibitem{Mayet:2016zxu}
F.~Mayet {\em et.~al.}, {\it {A review of the discovery reach of directional
  Dark Matter detection}},  {\em Phys. Rept.} {\bf 627} (2016) 1--49,
  [\href{http://arxiv.org/abs/1602.03781}{{\tt arXiv:1602.03781}}].

\bibitem{Bernabei:2008yh}
DAMA: R.~Bernabei {\em et.~al.}, {\it {The DAMA/LIBRA apparatus}},  {\em Nucl.
  Instrum. Meth.} {\bf A592} (2008) 297--315,
  [\href{http://arxiv.org/abs/0804.2738}{{\tt arXiv:0804.2738}}].

\bibitem{Bernabei:2013xsa}
R.~Bernabei {\em et.~al.}, {\it {Final model independent result of
  DAMA/LIBRA-phase1}},  {\em Eur. Phys. J.} {\bf C73} (2013) 2648,
  [\href{http://arxiv.org/abs/1308.5109}{{\tt arXiv:1308.5109}}].

\bibitem{Bernabei:2018yyw}
R.~Bernabei {\em et.~al.}, {\it {First model independent results from
  DAMA/LIBRA-phase2}},  \href{http://arxiv.org/abs/1805.10486}{{\tt
  arXiv:1805.10486}}.

\bibitem{Froborg:2016ova}
SABRE: F.~Froborg, {\it {SABRE: WIMP modulation detection in the northern and
  southern hemisphere}},  {\em J. Phys. Conf. Ser.} {\bf 718} (2016) 042021,
  [\href{http://arxiv.org/abs/1601.05307}{{\tt arXiv:1601.05307}}].

\bibitem{Adhikari:2017esn}
G.~Adhikari {\em et.~al.}, {\it {Initial Performance of the COSINE-100
  Experiment}},  {\em Eur. Phys. J.} {\bf C78} (2018) 107,
  [\href{http://arxiv.org/abs/1710.05299}{{\tt arXiv:1710.05299}}].

\bibitem{2012PhLB..711..264B}
J.~{Barreto}, H.~{Cease}, {\em et.~al.}, {\it {Direct search for low mass dark
  matter particles with CCDs}},  {\em Physics Letters B} {\bf 711} (2012)
  264--269, [\href{http://arxiv.org/abs/1105.5191}{{\tt arXiv:1105.5191}}].

\bibitem{deMelloNeto:2015mca}
DAMIC: J.~R.~T. de~Mello~Neto {\em et.~al.}, {\it {The DAMIC dark matter
  experiment}},  {\em PoS} {\bf ICRC2015} (2016) 1221,
  [\href{http://arxiv.org/abs/1510.02126}{{\tt arXiv:1510.02126}}].

\bibitem{Aalseth:2012if}
CoGeNT: C.~E. Aalseth {\em et.~al.}, {\it {CoGeNT: A Search for Low-Mass Dark
  Matter using p-type Point Contact Germanium Detectors}},  {\em Phys. Rev.}
  {\bf D88} (2013) 012002, [\href{http://arxiv.org/abs/1208.5737}{{\tt
  arXiv:1208.5737}}].

\bibitem{Aalseth:2014arxiv}
CoGeNT: C.~E. Aalseth {\em et.~al.}, {\it {Maximum Likelihood Signal Extraction
  Method Applied to 3.4 years of CoGeNT Data}},
  \href{http://arxiv.org/abs/1401.6234}{{\tt arXiv:1401.6234}}.

\bibitem{Behnke:2012ys}
COUPP: E.~Behnke {\em et.~al.}, {\it {First Dark Matter Search Results from a
  4-kg CF$_3$I Bubble Chamber Operated in a Deep Underground Site}},  {\em
  Phys. Rev.} {\bf D86} (2012) 052001,
  [\href{http://arxiv.org/abs/1204.3094}{{\tt arXiv:1204.3094}}]. [Erratum:
  Phys. Rev.D90,no.7,079902(2014)].

\bibitem{Archambault:2012pm}
PICASSO: S.~Archambault {\em et.~al.}, {\it {Constraints on Low-Mass WIMP
  Interactions on $^{19}F$ from PICASSO}},  {\em Phys. Lett.} {\bf B711} (2012)
  153--161, [\href{http://arxiv.org/abs/1202.1240}{{\tt arXiv:1202.1240}}].

\bibitem{Amole:2017dex}
PICO: C.~Amole {\em et.~al.}, {\it {Dark Matter Search Results from the PICO-60
  C$_3$F$_8$ Bubble Chamber}},  {\em Phys. Rev. Lett.} {\bf 118} (2017) 251301,
  [\href{http://arxiv.org/abs/1702.07666}{{\tt arXiv:1702.07666}}].

\bibitem{Agnese:2016cpb}
SuperCDMS: R.~Agnese {\em et.~al.}, {\it {Projected Sensitivity of the
  SuperCDMS SNOLAB experiment}},  {\em Phys. Rev.} {\bf D95} (2017) 082002,
  [\href{http://arxiv.org/abs/1610.00006}{{\tt arXiv:1610.00006}}].

\bibitem{Agnese:2017njq}
SuperCDMS: R.~Agnese {\em et.~al.}, {\it {Results from the Super Cryogenic Dark
  Matter Search Experiment at Soudan}},  {\em Phys. Rev. Lett.} {\bf 120}
  (2018) 061802, [\href{http://arxiv.org/abs/1708.08869}{{\tt
  arXiv:1708.08869}}].

\bibitem{Agnese:2018col}
SuperCDMS: R.~Agnese {\em et.~al.}, {\it {First Dark Matter Constraints from
  SuperCDMS Single-Charge Sensitive Detectors}},  {\em Submitted to: Phys. Rev.
  Lett.} (2018) [\href{http://arxiv.org/abs/1804.10697}{{\tt
  arXiv:1804.10697}}].

\bibitem{Agnese:2017jvy}
SuperCDMS: R.~Agnese {\em et.~al.}, {\it {Low-mass dark matter search with
  CDMSlite}},  {\em Phys. Rev.} {\bf D97} (2018) 022002,
  [\href{http://arxiv.org/abs/1707.01632}{{\tt arXiv:1707.01632}}].

\bibitem{Hehn:2014bya}
EDELWEISS: L.~Hehn, {\it {The EDELWEISS-III Dark Matter Search: Status and
  Perspectives}},  in {\em {Proceedings, 20th International Conference on
  Particles and Nuclei (PANIC 14): Hamburg, Germany, August 24-29, 2014}}
  (2014) 378--381.

\bibitem{Hehn:2016nll}
EDELWEISS: L.~Hehn {\em et.~al.}, {\it {Improved EDELWEISS-III sensitivity for
  low-mass WIMPs using a profile likelihood approach}},  {\em Eur. Phys. J.}
  {\bf C76} (2016) 548, [\href{http://arxiv.org/abs/1607.03367}{{\tt
  arXiv:1607.03367}}].

\bibitem{Angloher:2015ewa}
CRESST: G.~Angloher {\em et.~al.}, {\it {Results on light dark matter particles
  with a low-threshold CRESST-II detector}},  {\em Eur. Phys. J.} {\bf C76}
  (2016) 25, [\href{http://arxiv.org/abs/1509.01515}{{\tt arXiv:1509.01515}}].

\bibitem{Petricca:2017zdp}
CRESST: F.~Petricca {\em et.~al.}, {\it {First results on low-mass dark matter
  from the CRESST-III experiment}},  in {\em {15th International Conference on
  Topics in Astroparticle and Underground Physics (TAUP 2017) Sudbury, Ontario,
  Canada, July 24-28, 2017}} (2017)
  [\href{http://arxiv.org/abs/1711.07692}{{\tt arXiv:1711.07692}}].

\bibitem{Angloher:2016ooq}
G.~Angloher {\em et.~al.}, {\it {The COSINUS project - perspectives of a NaI
  scintillating calorimeter for dark matter search}},  {\em Eur. Phys. J.} {\bf
  C76} (2016) 441, [\href{http://arxiv.org/abs/1603.02214}{{\tt
  arXiv:1603.02214}}].

\bibitem{Arnaud:2017bjh}
NEWS-G: Q.~Arnaud {\em et.~al.}, {\it {First results from the NEWS-G direct
  dark matter search experiment at the LSM}},  {\em Astropart. Phys.} {\bf 97}
  (2018) 54--62, [\href{http://arxiv.org/abs/1706.04934}{{\tt
  arXiv:1706.04934}}].

\bibitem{Battat:2016xxe}
DRIFT: J.~B.~R. Battat {\em et.~al.}, {\it {Low Threshold Results and Limits
  from the DRIFT Directional Dark Matter Detector}},  {\em Astropart. Phys.}
  {\bf 91} (2017) 65--74, [\href{http://arxiv.org/abs/1701.00171}{{\tt
  arXiv:1701.00171}}].

\bibitem{Couturier:2016blf}
C.~Couturier {\em et.~al.}, {\it {Directional detection of Dark Matter with the
  MIcro-tpc MAtrix of Chambers}},  in {\em {Proceedings, 51st Rencontres de
  Moriond, Cosmology session: La Thuile, Italy, March 19-26, 2016}} (2016)
  165--170, [\href{http://arxiv.org/abs/1607.08765}{{\tt arXiv:1607.08765}}].

\bibitem{Leyton:2016nit}
DMTPC: M.~Leyton, {\it {Directional dark matter detection with the DMTPC m$^3$
  experiment}},  {\em J. Phys. Conf. Ser.} {\bf 718} (2016) 042035.

\bibitem{Nakamura:2015tna}
K.~Nakamura {\em et.~al.}, {\it {NEWAGE - Direction-sensitive Dark Matter
  Search Experiment}},  {\em Phys. Procedia} {\bf 61} (2015) 737--741.

\bibitem{Akerib:2016vxi}
LUX: D.~S. Akerib {\em et.~al.}, {\it {Results from a search for dark matter in
  the complete LUX exposure}},  {\em Phys. Rev. Lett.} {\bf 118} (2017) 021303,
  [\href{http://arxiv.org/abs/1608.07648}{{\tt arXiv:1608.07648}}].

\bibitem{Aprile:2017iyp}
XENON: E.~Aprile {\em et.~al.}, {\it {First Dark Matter Search Results from the
  XENON1T Experiment}},  {\em Phys. Rev. Lett.} {\bf 119} (2017) 181301,
  [\href{http://arxiv.org/abs/1705.06655}{{\tt arXiv:1705.06655}}].

\bibitem{Aprile:2018dbl}
XENON: E.~Aprile {\em et.~al.}, {\it {Dark Matter Search Results from a One
  Tonne$\times$Year Exposure of XENON1T}},
  \href{http://arxiv.org/abs/1805.12562}{{\tt arXiv:1805.12562}}.

\bibitem{Cui:2017nnn}
PandaX-II: X.~Cui {\em et.~al.}, {\it {Dark Matter Results From 54-Ton-Day
  Exposure of PandaX-II Experiment}},  {\em Phys. Rev. Lett.} {\bf 119} (2017)
  181302, [\href{http://arxiv.org/abs/1708.06917}{{\tt arXiv:1708.06917}}].

\bibitem{Agnes:2014bvk}
DarkSide: P.~Agnes {\em et.~al.}, {\it {First Results from the DarkSide-50 Dark
  Matter Experiment at Laboratori Nazionali del Gran Sasso}},  {\em Phys.
  Lett.} {\bf B743} (2015) 456--466,
  [\href{http://arxiv.org/abs/1410.0653}{{\tt arXiv:1410.0653}}].

\bibitem{Aalseth:2017fik}
C.~E. Aalseth {\em et.~al.}, {\it {DarkSide-20k: A 20 tonne two-phase LAr TPC
  for direct dark matter detection at LNGS}},  {\em Eur. Phys. J. Plus} {\bf
  133} (2018) 131, [\href{http://arxiv.org/abs/1707.08145}{{\tt
  arXiv:1707.08145}}].

\bibitem{Amaudruz:2017ekt}
DEAP-3600: P.~A. Amaudruz {\em et.~al.}, {\it {First results from the DEAP-3600
  dark matter search with argon at SNOLAB}},
  \href{http://arxiv.org/abs/1707.08042}{{\tt arXiv:1707.08042}}.

\bibitem{LZ:sensitivity}
LUX-ZEPLIN: D.~A. et~al., {\it {Projected {WIMP} sensitivity of the
  {LUX-ZEPLIN} ({LZ}) dark matter experiment}},
  \href{http://arxiv.org/abs/1802.06039}{{\tt arXiv:1802.06039}}.

\bibitem{XENONnT:sensitivity}
XENON: E.~et~al., {\it {Physics reach of the {XENON1T} dark matter
  experiment}},  {\em JCAP} {\bf 04} (2016) 027,
  [\href{http://arxiv.org/abs/1512.07501}{{\tt arXiv:1512.07501}}].

\bibitem{DS:sensitivity}
DarkSide-20k: C.~A. et~al., {\it {{DarkSide}-20k: A 20 Tonne Two-Phase {LAr
  TPC} for Direct Dark Matter Detection at {LNGS}}},
  \href{http://arxiv.org/abs/1707.08145}{{\tt arXiv:1707.08145}}.

\bibitem{PICO500:sensitivity}
PICO: C.~A. et~al., {\it {{PICO-500L}: Simulations for a 500L bubble chamber
  for dark matter search}},  {\em Proceedings of the XV International
  Conference on Topics in Astroparticle and Underground Physics, {TAUP2017}}
  (2018).

\bibitem{SuperCDMS:sensitivity}
SuperCDMS: R.~A. et~al., {\it {Projected Sensitivity of the {SuperCDMS SNOLAB}
  experiment}},  {\em Phys. Rev. D} {\bf 95} (2017) 082002,
  [\href{http://arxiv.org/abs/1601.00006}{{\tt arXiv:1601.00006}}].

\bibitem{Yellin:2002xd}
S.~Yellin, {\it {Finding an upper limit in the presence of unknown
  background}},  {\em Phys. Rev.} {\bf D66} (2002) 032005,
  [\href{http://arxiv.org/abs/physics/0203002}{{\tt physics/0203002}}].

\bibitem{Yellin:2008da}
S.~Yellin, {\it {Extending the optimum interval method}},
  \href{http://arxiv.org/abs/0709.2701}{{\tt arXiv:0709.2701}}.

\bibitem{Yellin:2011xf}
S.~Yellin, {\it {Some ways of combining optimum interval upper limits}},
  \href{http://arxiv.org/abs/1105.2928}{{\tt arXiv:1105.2928}}.

\bibitem{Gaskins:2016cha}
J.~M. Gaskins, {\it {A review of indirect searches for particle dark matter}},
  {\em Contemp. Phys.} {\bf 57} (2016) 496--525,
  [\href{http://arxiv.org/abs/1604.00014}{{\tt arXiv:1604.00014}}].

\bibitem{PAMELA}
PAMELA: S.~Orsi, {\it {PAMELA: A payload for antimatter matter exploration and
  light nuclei astrophysics}},  {\em Nucl. Instrum. Meth.} {\bf A580} (2007)
  880--883.

\bibitem{PAMELA2}
PAMELA: P.~Carlson, {\it {PAMELA science}},  {\em Int. J. Mod. Phys.} {\bf A20}
  (2005) 6731--6734.

\bibitem{Aguilar:2013qda}
AMS: M.~Aguilar {\em et.~al.}, {\it {First Result from the Alpha Magnetic
  Spectrometer on the International Space Station: Precision Measurement of the
  Positron Fraction in Primary Cosmic Rays of 0.5–350 GeV}},  {\em Phys. Rev.
  Lett.} {\bf 110} (2013) 141102.

\bibitem{ATIC4TALK}
ATIC: T.~G. Guzik, {\it {Talk given at 2009 APS April Meeting, May 2-5}}, .

\bibitem{ATIC2005}
ATIC: J.~Chang {\em et.~al.}, {\it {The Electron Spectrum above 20 GeV Measured
  by ATIC-2}}, . Prepared for 29th International Cosmic Ray Conferences (ICRC
  2005), Pune, India, 31 Aug 03 - 10 2005.

\bibitem{Gruber:1999yr}
D.~E. Gruber, J.~L. Matteson, L.~E. Peterson, and G.~V. Jung, {\it {The
  spectrum of diffuse cosmic hard x-rays measured with heao-1}},  {\em
  Astrophys. J.} {\bf 520} (1999) 124,
  [\href{http://arxiv.org/abs/astro-ph/9903492}{{\tt astro-ph/9903492}}].

\bibitem{Boyle:2007wi}
P.~J. Boyle, {\it {Cosmic ray composition at high energies: Results from the
  TRACER project}},  in {\em {36th COSPAR Scientific Assembly Beijing, China,
  July 16-23, 2006}} (2007) [\href{http://arxiv.org/abs/astro-ph/0703707}{{\tt
  astro-ph/0703707}}].

\bibitem{2011ApJ...728..122Y}
Y.~S. {Yoon}, H.~S. {Ahn}, {\em et.~al.}, {\it {Cosmic-ray Proton and Helium
  Spectra from the First CREAM Flight}},  {\em \apj} {\bf 728} (2011) 122,
  [\href{http://arxiv.org/abs/1102.2575}{{\tt arXiv:1102.2575}}].

\bibitem{2009ApJ...697.1071A}
W.~B. {Atwood}, A.~A. {Abdo}, {\em et.~al.}, {\it {The Large Area Telescope on
  the Fermi Gamma-Ray Space Telescope Mission}},  {\em \apj} {\bf 697} (2009)
  1071--1102, [\href{http://arxiv.org/abs/0902.1089}{{\tt arXiv:0902.1089}}].

\bibitem{TheDAMPE:2017dtc}
DAMPE: J.~Chang {\em et.~al.}, {\it {The DArk Matter Particle Explorer
  mission}},  {\em Astropart. Phys.} {\bf 95} (2017) 6--24,
  [\href{http://arxiv.org/abs/1706.08453}{{\tt arXiv:1706.08453}}].

\bibitem{Vedrenne:2003}
G.~{Vedrenne}, J.-P. {Roques}, {\em et.~al.}, {\it {SPI: The spectrometer
  aboard INTEGRAL}},  {\em A \& A} {\bf 411} (2003) L63--L70.

\bibitem{Weisskopf:2000tx}
M.~C. Weisskopf, H.~D. Tananbaum, L.~P. van Speybroeck, and S.~L. O'Dell, {\it
  {Chandra x-ray observatory (cxo):overview}},  {\em Proc. SPIE Int. Soc. Opt.
  Eng.} {\bf 4012} (2000) 2, [\href{http://arxiv.org/abs/astro-ph/0004127}{{\tt
  astro-ph/0004127}}].

\bibitem{MAGIC}
MAGIC: D.~Elsaesser and K.~Mannheim, {\it {MAGIC and the search for signatures
  of supersymmetric dark matter}},  {\em New Astron. Rev.} {\bf 49} (2005)
  297--301, [\href{http://arxiv.org/abs/astro-ph/0409563}{{\tt
  astro-ph/0409563}}].

\bibitem{Weekes:2001pd}
T.~C. Weekes {\em et.~al.}, {\it {VERITAS: The Very energetic radiation imaging
  telescope array system}},  {\em Astropart. Phys.} {\bf 17} (2002) 221--243,
  [\href{http://arxiv.org/abs/astro-ph/0108478}{{\tt astro-ph/0108478}}].

\bibitem{Hinton:2004eu}
HESS: J.~A. Hinton, {\it {The status of the HESS project}},  {\em New Astron.
  Rev.} {\bf 48} (2004) 331--337,
  [\href{http://arxiv.org/abs/astro-ph/0403052}{{\tt astro-ph/0403052}}].

\bibitem{1998PhRvL..81.1158F}
Y.~{Fukuda}, T.~{Hayakawa}, {\em et.~al.}, {\it {Measurements of the Solar
  Neutrino Flux from Super-Kamiokande's First 300 Days}},  {\em Physical Review
  Letters} {\bf 81} (1998) 1158--1162,
  [\href{http://arxiv.org/abs/hep-ex/9805021}{{\tt hep-ex/9805021}}].

\bibitem{Ahrens:2003ix}
IceCube: J.~Ahrens {\em et.~al.}, {\it {Sensitivity of the IceCube detector to
  astrophysical sources of high energy muon neutrinos}},  {\em Astropart.
  Phys.} {\bf 20} (2004) 507--532,
  [\href{http://arxiv.org/abs/astro-ph/0305196}{{\tt astro-ph/0305196}}].

\bibitem{Collaboration:2011nsa}
ANTARES: M.~Ageron {\em et.~al.}, {\it {ANTARES: the first undersea neutrino
  telescope}},  {\em Nucl. Instrum. Meth.} {\bf A656} (2011) 11--38,
  [\href{http://arxiv.org/abs/1104.1607}{{\tt arXiv:1104.1607}}].

\bibitem{Barger:2001ur}
V.~D. Barger, F.~Halzen, D.~Hooper, and C.~Kao, {\it {Indirect search for
  neutralino dark matter with high-energy neutrinos}},  {\em Phys. Rev.} {\bf
  D65} (2002) 075022, [\href{http://arxiv.org/abs/hep-ph/0105182}{{\tt
  hep-ph/0105182}}].

\bibitem{Padmanabhan:2005es}
N.~Padmanabhan and D.~P. Finkbeiner, {\it {Detecting Dark Matter Annihilation
  with CMB Polarization : Signatures and Experimental Prospects}},  {\em Phys.
  Rev.} {\bf D72} (2005) 023508,
  [\href{http://arxiv.org/abs/astro-ph/0503486}{{\tt astro-ph/0503486}}].

\bibitem{Fermi-LAT:2016uux}
DES, Fermi-LAT: A.~Albert {\em et.~al.}, {\it {Searching for Dark Matter
  Annihilation in Recently Discovered Milky Way Satellites with Fermi-LAT}},
  {\em Astrophys. J.} {\bf 834} (2017) 110,
  [\href{http://arxiv.org/abs/1611.03184}{{\tt arXiv:1611.03184}}].

\bibitem{Bergstrom:2013jra}
L.~Bergstrom, T.~Bringmann, I.~Cholis, D.~Hooper, and C.~Weniger, {\it {New
  limits on dark matter annihilation from AMS cosmic ray positron data}},  {\em
  Phys. Rev. Lett.} {\bf 111} (2013) 171101,
  [\href{http://arxiv.org/abs/1306.3983}{{\tt arXiv:1306.3983}}].

\bibitem{Choi:2015ara}
Super-Kamiokande: K.~Choi {\em et.~al.}, {\it {Search for neutrinos from
  annihilation of captured low-mass dark matter particles in the Sun by
  Super-Kamiokande}},  {\em Phys. Rev. Lett.} {\bf 114} (2015) 141301,
  [\href{http://arxiv.org/abs/1503.04858}{{\tt arXiv:1503.04858}}].

\bibitem{Aartsen:2016exj}
IceCube: M.~G. Aartsen {\em et.~al.}, {\it {Improved limits on dark matter
  annihilation in the Sun with the 79-string IceCube detector and implications
  for supersymmetry}},  {\em JCAP} {\bf 1604} (2016) 022,
  [\href{http://arxiv.org/abs/1601.00653}{{\tt arXiv:1601.00653}}].

\bibitem{1748-0221-3-08-S08001}
L.~Evans and P.~Bryant, {\it Lhc machine},  {\em Journal of Instrumentation}
  {\bf 3} (2008) S08001.

\bibitem{Aad:2008zzm}
ATLAS: G.~Aad {\em et.~al.}, {\it {The ATLAS Experiment at the CERN Large
  Hadron Collider}},  {\em JINST} {\bf 3} (2008) S08003.

\bibitem{Chatrchyan:2008aa}
CMS: S.~Chatrchyan {\em et.~al.}, {\it {The CMS Experiment at the CERN LHC}},
  {\em JINST} {\bf 3} (2008) S08004.

\bibitem{Alves:2008zz}
LHCb: A.~A. Alves, Jr. {\em et.~al.}, {\it {The LHCb Detector at the LHC}},
  {\em JINST} {\bf 3} (2008) S08005.

\bibitem{Aamodt:2008zz}
ALICE: K.~Aamodt {\em et.~al.}, {\it {The ALICE experiment at the CERN LHC}},
  {\em JINST} {\bf 3} (2008) S08002.

\bibitem{Abercrombie:2015wmb}
D.~Abercrombie {\em et.~al.}, {\it {Dark Matter Benchmark Models for Early LHC
  Run-2 Searches: Report of the ATLAS/CMS Dark Matter Forum}},
  \href{http://arxiv.org/abs/1507.00966}{{\tt arXiv:1507.00966}}.

\bibitem{Bauer:2017ota}
M.~Bauer, U.~Haisch, and F.~Kahlhoefer, {\it {Simplified dark matter models
  with two Higgs doublets: I. Pseudoscalar mediators}},  {\em JHEP} {\bf 05}
  (2017) 138, [\href{http://arxiv.org/abs/1701.07427}{{\tt arXiv:1701.07427}}].

\bibitem{Golfand:1971iw}
{\relax Yu}.~A. Golfand and E.~P. Likhtman, {\it {Extension of the Algebra of
  Poincare Group Generators and Violation of p Invariance}},  {\em JETP Lett.}
  {\bf 13} (1971) 323. [Pisma Zh. Eksp. Teor. Fiz. \textbf{13} (1971) 452].

\bibitem{Volkov:1973ix}
D.~V. Volkov and V.~P. Akulov, {\it {Is the Neutrino a Goldstone Particle?}},
  {\em Phys. Lett. B} {\bf 46} (1973) 109.

\bibitem{Wess:1974tw}
J.~Wess and B.~Zumino, {\it {Supergauge Transformations in Four-Dimensions}},
  {\em Nucl. Phys. B} {\bf 70} (1974) 39.

\bibitem{Wess:1974jb}
J.~Wess and B.~Zumino, {\it {Supergauge Invariant Extension of Quantum
  Electrodynamics}},  {\em Nucl. Phys. B} {\bf 78} (1974) 1.

\bibitem{Ferrara:1974pu}
S.~Ferrara and B.~Zumino, {\it {Supergauge Invariant Yang-Mills Theories}},
  {\em Nucl. Phys. B} {\bf 79} (1974) 413.

\bibitem{Salam:1974ig}
A.~Salam and J.~A. Strathdee, {\it {Supersymmetry and Nonabelian Gauges}},
  {\em Phys. Lett. B} {\bf 51} (1974) 353.

\bibitem{DAmbrosio:2002vsn}
G.~D'Ambrosio, G.~F. Giudice, G.~Isidori, and A.~Strumia, {\it {Minimal flavor
  violation: An Effective field theory approach}},  {\em Nucl. Phys.} {\bf
  B645} (2002) 155--187, [\href{http://arxiv.org/abs/hep-ph/0207036}{{\tt
  hep-ph/0207036}}].

\bibitem{Boveia:2016mrp}
G.~Busoni {\em et.~al.}, {\it {Recommendations on presenting LHC searches for
  missing transverse energy signals using simplified $s$-channel models of dark
  matter}},  \href{http://arxiv.org/abs/1603.04156}{{\tt arXiv:1603.04156}}.

\bibitem{Baak:2014wma}
M.~Baak, G.~J. Besjes, {\em et.~al.}, {\it {HistFitter software framework for
  statistical data analysis}},  {\em Eur. Phys. J.} {\bf C75} (2015) 153,
  [\href{http://arxiv.org/abs/1410.1280}{{\tt arXiv:1410.1280}}].

\bibitem{Cowan:2010js}
G.~Cowan, K.~Cranmer, E.~Gross, and O.~Vitells, {\it {Asymptotic formulae for
  likelihood-based tests of new physics}},  {\em Eur. Phys. J.} {\bf C71}
  (2011) 1554, [\href{http://arxiv.org/abs/1007.1727}{{\tt arXiv:1007.1727}}].
  [Erratum: Eur. Phys. J.C73,2501(2013)].

\bibitem{Wilks:1938dza}
S.~S. Wilks, {\it {The Large-Sample Distribution of the Likelihood Ratio for
  Testing Composite Hypotheses}},  {\em Annals Math. Statist.} {\bf 9} (1938)
  60--62.

\bibitem{0954-3899-28-10-313}
A.~L. Read, {\it Presentation of search results: the cl s technique},  {\em
  Journal of Physics G: Nuclear and Particle Physics} {\bf 28} (2002) 2693.

\bibitem{Hastie:2009fk}
{T. Hastie et al.}, {\em {Elements of Statistical Learning}}.
\newblock Springer, 2009.

\bibitem{0483bd9444a348c8b59d54a190839ec9}
Y.~Lecun, Y.~Bengio, and G.~Hinton, {\it Deep learning},  {\em Nature} {\bf
  521} (2015) 436--444.

\bibitem{DBLP:journals/corr/GoodfellowPMXWOCB14}
I.~J. Goodfellow, J.~Pouget{-}Abadie, {\em et.~al.}, {\it Generative
  adversarial networks},  {\em CoRR} {\bf abs/1406.2661} (2014)
  [\href{http://arxiv.org/abs/1406.2661}{{\tt arXiv:1406.2661}}].

\bibitem{Chala:2015ama}
M.~Chala, F.~Kahlhoefer, M.~McCullough, G.~Nardini, and K.~Schmidt-Hoberg, {\it
  {Constraining Dark Sectors with Monojets and Dijets}},  {\em JHEP} {\bf 07}
  (2015) 089, [\href{http://arxiv.org/abs/1503.05916}{{\tt arXiv:1503.05916}}].

\bibitem{Hlozek:2011pc}
R.~Hlozek {\em et.~al.}, {\it {The Atacama Cosmology Telescope: a measurement
  of the primordial power spectrum}},  {\em Astrophys. J.} {\bf 749} (2012) 90,
  [\href{http://arxiv.org/abs/1105.4887}{{\tt arXiv:1105.4887}}].

\bibitem{Brooks:2014qya}
A.~Brooks, {\it {Re-Examining Astrophysical Constraints on the Dark Matter
  Model}},  {\em Annalen Phys.} {\bf 526} (2014) 294--308,
  [\href{http://arxiv.org/abs/1407.7544}{{\tt arXiv:1407.7544}}].

\bibitem{vandenBosch:2001bp}
F.~C. van~den Bosch, A.~Burkert, and R.~A. Swaters, {\it {The angular momentum
  content of dwarf galaxies: new challenges for the theory of galaxy
  formation}},  {\em Mon. Not. Roy. Astron. Soc.} {\bf 326} (2001) 1205,
  [\href{http://arxiv.org/abs/astro-ph/0105082}{{\tt astro-ph/0105082}}].

\bibitem{deBlok:2009sp}
W.~J.~G. de~Blok, {\it {The Core-Cusp Problem}},  {\em Adv. Astron.} {\bf 2010}
  (2010) 789293, [\href{http://arxiv.org/abs/0910.3538}{{\tt
  arXiv:0910.3538}}].

\bibitem{Oh:2010mc}
S.-H. Oh, C.~Brook, {\em et.~al.}, {\it {The central slope of dark matter cores
  in dwarf galaxies: Simulations vs. THINGS}},  {\em Astron. J.} {\bf 142}
  (2011) 24, [\href{http://arxiv.org/abs/1011.2777}{{\tt arXiv:1011.2777}}].

\bibitem{Klypin:1999uc}
A.~A. Klypin, A.~V. Kravtsov, O.~Valenzuela, and F.~Prada, {\it {Where are the
  missing Galactic satellites?}},  {\em Astrophys. J.} {\bf 522} (1999) 82--92,
  [\href{http://arxiv.org/abs/astro-ph/9901240}{{\tt astro-ph/9901240}}].

\bibitem{BoylanKolchin:2011de}
M.~Boylan-Kolchin, J.~S. Bullock, and M.~Kaplinghat, {\it {Too big to fail? The
  puzzling darkness of massive Milky Way subhaloes}},  {\em Mon. Not. Roy.
  Astron. Soc.} {\bf 415} (2011) L40,
  [\href{http://arxiv.org/abs/1103.0007}{{\tt arXiv:1103.0007}}].

\bibitem{BoylanKolchin:2011dk}
M.~Boylan-Kolchin, J.~S. Bullock, and M.~Kaplinghat, {\it {The Milky Way's
  bright satellites as an apparent failure of LCDM}},  {\em Mon. Not. Roy.
  Astron. Soc.} {\bf 422} (2012) 1203--1218,
  [\href{http://arxiv.org/abs/1111.2048}{{\tt arXiv:1111.2048}}].

\bibitem{Brook:2010bn}
C.~B. Brook {\em et.~al.}, {\it {Hierarchical formation of bulgeless galaxies:
  Why outflows have low angular momentum}},  {\em Mon. Not. Roy. Astron. Soc.}
  {\bf 415} (2011) 1051, [\href{http://arxiv.org/abs/1010.1004}{{\tt
  arXiv:1010.1004}}].

\bibitem{Brook:2011nz}
C.~B. Brook, G.~Stinson, {\em et.~al.}, {\it {Hierarchical formation of
  bulgeless galaxies II: Redistribution of angular momentum via galactic
  fountains}},  {\em Mon. Not. Roy. Astron. Soc.} {\bf 419} (2012) 771,
  [\href{http://arxiv.org/abs/1105.2562}{{\tt arXiv:1105.2562}}].

\bibitem{Pontzen:2011ty}
A.~Pontzen and F.~Governato, {\it {How supernova feedback turns dark matter
  cusps into cores}},  {\em Mon. Not. Roy. Astron. Soc.} {\bf 421} (2012) 3464,
  [\href{http://arxiv.org/abs/1106.0499}{{\tt arXiv:1106.0499}}].

\bibitem{Brooks2013}
A.~M. {Brooks}, M.~{Kuhlen}, A.~{Zolotov}, and D.~{Hooper}, {\it {A Baryonic
  Solution to the Missing Satellites Problem}},  {\em \apj} {\bf 765} (2013)
  22, [\href{http://arxiv.org/abs/1209.5394}{{\tt arXiv:1209.5394}}].

\bibitem{DiCintio:2014xia}
A.~Di~Cintio, C.~B. Brook, {\em et.~al.}, {\it {A mass-dependent density
  profile for dark matter haloes including the influence of galaxy formation}},
   {\em Mon. Not. Roy. Astron. Soc.} {\bf 441} (2014) 2986--2995,
  [\href{http://arxiv.org/abs/1404.5959}{{\tt arXiv:1404.5959}}].

\bibitem{BZ2014}
A.~M. {Brooks} and A.~{Zolotov}, {\it {Why Baryons Matter: The Kinematics of
  Dwarf Spheroidal Satellites}},  {\em \apj} {\bf 786} (2014) 87,
  [\href{http://arxiv.org/abs/1207.2468}{{\tt arXiv:1207.2468}}].

\bibitem{Chan2015}
T.~K. {Chan}, D.~{Kere{\v s}}, {\em et.~al.}, {\it {The impact of baryonic
  physics on the structure of dark matter haloes: the view from the FIRE
  cosmological simulations}},  {\em \mnras} {\bf 454} (2015) 2981--3001,
  [\href{http://arxiv.org/abs/1507.02282}{{\tt arXiv:1507.02282}}].

\bibitem{Wetzel2016}
A.~R. {Wetzel}, P.~F. {Hopkins}, {\em et.~al.}, {\it {Reconciling Dwarf
  Galaxies with {$\Lambda$}CDM Cosmology: Simulating a Realistic Population of
  Satellites around a Milky Way-mass Galaxy}},  {\em \apjl} {\bf 827} (2016)
  L23, [\href{http://arxiv.org/abs/1602.05957}{{\tt arXiv:1602.05957}}].

\bibitem{Vogelsberger2014}
M.~{Vogelsberger}, J.~{Zavala}, C.~{Simpson}, and A.~{Jenkins}, {\it {Dwarf
  galaxies in CDM and SIDM with baryons: observational probes of the nature of
  dark matter}},  {\em \mnras} {\bf 444} (2014) 3684--3698,
  [\href{http://arxiv.org/abs/1405.5216}{{\tt arXiv:1405.5216}}].

\bibitem{Elbert:2014bma}
O.~D. Elbert, J.~S. Bullock, {\em et.~al.}, {\it {Core formation in dwarf
  haloes with self-interacting dark matter: no fine-tuning necessary}},  {\em
  Mon. Not. Roy. Astron. Soc.} {\bf 453} (2015) 29--37,
  [\href{http://arxiv.org/abs/1412.1477}{{\tt arXiv:1412.1477}}].

\bibitem{Fry:2015rta}
A.~B. Fry, F.~Governato, {\em et.~al.}, {\it {All about baryons: revisiting
  SIDM predictions at small halo masses}},  {\em Mon. Not. Roy. Astron. Soc.}
  {\bf 452} (2015) 1468--1479, [\href{http://arxiv.org/abs/1501.00497}{{\tt
  arXiv:1501.00497}}].

\bibitem{Elbert:2016dbb}
O.~D. Elbert, J.~S. Bullock, {\em et.~al.}, {\it {A Testable Conspiracy:
  Simulating Baryonic Effects on Self-Interacting Dark Matter Halos}},  {\em
  Astrophys. J.} {\bf 853} (2018) 109,
  [\href{http://arxiv.org/abs/1609.08626}{{\tt arXiv:1609.08626}}].

\bibitem{spergel00}
D.~N. {Spergel} and P.~J. {Steinhardt}, {\it {Observational Evidence for
  Self-Interacting Cold Dark Matter}},  {\em Physical Review Letters} {\bf 84}
  (2000) 3760--3763, [\href{http://arxiv.org/abs/astro-ph/9}{{\tt
  astro-ph/9}}].

\bibitem{Loeb2011}
A.~{Loeb} and N.~{Weiner}, {\it {Cores in Dwarf Galaxies from Dark Matter with
  a Yukawa Potential}},  {\em Physical Review Letters} {\bf 106} (2011)
  171302--+, [\href{http://arxiv.org/abs/1011.6374}{{\tt arXiv:1011.6374}}].

\bibitem{Governato:2014gja}
F.~Governato {\em et.~al.}, {\it {Faint dwarfs as a test of DM models: WDM
  versus CDM}},  {\em Mon. Not. Roy. Astron. Soc.} {\bf 448} (2015) 792--803,
  [\href{http://arxiv.org/abs/1407.0022}{{\tt arXiv:1407.0022}}].

\bibitem{Herpich2014}
J.~{Herpich}, G.~S. {Stinson}, {\em et.~al.}, {\it {MaGICC-WDM: the effects of
  warm dark matter in hydrodynamical simulations of disc galaxy formation}},
  {\em \mnras} {\bf 437} (2014) 293--304,
  [\href{http://arxiv.org/abs/1308.1088}{{\tt arXiv:1308.1088}}].

\bibitem{Lovell:2016fec}
M.~R. Lovell {\em et.~al.}, {\it {Properties of Local Group galaxies in
  hydrodynamical simulations of sterile neutrino dark matter cosmologies}},
  {\em Mon. Not. Roy. Astron. Soc.} {\bf 468} (2017) 4285--4298,
  [\href{http://arxiv.org/abs/1611.00010}{{\tt arXiv:1611.00010}}].

\bibitem{Zhang:2016uiy}
J.~Zhang, Y.-L.~S. Tsai, J.-L. Kuo, K.~Cheung, and M.-C. Chu, {\it {Ultralight
  Axion Dark Matter and Its Impact on Dark Halo Structure in $N$-body
  Simulations}},  {\em Astrophys. J.} {\bf 853} (2018) 51,
  [\href{http://arxiv.org/abs/1611.00892}{{\tt arXiv:1611.00892}}].

\bibitem{Zhang:2017chj}
J.~Zhang, J.-L. Kuo, {\em et.~al.}, {\it {Is Fuzzy Dark Matter in tension with
  Lyman-alpha forest?}},  \href{http://arxiv.org/abs/1708.04389}{{\tt
  arXiv:1708.04389}}.

\bibitem{Tollerud:2008ze}
E.~J. Tollerud, J.~S. Bullock, L.~E. Strigari, and B.~Willman, {\it {Hundreds
  of Milky Way Satellites? Luminosity Bias in the Satellite Luminosity
  Function}},  {\em Astrophys. J.} {\bf 688} (2008) 277--289,
  [\href{http://arxiv.org/abs/0806.4381}{{\tt arXiv:0806.4381}}].

\bibitem{Tollerud:2010bj}
E.~J. Tollerud, J.~S. Bullock, G.~J. Graves, and J.~Wolf, {\it {From Galaxy
  Clusters to Ultra-Faint Dwarf Spheroidals: A Fundamental Curve Connecting
  Dispersion-supported Galaxies to Their Dark Matter Halos}},  {\em Astrophys.
  J.} {\bf 726} (2011) 108, [\href{http://arxiv.org/abs/1007.5311}{{\tt
  arXiv:1007.5311}}].

\bibitem{Walsh2009}
S.~M. {Walsh}, B.~{Willman}, and H.~{Jerjen}, {\it {The Invisibles: A Detection
  Algorithm to Trace the Faintest Milky Way Satellites}},  {\em \aj} {\bf 137}
  (2009) 450--469, [\href{http://arxiv.org/abs/0807.3345}{{\tt
  arXiv:0807.3345}}].

\bibitem{Governato:2012fa}
F.~Governato, A.~Zolotov, {\em et.~al.}, {\it {Cuspy No More: How Outflows
  Affect the Central Dark Matter and Baryon Distribution in Lambda CDM
  Galaxies}},  {\em Mon. Not. Roy. Astron. Soc.} {\bf 422} (2012) 1231--1240,
  [\href{http://arxiv.org/abs/1202.0554}{{\tt arXiv:1202.0554}}].

\bibitem{Tollet:2015gqa}
E.~Tollet {\em et.~al.}, {\it {NIHAO – IV: core creation and destruction in
  dark matter density profiles across cosmic time}},  {\em Mon. Not. Roy.
  Astron. Soc.} {\bf 456} (2016) 3542--3552,
  [\href{http://arxiv.org/abs/1507.03590}{{\tt arXiv:1507.03590}}].

\bibitem{Walker:2011zu}
M.~G. Walker and J.~Penarrubia, {\it {A Method for Measuring (Slopes of) the
  Mass Profiles of Dwarf Spheroidal Galaxies}},  {\em Astrophys. J.} {\bf 742}
  (2011) 20, [\href{http://arxiv.org/abs/1108.2404}{{\tt arXiv:1108.2404}}].

\bibitem{Bonnivard:2015xpq}
V.~Bonnivard {\em et.~al.}, {\it {Dark matter annihilation and decay in dwarf
  spheroidal galaxies: The classical and ultrafaint dSphs}},  {\em Mon. Not.
  Roy. Astron. Soc.} {\bf 453} (2015) 849--867,
  [\href{http://arxiv.org/abs/1504.02048}{{\tt arXiv:1504.02048}}].

\bibitem{Dooley:2016xkj}
G.~A. Dooley, A.~H.~G. Peter, {\em et.~al.}, {\it {An observer's guide to the
  (Local Group) dwarf galaxies: predictions for their own dwarf satellite
  populations}},  {\em Mon. Not. Roy. Astron. Soc.} {\bf 471} (2017)
  4894--4909, [\href{http://arxiv.org/abs/1610.00708}{{\tt arXiv:1610.00708}}].

\bibitem{Carlberg:2009ae}
R.~G. Carlberg, {\it {Star Stream Folding by Dark Galactic Sub-Halos}},  {\em
  Astrophys. J.} {\bf 705} (2009) L223--L226,
  [\href{http://arxiv.org/abs/0908.4345}{{\tt arXiv:0908.4345}}].

\bibitem{Carlberg:2012ur}
R.~G. Carlberg, C.~J. Grillmair, and N.~Hetherington, {\it {The Pal 5 Star
  Stream Gaps}},  {\em Astrophys. J.} {\bf 760} (2012) 75,
  [\href{http://arxiv.org/abs/1209.1741}{{\tt arXiv:1209.1741}}].

\bibitem{Carlberg:2011xj}
R.~G. Carlberg, {\it {Dark Matter Sub-Halo Counts via Star Stream Crossings}},
  {\em Astrophys. J.} {\bf 748} (2012) 20,
  [\href{http://arxiv.org/abs/1109.6022}{{\tt arXiv:1109.6022}}].

\bibitem{Carlberg:2013eya}
R.~G. Carlberg, {\it {The Dynamics of Star Stream Gaps}},  {\em Astrophys. J.}
  {\bf 775} (2013) 90, [\href{http://arxiv.org/abs/1307.1929}{{\tt
  arXiv:1307.1929}}].

\bibitem{Carlberg:2013gxa}
R.~G. Carlberg and C.~J. Grillmair, {\it {Gaps in the GD-1 Star Stream}},  {\em
  Astrophys. J.} {\bf 768} (2013) 171,
  [\href{http://arxiv.org/abs/1303.4342}{{\tt arXiv:1303.4342}}].

\bibitem{2014ApJ...788..181N}
W.~H.~W. {Ngan} and R.~G. {Carlberg}, {\it {Using Gaps in N-body Tidal Streams
  to Probe Missing Satellites}},  {\em \apj} {\bf 788} (2014) 181,
  [\href{http://arxiv.org/abs/1311.1710}{{\tt arXiv:1311.1710}}].

\bibitem{2016ApJ...820...45C}
R.~G. {Carlberg}, {\it {Modeling GD-1 Gaps in a Milky Way Potential}},  {\em
  \apj} {\bf 820} (2016) 45, [\href{http://arxiv.org/abs/1512.01620}{{\tt
  arXiv:1512.01620}}].

\bibitem{Erkal:2014tda}
D.~Erkal and V.~Belokurov, {\it {Forensics of subhalo--stream encounters: the
  three phases of gap growth}},  {\em Mon. Not. Roy. Astron. Soc.} {\bf 450}
  (2015) 1136--1149, [\href{http://arxiv.org/abs/1412.6035}{{\tt
  arXiv:1412.6035}}].

\bibitem{Erkal:2015kqa}
D.~Erkal and V.~Belokurov, {\it {Properties of Dark Subhaloes from Gaps in
  Tidal Streams}},  {\em Mon. Not. Roy. Astron. Soc.} {\bf 454} (2015)
  3542--3558, [\href{http://arxiv.org/abs/1507.05625}{{\tt arXiv:1507.05625}}].

\bibitem{2016MNRAS.457.3817S}
J.~L. {Sanders}, J.~{Bovy}, and D.~{Erkal}, {\it {Dynamics of stream-subhalo
  interactions}},  {\em \mnras} {\bf 457} (2016) 3817--3835,
  [\href{http://arxiv.org/abs/1510.03426}{{\tt arXiv:1510.03426}}].

\bibitem{2016MNRAS.463..102E}
D.~{Erkal}, V.~{Belokurov}, J.~{Bovy}, and J.~L. {Sanders}, {\it {The number
  and size of subhalo-induced gaps in stellar streams}},  {\em \mnras} {\bf
  463} (2016) 102--119, [\href{http://arxiv.org/abs/1606.04946}{{\tt
  arXiv:1606.04946}}].

\bibitem{2017MNRAS.466..628B}
J.~{Bovy}, D.~{Erkal}, and J.~L. {Sanders}, {\it {Linear perturbation theory
  for tidal streams and the small-scale CDM power spectrum}},  {\em \mnras}
  {\bf 466} (2017) 628--668, [\href{http://arxiv.org/abs/1606.03470}{{\tt
  arXiv:1606.03470}}].

\bibitem{2017MNRAS.470...60E}
D.~{Erkal}, S.~E. {Koposov}, and V.~{Belokurov}, {\it {A sharper view of Pal
  5's tails: discovery of stream perturbations with a novel non-parametric
  technique}},  {\em \mnras} {\bf 470} (2017) 60--84,
  [\href{http://arxiv.org/abs/1609.01282}{{\tt arXiv:1609.01282}}].

\bibitem{2016PhRvL.116l1301B}
J.~{Bovy}, {\it {Detecting the Disruption of Dark-Matter Halos with Stellar
  Streams}},  {\em Physical Review Letters} {\bf 116} (2016) 121301,
  [\href{http://arxiv.org/abs/1512.00452}{{\tt arXiv:1512.00452}}].

\bibitem{Banik:2018pjp}
N.~Banik, G.~Bertone, J.~Bovy, and N.~Bozorgnia, {\it {Probing the nature of
  dark matter particles with stellar streams}},
  \href{http://arxiv.org/abs/1804.04384}{{\tt arXiv:1804.04384}}.

\bibitem{2012ApJ...750L..41W}
L.~M. {Widrow}, S.~{Gardner}, B.~{Yanny}, S.~{Dodelson}, and H.-Y. {Chen}, {\it
  {Galactoseismology: Discovery of Vertical Waves in the Galactic Disk}},  {\em
  \apjl} {\bf 750} (2012) L41, [\href{http://arxiv.org/abs/1203.6861}{{\tt
  arXiv:1203.6861}}].

\bibitem{Feldmann:2013hqa}
R.~Feldmann and D.~Spolyar, {\it {Detecting Dark Matter Substructures around
  the Milky Way with Gaia}},  {\em Mon. Not. Roy. Astron. Soc.} {\bf 446}
  (2015) 1000--1012, [\href{http://arxiv.org/abs/1310.2243}{{\tt
  arXiv:1310.2243}}].

\bibitem{Buschmann:2017ams}
M.~Buschmann, J.~Kopp, B.~R. Safdi, and C.-L. Wu, {\it {Stellar Wakes from Dark
  Matter Subhalos}},  {\em Phys. Rev. Lett.} {\bf 120} (2018) 211101,
  [\href{http://arxiv.org/abs/1711.03554}{{\tt arXiv:1711.03554}}].

\bibitem{Erickcek:2010fc}
A.~L. Erickcek and N.~M. Law, {\it {Astrometric Microlensing by Local Dark
  Matter Subhalos}},  {\em Astrophys. J.} {\bf 729} (2011) 49,
  [\href{http://arxiv.org/abs/1007.4228}{{\tt arXiv:1007.4228}}].

\bibitem{Li:2012qha}
F.~Li, A.~L. Erickcek, and N.~M. Law, {\it {A new probe of the small-scale
  primordial power spectrum: astrometric microlensing by ultracompact
  minihalos}},  {\em Phys. Rev.} {\bf D86} (2012) 043519,
  [\href{http://arxiv.org/abs/1202.1284}{{\tt arXiv:1202.1284}}].

\bibitem{VanTilburg:2018ykj}
K.~Van~Tilburg, A.-M. Taki, and N.~Weiner, {\it {Halometry from Astrometry}},
  \href{http://arxiv.org/abs/1804.01991}{{\tt arXiv:1804.01991}}.

\bibitem{Mao:1998aa}
S.~Mao and P.~Schneider, {\it Evidence for substructure in lens galaxies?},
  {\em Mon. Not. R. Astron. Soc.} {\bf 295} (1998) 587--594,
  [\href{http://arxiv.org/abs/astro-ph/9707187}{{\tt astro-ph/9707187}}].

\bibitem{Metcalf:ad}
R.~B. {Metcalf} and P.~{Madau}, {\it {Compound Gravitational Lensing as a Probe
  of Dark Matter Substructure within Galaxy Halos}},  {\em \apj} {\bf 563}
  (2001) 9--20, [\href{http://arxiv.org/abs/astro-ph/0108224}{{\tt
  astro-ph/0108224}}].

\bibitem{Dalal:2002aa}
N.~Dalal and C.~Kochanek, {\it Direct detection of cdm substructure},  {\em
  Astrophys. J.} {\bf 572} (2002) 25--33,
  [\href{http://arxiv.org/abs/astro-ph/0111456}{{\tt astro-ph/0111456}}].

\bibitem{Koopmans:aa}
L.~V.~E. {Koopmans}, {\it {Gravitational imaging of cold dark matter
  substructures}},  {\em \mnras} {\bf 363} (2005) 1136--1144,
  [\href{http://arxiv.org/abs/astro-ph/0501324}{{\tt astro-ph/0501324}}].

\bibitem{Vegetti:2008aa}
S.~{Vegetti} and L.~V.~E. {Koopmans}, {\it {Bayesian strong gravitational-lens
  modelling on adaptive grids: objective detection of mass substructure in
  Galaxies}},  {\em \mnras} {\bf 392} (2009) 945--963,
  [\href{http://arxiv.org/abs/0805.0201}{{\tt arXiv:0805.0201}}].

\bibitem{Vegetti:2009aa}
S.~{Vegetti} and L.~V.~E. {Koopmans}, {\it {Statistics of mass substructure
  from strong gravitational lensing: quantifying the mass fraction and mass
  function}},  {\em \mnras} {\bf 400} (2009) 1583--1592,
  [\href{http://arxiv.org/abs/0903.4752}{{\tt arXiv:0903.4752}}].

\bibitem{Vegetti_2010_1}
S.~{Vegetti}, O.~{Czoske}, and L.~V.~E. {Koopmans}, {\it {Quantifying dwarf
  satellites through gravitational imaging: the case of
  SDSSJ120602.09+514229.5}},  {\em \mnras} {\bf 407} (2010) 225--231,
  [\href{http://arxiv.org/abs/1002.4708}{{\tt arXiv:1002.4708}}].

\bibitem{Vegetti_2010_2}
S.~{Vegetti}, L.~V.~E. {Koopmans}, A.~{Bolton}, T.~{Treu}, and R.~{Gavazzi},
  {\it {Detection of a dark substructure through gravitational imaging}},  {\em
  \mnras} {\bf 408} (2010) 1969--1981,
  [\href{http://arxiv.org/abs/0910.0760}{{\tt arXiv:0910.0760}}].

\bibitem{Vegetti_2012}
S.~{Vegetti}, D.~J. {Lagattuta}, {\em et.~al.}, {\it {Gravitational detection
  of a low-mass dark satellite galaxy at cosmological distance}},  {\em \nat}
  {\bf 481} (2012) 341--343, [\href{http://arxiv.org/abs/1201.3643}{{\tt
  arXiv:1201.3643}}].

\bibitem{2014MNRAS.442.2017V}
S.~{Vegetti}, L.~V.~E. {Koopmans}, M.~W. {Auger}, T.~{Treu}, and A.~S.
  {Bolton}, {\it {Inference of the cold dark matter substructure mass function
  at z = 0.2 using strong gravitational lenses}},  {\em \mnras} {\bf 442}
  (2014) 2017--2035, [\href{http://arxiv.org/abs/1405.3666}{{\tt
  arXiv:1405.3666}}].

\bibitem{Hezaveh:2012ai}
Y.~Hezaveh, N.~Dalal, {\em et.~al.}, {\it {Dark Matter Substructure Detection
  Using Spatially Resolved Spectroscopy of Lensed Dusty Galaxies}},  {\em
  Astrophys. J.} {\bf 767} (2013) 9,
  [\href{http://arxiv.org/abs/1210.4562}{{\tt arXiv:1210.4562}}].

\bibitem{Hezaveh_2016_2}
Y.~D. Hezaveh {\em et.~al.}, {\it {Detection of lensing substructure using ALMA
  observations of the dusty galaxy SDP.81}},  {\em Astrophys. J.} {\bf 823}
  (2016) 37, [\href{http://arxiv.org/abs/1601.01388}{{\tt arXiv:1601.01388}}].

\bibitem{Hezaveh_2014}
Y.~Hezaveh, N.~Dalal, {\em et.~al.}, {\it {Measuring the power spectrum of dark
  matter substructure using strong gravitational lensing}},  {\em \jcap} {\bf
  1611} (2016) 048, [\href{http://arxiv.org/abs/1403.2720}{{\tt
  arXiv:1403.2720}}].

\bibitem{Fadely:2009aa}
R.~{Fadely} and C.~R. {Keeton}, {\it {Substructure in the lens HE 0435-1223}},
  {\em \mnras} {\bf 419} (2012) 936--951,
  [\href{http://arxiv.org/abs/1109.0548}{{\tt arXiv:1109.0548}}].

\bibitem{Daylan:2017kfh}
T.~Daylan, F.-Y. Cyr-Racine, A.~Diaz~Rivero, C.~Dvorkin, and D.~P. Finkbeiner,
  {\it {Probing the small-scale structure in strongly lensed systems via
  transdimensional inference}},  {\em Astrophys. J.} {\bf 854} (2018) 141,
  [\href{http://arxiv.org/abs/1706.06111}{{\tt arXiv:1706.06111}}].

\bibitem{Cyr-Racine:2018htu}
F.-Y. Cyr-Racine, C.~R. Keeton, and L.~A. Moustakas, {\it {Beyond subhalos:
  Probing the collective effect of the Universe's small-scale structure with
  gravitational lensing}},  \href{http://arxiv.org/abs/1806.07897}{{\tt
  arXiv:1806.07897}}.

\bibitem{wilks}
S.~Wilks, {\it The large-sample distribution of the likelihood ratio for
  testing composite hypotheses},  {\em The Annals of Mathematical Statistics}
  {\bf 9} (1938) 60--62.

\bibitem{Lewis:2002ah}
A.~Lewis and S.~Bridle, {\it {Cosmological parameters from CMB and other data:
  A Monte Carlo approach}},  {\em Phys. Rev.} {\bf D66} (2002) 103511,
  [\href{http://arxiv.org/abs/astro-ph/0205436}{{\tt astro-ph/0205436}}].

\bibitem{Putze:2014aba}
A.~Putze and L.~Derome, {\it {The Grenoble Analysis Toolkit (GreAT)—A
  statistical analysis framework}},  {\em Phys. Dark Univ.} {\bf 5-6} (2014)
  29--34.

\bibitem{2013PASP..125..306F}
D.~{Foreman-Mackey}, D.~W. {Hogg}, D.~{Lang}, and J.~{Goodman}, {\it {emcee:
  The MCMC Hammer}},  {\em \pasp} {\bf 125} (2013) 306,
  [\href{http://arxiv.org/abs/1202.3665}{{\tt arXiv:1202.3665}}].

\bibitem{Feroz:2007kg}
F.~Feroz and M.~P. Hobson, {\it {Multimodal nested sampling: an efficient and
  robust alternative to MCMC methods for astronomical data analysis}},  {\em
  Mon. Not. Roy. Astron. Soc.} {\bf 384} (2008) 449,
  [\href{http://arxiv.org/abs/0704.3704}{{\tt arXiv:0704.3704}}].

\bibitem{Feroz:2008xx}
F.~Feroz, M.~P. Hobson, and M.~Bridges, {\it {MultiNest: an efficient and
  robust Bayesian inference tool for cosmology and particle physics}},  {\em
  Mon. Not. Roy. Astron. Soc.} {\bf 398} (2009) 1601--1614,
  [\href{http://arxiv.org/abs/0809.3437}{{\tt arXiv:0809.3437}}].

\bibitem{Feroz:2013hea}
F.~Feroz, M.~P. Hobson, E.~Cameron, and A.~N. Pettitt, {\it {Importance Nested
  Sampling and the MultiNest Algorithm}},
  \href{http://arxiv.org/abs/1306.2144}{{\tt arXiv:1306.2144}}.

\bibitem{2015MNRAS.453.4384H}
W.~J. {Handley}, M.~P. {Hobson}, and A.~N. {Lasenby}, {\it {POLYCHORD:
  next-generation nested sampling}},  {\em \mnras} {\bf 453} (2015) 4384--4398,
  [\href{http://arxiv.org/abs/1506.00171}{{\tt arXiv:1506.00171}}].

\bibitem{Workgroup:2017htr}
GAMBIT: G.~D. Martinez, J.~McKay, {\em et.~al.}, {\it {Comparison of
  statistical sampling methods with ScannerBit, the GAMBIT scanning module}},
  {\em Eur. Phys. J.} {\bf C77} (2017) 761,
  [\href{http://arxiv.org/abs/1705.07959}{{\tt arXiv:1705.07959}}].

\bibitem{Lasdon2010}
L.~Lasdon, A.~Duarte, F.~Glover, M.~Laguna, and R.~Mart{\'{\i}}, {\it Adaptive
  memory programming for constrained global optimization},  {\em Computers {\&}
  Operations Research} {\bf 37} (2010) 1500--1509.

\bibitem{algeri18}
S.~Algeri and D.~van Dyk, {\it Testing one hypothesis multiple times: The
  multidimensional case},  {\em In Preparation} (2018).

\bibitem{algeri16}
S.~Algeri, J.~Conrad, and D.~van Dyk, {\it A method for comparing non-nested
  models with application to astrophysical searches for new physics},  {\em
  Monthly Notices of the Royal Astronomical Society: Letters} {\bf 458} (2016)
  L84--L88.

\bibitem{gv10}
E.~Gross and O.~Vitells, {\it Trial factors for the look elsewhere effect in
  high energy physics},  {\em The European Physical Journal C} {\bf 70} (2010)
  525--530.

\bibitem{vg11}
O.~Vitells and E.~Gross, {\it Estimating the significance of a signal in a
  multi-dimensional search},  {\em Astroparticle Physics} {\bf 35} (2011)
  230--234.

\bibitem{chernoff}
H.~Chernoff, {\it On the distribution of the likelihood ratio},  {\em The
  Annals of Mathematical Statistics} (1954) 573--578.

\bibitem{taylor2003}
J.~Taylor and R.~Adler, {\it Euler characteristics for gaussian fields on
  manifolds},  {\em Annals of Probability} (2003) 533--563.

\bibitem{Malyshev:2011zi}
D.~Malyshev and D.~W. Hogg, {\it {Statistics of gamma-ray point sources below
  the Fermi detection limit}},  {\em Astrophys. J.} {\bf 738} (2011) 181,
  [\href{http://arxiv.org/abs/1104.0010}{{\tt arXiv:1104.0010}}].

\bibitem{Lee:2015fea}
S.~K. Lee, M.~Lisanti, B.~R. Safdi, T.~R. Slatyer, and W.~Xue, {\it {Evidence
  for Unresolved $\gamma$-Ray Point Sources in the Inner Galaxy}},  {\em Phys.
  Rev. Lett.} {\bf 116} (2016) 051103,
  [\href{http://arxiv.org/abs/1506.05124}{{\tt arXiv:1506.05124}}].

\bibitem{Lee:2014mza}
S.~K. Lee, M.~Lisanti, and B.~R. Safdi, {\it {Distinguishing Dark Matter from
  Unresolved Point Sources in the Inner Galaxy with Photon Statistics}},  {\em
  JCAP} {\bf 1505} (2015) 056, [\href{http://arxiv.org/abs/1412.6099}{{\tt
  arXiv:1412.6099}}].

\bibitem{Mishra-Sharma:2016gis}
S.~Mishra-Sharma, N.~L. Rodd, and B.~R. Safdi, {\it {NPTFit: A code package for
  Non-Poissonian Template Fitting}},  {\em Astron. J.} {\bf 153} (2017) 253,
  [\href{http://arxiv.org/abs/1612.03173}{{\tt arXiv:1612.03173}}].

\bibitem{2010ApJ...717..825D}
G.~{Dobler}, D.~P. {Finkbeiner}, I.~{Cholis}, T.~{Slatyer}, and N.~{Weiner},
  {\it {The Fermi Haze: A Gamma-ray Counterpart to the Microwave Haze}},  {\em
  \apj} {\bf 717} (2010) 825--842, [\href{http://arxiv.org/abs/0910.4583}{{\tt
  arXiv:0910.4583}}].

\bibitem{2012ApJ...761...91A}
M.~{Ackermann}, M.~{Ajello}, {\em et.~al.}, {\it {Constraints on the Galactic
  Halo Dark Matter from Fermi-LAT Diffuse Measurements}},  {\em \apj} {\bf 761}
  (2012) 91, [\href{http://arxiv.org/abs/1205.6474}{{\tt arXiv:1205.6474}}].

\bibitem{Hooper:2013rwa}
D.~Hooper and T.~R. Slatyer, {\it {Two Emission Mechanisms in the Fermi
  Bubbles: A Possible Signal of Annihilating Dark Matter}},  {\em Phys.Dark
  Univ.} {\bf 2} (2013) 118--138, [\href{http://arxiv.org/abs/1302.6589}{{\tt
  arXiv:1302.6589}}].

\bibitem{Chang:2018bpt}
L.~J. Chang, M.~Lisanti, and S.~Mishra-Sharma, {\it {A Search for Dark Matter
  Annihilation in the Milky Way Halo}},
  \href{http://arxiv.org/abs/1804.04132}{{\tt arXiv:1804.04132}}.

\bibitem{Lisanti:2017qoz}
M.~Lisanti, S.~Mishra-Sharma, N.~L. Rodd, B.~R. Safdi, and R.~H. Wechsler, {\it
  {Mapping Extragalactic Dark Matter Annihilation with Galaxy Surveys: A
  Systematic Study of Stacked Group Searches}},
  \href{http://arxiv.org/abs/1709.00416}{{\tt arXiv:1709.00416}}.

\bibitem{Lisanti:2017qlb}
M.~Lisanti, S.~Mishra-Sharma, N.~L. Rodd, and B.~R. Safdi, {\it {A Search for
  Dark Matter Annihilation in Galaxy Groups}},
  \href{http://arxiv.org/abs/1708.09385}{{\tt arXiv:1708.09385}}.

\bibitem{Edwards:2018lsl}
T.~D.~P. Edwards, B.~J. Kavanagh, and C.~Weniger, {\it {Dark Matter Model or
  Mass, but Not Both: Assessing Near-Future Direct Searches with Benchmark-free
  Forecasting}},  \href{http://arxiv.org/abs/1805.04117}{{\tt
  arXiv:1805.04117}}.

\bibitem{Edwards:2017kqw}
T.~D.~P. Edwards and C.~Weniger, {\it {swordfish: Efficient Forecasting of New
  Physics Searches without Monte Carlo}},
  \href{http://arxiv.org/abs/1712.05401}{{\tt arXiv:1712.05401}}.

\bibitem{Fitzpatrick:2012ix}
A.~L. Fitzpatrick, W.~Haxton, E.~Katz, N.~Lubbers, and Y.~Xu, {\it {The
  Effective Field Theory of Dark Matter Direct Detection}},  {\em JCAP} {\bf
  1302} (2013) 004, [\href{http://arxiv.org/abs/1203.3542}{{\tt
  arXiv:1203.3542}}].

\bibitem{beaumont2002approximate}
M.~A. Beaumont, W.~Zhang, and D.~J. Balding, {\it Approximate bayesian
  computation in population genetics},  {\em Genetics} {\bf 162} (2002)
  2025--2035.

\bibitem{csillery2010approximate}
K.~Csill{\'e}ry, M.~G. Blum, O.~E. Gaggiotti, and O.~Fran{\c{c}}ois, {\it
  Approximate bayesian computation (abc) in practice},  {\em Trends in ecology
  \& evolution} {\bf 25} (2010) 410--418.

\bibitem{tavare1997inferring}
S.~Tavar{\'e}, D.~J. Balding, R.~C. Griffiths, and P.~Donnelly, {\it Inferring
  coalescence times from dna sequence data},  {\em Genetics} {\bf 145} (1997)
  505--518.

\bibitem{pritchard1999population}
J.~K. Pritchard, M.~T. Seielstad, A.~Perez-Lezaun, and M.~W. Feldman, {\it
  Population growth of human y chromosomes: a study of y chromosome
  microsatellites.},  {\em Molecular biology and evolution} {\bf 16} (1999)
  1791--1798.

\bibitem{rubin1984bayesianly}
D.~B. Rubin {\em et.~al.}, {\it Bayesianly justifiable and relevant frequency
  calculations for the applied statistician},  {\em The Annals of Statistics}
  {\bf 12} (1984) 1151--1172.

\bibitem{cameron2012approximate}
E.~Cameron and A.~Pettitt, {\it Approximate bayesian computation for
  astronomical model analysis: a case study in galaxy demographics and
  morphological transformation at high redshift},  {\em Monthly Notices of the
  Royal Astronomical Society} {\bf 425} (2012) 44--65.

\bibitem{weyant2013likelihood}
A.~Weyant, C.~Schafer, and W.~M. Wood-Vasey, {\it Likelihood-free cosmological
  inference with type ia supernovae: approximate bayesian computation for a
  complete treatment of uncertainty},  {\em The Astrophysical Journal} {\bf
  764} (2013) 116.

\bibitem{akeret2015approximate}
J.~Akeret, A.~Refregier, A.~Amara, S.~Seehars, and C.~Hasner, {\it Approximate
  bayesian computation for forward modeling in cosmology},  {\em Journal of
  Cosmology and Astroparticle Physics} {\bf 2015} (2015) 043.

\bibitem{ishida2015cosmoabc}
E.~Ishida, S.~Vitenti, {\em et.~al.}, {\it Cosmoabc: likelihood-free inference
  via population monte carlo approximate bayesian computation},  {\em Astronomy
  and Computing} {\bf 13} (2015) 1--11.

\bibitem{fearnhead2012constructing}
P.~Fearnhead and D.~Prangle, {\it Constructing summary statistics for
  approximate bayesian computation: semi-automatic approximate bayesian
  computation},  {\em Journal of the Royal Statistical Society: Series B
  (Statistical Methodology)} {\bf 74} (2012) 419--474.

\bibitem{blum2013comparative}
M.~G. Blum, M.~A. Nunes, D.~Prangle, S.~A. Sisson, {\em et.~al.}, {\it A
  comparative review of dimension reduction methods in approximate bayesian
  computation},  {\em Statistical Science} {\bf 28} (2013) 189--208.

\bibitem{beaumont2009adaptive}
M.~A. Beaumont, J.-M. Cornuet, J.-M. Marin, and C.~P. Robert, {\it Adaptive
  approximate bayesian computation},  {\em Biometrika} {\bf 96} (2009)
  983--990.

\bibitem{bonassi2015sequential}
F.~V. Bonassi, M.~West, {\em et.~al.}, {\it Sequential monte carlo with
  adaptive weights for approximate bayesian computation},  {\em Bayesian
  Analysis} {\bf 10} (2015) 171--187.

\bibitem{del2012adaptive}
P.~Del~Moral, A.~Doucet, and A.~Jasra, {\it An adaptive sequential monte carlo
  method for approximate bayesian computation},  {\em Statistics and Computing}
  {\bf 22} (2012) 1009--1020.

\bibitem{Pato:2015dua}
M.~Pato, F.~Iocco, and G.~Bertone, {\it {Dynamical constraints on the dark
  matter distribution in the Milky Way}},  {\em JCAP} {\bf 1512} (2015) 001,
  [\href{http://arxiv.org/abs/1504.06324}{{\tt arXiv:1504.06324}}].

\bibitem{Catena:2009mf}
R.~Catena and P.~Ullio, {\it {A novel determination of the local dark matter
  density}},  {\em JCAP} {\bf 1008} (2010) 004,
  [\href{http://arxiv.org/abs/0907.0018}{{\tt arXiv:0907.0018}}].

\bibitem{Bozorgnia:2016ogo}
N.~Bozorgnia, F.~Calore, {\em et.~al.}, {\it {Simulated Milky Way analogues:
  implications for dark matter direct searches}},  {\em JCAP} {\bf 1605} (2016)
  024, [\href{http://arxiv.org/abs/1601.04707}{{\tt arXiv:1601.04707}}].

\bibitem{Kelso:2016qqj}
C.~Kelso, C.~Savage, {\em et.~al.}, {\it {The impact of baryons on the direct
  detection of dark matter}},  {\em JCAP} {\bf 1608} (2016) 071,
  [\href{http://arxiv.org/abs/1601.04725}{{\tt arXiv:1601.04725}}].

\bibitem{Sloane:2016kyi}
J.~D. Sloane, M.~R. Buckley, A.~M. Brooks, and F.~Governato, {\it {Assessing
  Astrophysical Uncertainties in Direct Detection with Galaxy Simulations}},
  {\em Astrophys. J.} {\bf 831} (2016) 93,
  [\href{http://arxiv.org/abs/1601.05402}{{\tt arXiv:1601.05402}}].

\bibitem{Bozorgnia:2017brl}
N.~Bozorgnia and G.~Bertone, {\it {Implications of hydrodynamical simulations
  for the interpretation of direct dark matter searches}},  {\em Int. J. Mod.
  Phys.} {\bf A32} (2017) 1730016, [\href{http://arxiv.org/abs/1705.05853}{{\tt
  arXiv:1705.05853}}].

\bibitem{Calore:2015oya}
F.~Calore, N.~Bozorgnia, {\em et.~al.}, {\it {Simulated Milky Way analogues:
  implications for dark matter indirect searches}},  {\em JCAP} {\bf 1512}
  (2015) 053, [\href{http://arxiv.org/abs/1509.02164}{{\tt arXiv:1509.02164}}].

\bibitem{Iocco:2015xga}
F.~Iocco, M.~Pato, and G.~Bertone, {\it {Evidence for dark matter in the inner
  Milky Way}},  {\em Nature Phys.} {\bf 11} (2015) 245--248,
  [\href{http://arxiv.org/abs/1502.03821}{{\tt arXiv:1502.03821}}].

\bibitem{Benito:2016kyp}
M.~Benito, N.~Bernal, N.~Bozorgnia, F.~Calore, and F.~Iocco, {\it {Particle
  Dark Matter Constraints: the Effect of Galactic Uncertainties}},  {\em JCAP}
  {\bf 1702} (2017) 007, [\href{http://arxiv.org/abs/1612.02010}{{\tt
  arXiv:1612.02010}}].

\bibitem{Akerib:2015rjg}
LUX: D.~S. Akerib {\em et.~al.}, {\it {Improved Limits on Scattering of Weakly
  Interacting Massive Particles from Reanalysis of 2013 LUX Data}},  {\em Phys.
  Rev. Lett.} {\bf 116} (2016) 161301,
  [\href{http://arxiv.org/abs/1512.03506}{{\tt arXiv:1512.03506}}].

\bibitem{TheFermi-LAT:2015kwa}
Fermi-LAT: M.~Ajello {\em et.~al.}, {\it {Fermi-LAT Observations of High-Energy
  $\gamma$-Ray Emission Toward the Galactic Center}},  {\em Astrophys. J.} {\bf
  819} (2016) 44, [\href{http://arxiv.org/abs/1511.02938}{{\tt
  arXiv:1511.02938}}].

\bibitem{Calore:2014nla}
F.~Calore, I.~Cholis, C.~McCabe, and C.~Weniger, {\it {A Tale of Tails: Dark
  Matter Interpretations of the Fermi GeV Excess in Light of Background Model
  Systematics}},  {\em Phys. Rev.} {\bf D91} (2015) 063003,
  [\href{http://arxiv.org/abs/1411.4647}{{\tt arXiv:1411.4647}}].

\bibitem{2010JCAP...01..031S}
P.~{Scott}, J.~{Conrad}, {\em et.~al.}, {\it {Direct constraints on minimal
  supersymmetry from Fermi-LAT observations of the dwarf galaxy Segue 1}},
  {\em \jcap} {\bf 1} (2010) 031, [\href{http://arxiv.org/abs/0909.3300}{{\tt
  arXiv:0909.3300}}].

\bibitem{2012JCAP...11..057S}
P.~{Scott}, C.~{Savage}, J.~{Edsj{\"o}}, and {IceCube Collaboration}, {\it {Use
  of event-level neutrino telescope data in global fits for theories of new
  physics}},  {\em \jcap} {\bf 11} (2012) 057,
  [\href{http://arxiv.org/abs/1207.0810}{{\tt arXiv:1207.0810}}].

\bibitem{2016JCAP...04..022A}
M.~G. {Aartsen}, K.~{Abraham}, {\em et.~al.}, {\it {Improved limits on dark
  matter annihilation in the Sun with the 79-string IceCube detector and
  implications for supersymmetry}},  {\em \jcap} {\bf 4} (2016) 022,
  [\href{http://arxiv.org/abs/1601.00653}{{\tt arXiv:1601.00653}}].

\bibitem{Collaboration:2242860}
CMS Collaboration, {\it {Simplified likelihood for the re-interpretation of
  public CMS results}},  Tech. Rep. CMS-NOTE-2017-001, CERN, Geneva, 2017.

\bibitem{2006JHEP...05..002R}
R.~{Ruiz de Austri}, R.~{Trotta}, and L.~{Roszkowski}, {\it {A Markov chain
  Monte Carlo analysis of the CMSSM}},  {\em Journal of High Energy Physics}
  {\bf 5} (2006) 002, [\href{http://arxiv.org/abs/hep-ph/0602028}{{\tt
  hep-ph/0602028}}].

\bibitem{2006PhRvD..73a5013A}
B.~C. {Allanach} and C.~G. {Lester}, {\it {Multidimensional mSUGRA likelihood
  maps}},  {\em \prd} {\bf 73} (2006) 015013,
  [\href{http://arxiv.org/abs/hep-ph/0507283}{{\tt hep-ph/0507283}}].

\bibitem{2007JHEP...08..023A}
B.~C. {Allanach}, K.~{Cranmer}, C.~G. {Lester}, and A.~M. {Weber}, {\it
  {Natural priors, CMSSM fits and LHC weather forecasts}},  {\em Journal of
  High Energy Physics} {\bf 8} (2007) 023,
  [\href{http://arxiv.org/abs/0705.0487}{{\tt arXiv:0705.0487}}].

\bibitem{2008JHEP...12..024T}
R.~{Trotta}, F.~{Feroz}, M.~{Hobson}, L.~{Roszkowski}, and R.~{Ruiz de Austri},
  {\it {The impact of priors and observables on parameter inferences in the
  constrained MSSM}},  {\em Journal of High Energy Physics} {\bf 12} (2008)
  024, [\href{http://arxiv.org/abs/0809.3792}{{\tt arXiv:0809.3792}}].

\bibitem{2009PhRvD..80i5013L}
D.~E. {L{\'o}pez-Fogliani}, L.~{Roszkowski}, R.~R. {de Austri}, and T.~A.
  {Varley}, {\it {Bayesian analysis of the constrained next-to-minimal
  supersymmetric standard model}},  {\em \prd} {\bf 80} (2009) 095013,
  [\href{http://arxiv.org/abs/0906.4911}{{\tt arXiv:0906.4911}}].

\bibitem{2012PhRvD..85g5012F}
A.~{Fowlie}, A.~{Kalinowski}, M.~{Kazana}, L.~{Roszkowski}, and Y.-L.~S.
  {Tsai}, {\it {Bayesian implications of current LHC and XENON100 search limits
  for the CMSSM}},  {\em \prd} {\bf 85} (2012) 075012,
  [\href{http://arxiv.org/abs/1111.6098}{{\tt arXiv:1111.6098}}].

\bibitem{2012JHEP...06..098B}
P.~{Bechtle}, T.~{Bringmann}, {\em et.~al.}, {\it {Constrained supersymmetry
  after two years of LHC data: a global view with Fittino}},  {\em Journal of
  High Energy Physics} {\bf 6} (2012) 98,
  [\href{http://arxiv.org/abs/1204.4199}{{\tt arXiv:1204.4199}}].

\bibitem{2013PhRvD..87k5010K}
K.~{Kowalska}, S.~{Munir}, {\em et.~al.}, {\it {Constrained next-to-minimal
  supersymmetric standard model with a 126 GeV Higgs boson: A global
  analysis}},  {\em \prd} {\bf 87} (2013) 115010,
  [\href{http://arxiv.org/abs/1211.1693}{{\tt arXiv:1211.1693}}].

\bibitem{2013PhRvD..88e5012F}
A.~{Fowlie}, K.~{Kowalska}, L.~{Roszkowski}, E.~M. {Sessolo}, and Y.-L.~S.
  {Tsai}, {\it {Dark matter and collider signatures of the MSSM}},  {\em \prd}
  {\bf 88} (2013) 055012, [\href{http://arxiv.org/abs/1306.1567}{{\tt
  arXiv:1306.1567}}].

\bibitem{2014PhRvD..89e5017H}
S.~{Henrot-Versill{\'e}}, R.~{Lafaye}, {\em et.~al.}, {\it {Constraining
  supersymmetry using the relic density and the Higgs boson}},  {\em \prd} {\bf
  89} (2014) 055017, [\href{http://arxiv.org/abs/1309.6958}{{\tt
  arXiv:1309.6958}}].

\bibitem{2014JHEP...09..081S}
C.~{Strege}, G.~{Bertone}, {\em et.~al.}, {\it {Profile likelihood maps of a
  15-dimensional MSSM}},  {\em Journal of High Energy Physics} {\bf 9} (2014)
  81, [\href{http://arxiv.org/abs/1405.0622}{{\tt arXiv:1405.0622}}].

\bibitem{2015EPJC...75..422D}
K.~J. {de Vries}, E.~A. {Bagnaschi}, {\em et.~al.}, {\it {The pMSSM10 after LHC
  run 1}},  {\em European Physical Journal C} {\bf 75} (2015) 422,
  [\href{http://arxiv.org/abs/1504.03260}{{\tt arXiv:1504.03260}}].

\bibitem{2015EPJC...75..500B}
E.~A. {Bagnaschi}, O.~{Buchmueller}, {\em et.~al.}, {\it {Supersymmetric dark
  matter after LHC run 1}},  {\em European Physical Journal C} {\bf 75} (2015)
  500, [\href{http://arxiv.org/abs/1508.01173}{{\tt arXiv:1508.01173}}].

\bibitem{2016JCAP...06..050C}
A.~{Cuoco}, B.~{Eiteneuer}, J.~{Heisig}, and M.~{Kr{\"a}mer}, {\it {A global
  fit of the {$\gamma$}-ray galactic center excess within the scalar singlet
  Higgs portal model}},  {\em \jcap} {\bf 6} (2016) 050,
  [\href{http://arxiv.org/abs/1603.08228}{{\tt arXiv:1603.08228}}].

\bibitem{2016EPJC...76...96B}
P.~{Bechtle}, J.~E. {Camargo-Molina}, {\em et.~al.}, {\it {Killing the cMSSM
  softly}},  {\em European Physical Journal C} {\bf 76} (2016) 96,
  [\href{http://arxiv.org/abs/1508.05951}{{\tt arXiv:1508.05951}}].

\bibitem{GAMBIT_GUT}
GAMBIT Collaboration: P.~{Athron}, C.~{Bal{\'a}zs}, {\em et.~al.}, {\it {Global
  fits of GUT-scale SUSY models with GAMBIT}},  {\em \epjc} {\bf 77} (2017)
  824, [\href{http://arxiv.org/abs/1705.07935}{{\tt arXiv:1705.07935}}].

\bibitem{GAMBIT_MSSM7}
GAMBIT Collaboration: P.~{Athron}, C.~{Bal{\'a}zs}, {\em et.~al.}, {\it {A
  global fit of the MSSM with GAMBIT}},  {\em \epjc\ in press} (2017)
  [\href{http://arxiv.org/abs/1705.07917}{{\tt arXiv:1705.07917}}].

\bibitem{2018EPJC...78..256B}
E.~{Bagnaschi}, K.~{Sakurai}, {\em et.~al.}, {\it {Likelihood analysis of the
  pMSSM11 in light of LHC 13-TeV data}},  {\em European Physical Journal C}
  {\bf 78} (2018) 256, [\href{http://arxiv.org/abs/1710.11091}{{\tt
  arXiv:1710.11091}}].

\bibitem{2011PhRvD..83c6008B}
G.~{Bertone}, K.~{Kong}, R.~R. {de Austri}, and R.~{Trotta}, {\it {Global fits
  of the minimal universal extra dimensions scenario}},  {\em \prd} {\bf 83}
  (2011) 036008, [\href{http://arxiv.org/abs/1010.2023}{{\tt
  arXiv:1010.2023}}].

\bibitem{2012JCAP...10..042C}
K.~{Cheung}, Y.-L.~S. {Tsai}, P.-Y. {Tseng}, T.-C. {Yuan}, and A.~{Zee}, {\it
  {Global study of the simplest scalar phantom dark matter model}},  {\em
  \jcap} {\bf 10} (2012) 042, [\href{http://arxiv.org/abs/1207.4930}{{\tt
  arXiv:1207.4930}}].

\bibitem{2014JCAP...06..030A}
A.~{Arhrib}, Y.-L. {Sming Tsai}, Q.~{Yuan}, and T.-C. {Yuan}, {\it {An updated
  analysis of Inert Higgs Doublet Model in light of the recent results from
  LUX, PLANCK, AMS-02 and LHC}},  {\em \jcap} {\bf 6} (2014) 030,
  [\href{http://arxiv.org/abs/1310.0358}{{\tt arXiv:1310.0358}}].

\bibitem{2016JHEP...11..070B}
S.~{Banerjee}, S.~{Matsumoto}, K.~{Mukaida}, and Y.-L. {Sming Tsai}, {\it {WIMP
  dark matter in a well-tempered regime -- A case study on singlet-doublets
  fermionic WIMP}},  {\em Journal of High Energy Physics} {\bf 11} (2016) 70,
  [\href{http://arxiv.org/abs/1603.07387}{{\tt arXiv:1603.07387}}].

\bibitem{2016PhRvD..94f5034M}
S.~{Matsumoto}, S.~{Mukhopadhyay}, and Y.-L.~S. {Tsai}, {\it {Effective theory
  of WIMP dark matter supplemented by simplified models: Singlet-like Majorana
  fermion case}},  {\em \prd} {\bf 94} (2016) 065034,
  [\href{http://arxiv.org/abs/1604.02230}{{\tt arXiv:1604.02230}}].

\bibitem{GAMBIT_SS}
GAMBIT Collaboration: P.~{Athron}, C.~{Bal{\'a}zs}, {\em et.~al.}, {\it {Status
  of the scalar singlet dark matter model}},  {\em \epjc} {\bf 77} (2017) 568,
  [\href{http://arxiv.org/abs/1705.07931}{{\tt arXiv:1705.07931}}].

\bibitem{2017arXiv171011138H}
S.~{Hoof} and {for the GAMBIT Collaboration}, {\it {A Preview of Global Fits of
  Axion Models in GAMBIT}},  \href{http://arxiv.org/abs/1710.11138}{{\tt
  arXiv:1710.11138}}.

\bibitem{2014JHEP...10..155M}
S.~{Matsumoto}, S.~{Mukhopadhyay}, and Y.-L.~S. {Tsai}, {\it {Singlet Majorana
  fermion dark matter: a comprehensive analysis in effective field theory}},
  {\em Journal of High Energy Physics} {\bf 10} (2014) 155,
  [\href{http://arxiv.org/abs/1407.1859}{{\tt arXiv:1407.1859}}].

\bibitem{2016JHEP...09..077L}
S.~{Liem}, G.~{Bertone}, {\em et.~al.}, {\it {Effective field theory of dark
  matter: a global analysis}},  {\em \jhep} {\bf 9} (2016) 77,
  [\href{http://arxiv.org/abs/1603.05994}{{\tt arXiv:1603.05994}}].

\bibitem{darkbit}
GAMBIT Dark Matter Workgroup: T.~{Bringmann}, J.~{Conrad}, {\em et.~al.}, {\it
  {DarkBit: A GAMBIT module for computing dark matter observables and
  likelihoods}},  {\em \epjc} {\bf 77} (2017) 831,
  [\href{http://arxiv.org/abs/1705.07920}{{\tt arXiv:1705.07920}}].

\bibitem{2017CoPhC.213..252H}
X.~{Huang}, Y.-L.~S. {Tsai}, and Q.~{Yuan}, {\it {LIKEDM: Likelihood calculator
  of dark matter detection}},  {\em Computer Physics Communications} {\bf 213}
  (2017) 252--263, [\href{http://arxiv.org/abs/1603.07119}{{\tt
  arXiv:1603.07119}}].

\bibitem{2018arXiv180400044A}
F.~{Ambrogi}, C.~{Arina}, {\em et.~al.}, {\it {MadDM v.3.0: a Comprehensive
  Tool for Dark Matter Studies}},  \href{http://arxiv.org/abs/1804.00044}{{\tt
  arXiv:1804.00044}}.

\bibitem{2010JHEP...04..057A}
Y.~{Akrami}, P.~{Scott}, J.~{Edsj{\"o}}, J.~{Conrad}, and L.~{Bergstr{\"o}m},
  {\it {A profile likelihood analysis of the constrained MSSM with genetic
  algorithms}},  {\em Journal of High Energy Physics} {\bf 4} (2010) 57,
  [\href{http://arxiv.org/abs/0910.3950}{{\tt arXiv:0910.3950}}].

\bibitem{2011JHEP...06..042F}
F.~{Feroz}, K.~{Cranmer}, M.~{Hobson}, R.~{Ruiz de Austri}, and R.~{Trotta},
  {\it {Challenges of profile likelihood evaluation in multi-dimensional SUSY
  scans}},  {\em Journal of High Energy Physics} {\bf 6} (2011) 42,
  [\href{http://arxiv.org/abs/1101.3296}{{\tt arXiv:1101.3296}}].

\bibitem{scannerbit}
GAMBIT Scanner Workgroup: G.~D. {Martinez}, J.~{McKay}, {\em et.~al.}, {\it
  {Comparison of statistical sampling methods with ScannerBit, the GAMBIT
  scanning module}},  {\em \epjc\ in press} {\bf 77} (2017) 761,
  [\href{http://arxiv.org/abs/1705.07959}{{\tt arXiv:1705.07959}}].

\bibitem{2011JHEP...03..012B}
M.~{Bridges}, K.~{Cranmer}, {\em et.~al.}, {\it {A coverage study of the CMSSM
  based on ATLAS sensitivity using fast neural networks techniques}},  {\em
  Journal of High Energy Physics} {\bf 3} (2011) 12,
  [\href{http://arxiv.org/abs/1011.4306}{{\tt arXiv:1011.4306}}].

\bibitem{2011JCAP...07..002A}
Y.~{Akrami}, C.~{Savage}, P.~{Scott}, J.~{Conrad}, and J.~{Edsj{\"o}}, {\it
  {Statistical coverage for supersymmetric parameter estimation: a case study
  with direct detection of dark matter}},  {\em \jcap} {\bf 7} (2011) 002,
  [\href{http://arxiv.org/abs/1011.4297}{{\tt arXiv:1011.4297}}].

\bibitem{2012PhRvD..86b3507S}
C.~{Strege}, R.~{Trotta}, G.~{Bertone}, A.~H.~G. {Peter}, and P.~{Scott}, {\it
  {Fundamental statistical limitations of future dark matter direct detection
  experiments}},  {\em \prd} {\bf 86} (2012) 023507,
  [\href{http://arxiv.org/abs/1201.3631}{{\tt arXiv:1201.3631}}].

\bibitem{gambit}
GAMBIT Collaboration: P.~{Athron}, C.~{Bal{\'a}zs}, {\em et.~al.}, {\it
  {GAMBIT: The Global and Modular Beyond-the-Standard-Model Inference Tool}},
  {\em \epjc} {\bf 77} (2017) 784, [\href{http://arxiv.org/abs/1705.07908}{{\tt
  arXiv:1705.07908}}].

\bibitem{Planck15cosmo}
{Planck Collaboration}, {\it {Planck 2015 results. XIII. Cosmological
  parameters}},  \href{http://arxiv.org/abs/1502.01589}{{\tt
  arXiv:1502.01589}}.

\bibitem{ColliderBit}
GAMBIT Collider Workgroup: C.~{Bal{\'a}zs}, A.~{Buckley}, {\em et.~al.}, {\it
  {ColliderBit: a GAMBIT module for the calculation of high-energy collider
  observables and likelihoods}},  {\em \epjc} {\bf 77} (2017) 795,
  [\href{http://arxiv.org/abs/1705.07919}{{\tt arXiv:1705.07919}}].

\bibitem{Tsai:2012cs}
Y.-L.~S. Tsai, Q.~Yuan, and X.~Huang, {\it {A generic method to constrain the
  dark matter model parameters from Fermi observations of dwarf spheroids}},
  {\em JCAP} {\bf 1303} (2013) 018, [\href{http://arxiv.org/abs/1212.3990}{{\tt
  arXiv:1212.3990}}].

\bibitem{Cui:2016ppb}
M.-Y. Cui, Q.~Yuan, Y.-L.~S. Tsai, and Y.-Z. Fan, {\it {Possible dark matter
  annihilation signal in the AMS-02 antiproton data}},  {\em Phys. Rev. Lett.}
  {\bf 118} (2017) 191101, [\href{http://arxiv.org/abs/1610.03840}{{\tt
  arXiv:1610.03840}}].

\bibitem{Cheng:2016slx}
H.-C. Cheng, W.-C. Huang, {\em et.~al.}, {\it {AMS-02 Positron Excess and
  Indirect Detection of Three-body Decaying Dark Matter}},  {\em JCAP} {\bf
  1703} (2017) 041, [\href{http://arxiv.org/abs/1608.06382}{{\tt
  arXiv:1608.06382}}].

\bibitem{Liu:2017kmx}
Z.~Liu, Y.~Su, Y.-L. Sming~Tsai, B.~Yu, and Q.~Yuan, {\it {A combined analysis
  of PandaX, LUX, and XENON1T experiments within the framework of dark matter
  effective theory}},  {\em JHEP} {\bf 11} (2017) 024,
  [\href{http://arxiv.org/abs/1708.04630}{{\tt arXiv:1708.04630}}].

\bibitem{Albertsson:2018maf}
K.~Albertsson {\em et.~al.}, {\it {Machine Learning in High Energy Physics
  Community White Paper}},  \href{http://arxiv.org/abs/1807.02876}{{\tt
  arXiv:1807.02876}}.

\bibitem{Guest:2018yhq}
D.~Guest, K.~Cranmer, and D.~Whiteson, {\it {Deep Learning and its Application
  to LHC Physics}},  \href{http://arxiv.org/abs/1806.11484}{{\tt
  arXiv:1806.11484}}.

\bibitem{arinaBayesianAnalysisMultiple2014}
C.~Arina, {\it Bayesian analysis of multiple direct detection experiments},
  {\em Physics of the Dark Universe} {\bf 5-6} (2014) 1--17.

\bibitem{Arnaud:2017usi}
EDELWEISS: Q.~Arnaud {\em et.~al.}, {\it {Optimizing EDELWEISS detectors for
  low-mass WIMP searches}},  {\em Phys. Rev.} {\bf D97} (2018) 022003,
  [\href{http://arxiv.org/abs/1707.04308}{{\tt arXiv:1707.04308}}].

\bibitem{Tan:2016zwf}
PandaX-II: A.~Tan {\em et.~al.}, {\it {Dark Matter Results from First 98.7 Days
  of Data from the PandaX-II Experiment}},  {\em Phys. Rev. Lett.} {\bf 117}
  (2016) 121303, [\href{http://arxiv.org/abs/1607.07400}{{\tt
  arXiv:1607.07400}}].

\bibitem{Caron:2017udl}
S.~Caron, G.~A. Gómez-Vargas, L.~Hendriks, and R.~Ruiz~de Austri, {\it
  {Analyzing $\gamma$-rays of the Galactic Center with Deep Learning}},  {\em
  JCAP} {\bf 1805} (2018) 058, [\href{http://arxiv.org/abs/1708.06706}{{\tt
  arXiv:1708.06706}}].

\bibitem{Schinzel:2017irp}
F.~K. Schinzel, L.~Petrov, G.~B. Taylor, and P.~G. Edwards, {\it {Radio
  Follow-up on all Unassociated Gamma-ray Sources from the Third Fermi Large
  Area Telescope Source Catalog}},  {\em Astrophys. J.} {\bf 838} (2017) 139,
  [\href{http://arxiv.org/abs/1702.07036}{{\tt arXiv:1702.07036}}].

\bibitem{Ackermann:2013yma}
Fermi-LAT: M.~Ackermann {\em et.~al.}, {\it {Determination of the Point-Spread
  Function for the Fermi Large Area Telescope from On-orbit Data and Limits on
  Pair Halos of Active Galactic Nuclei}},  {\em Astrophys. J.} {\bf 765} (2013)
  54, [\href{http://arxiv.org/abs/1309.5416}{{\tt arXiv:1309.5416}}].

\bibitem{Parkinson:2016oab}
P.~M. Saz~Parkinson, H.~Xu, {\em et.~al.}, {\it {Classification and Ranking of
  Fermi LAT Gamma-ray Sources from the 3FGL Catalog using Machine Learning
  Techniques}},  {\em Astrophys. J.} {\bf 820} (2016) 8,
  [\href{http://arxiv.org/abs/1602.00385}{{\tt arXiv:1602.00385}}].

\bibitem{Mirabal:2016huj}
N.~Mirabal, E.~Charles, {\em et.~al.}, {\it {3FGL Demographics Outside the
  Galactic Plane using Supervised Machine Learning: Pulsar and Dark Matter
  Subhalo Interpretations}},  {\em Astrophys. J.} {\bf 825} (2016) 69,
  [\href{http://arxiv.org/abs/1605.00711}{{\tt arXiv:1605.00711}}].

\bibitem{2017MNRAS.472.1129P}
C.~E. {Petrillo}, C.~{Tortora}, {\em et.~al.}, {\it {Finding strong
  gravitational lenses in the Kilo Degree Survey with Convolutional Neural
  Networks}},  {\em \mnras} {\bf 472} (2017) 1129--1150,
  [\href{http://arxiv.org/abs/1702.07675}{{\tt arXiv:1702.07675}}].

\bibitem{Bertone:2017adx}
G.~Bertone, N.~Bozorgnia, {\em et.~al.}, {\it {Identifying WIMP dark matter
  from particle and astroparticle data}},  {\em JCAP} {\bf 1803} (2018) 026,
  [\href{http://arxiv.org/abs/1712.04793}{{\tt arXiv:1712.04793}}].

\bibitem{BARBIERI198863}
R.~Barbieri and G.~Giudice, {\it Upper bounds on supersymmetric particle
  masses},  {\em Nuclear Physics B} {\bf 306} (1988) 63 -- 76.

\bibitem{Barbieri:1998uv}
R.~Barbieri and A.~Strumia, {\it {About the fine tuning price of LEP}},  {\em
  Phys. Lett.} {\bf B433} (1998) 63--66,
  [\href{http://arxiv.org/abs/hep-ph/9801353}{{\tt hep-ph/9801353}}].

\bibitem{Casas:2014eca}
J.~A. Casas, J.~M. Moreno, S.~Robles, K.~Rolbiecki, and B.~Zaldívar, {\it
  {What is a Natural SUSY scenario?}},  {\em JHEP} {\bf 06} (2015) 070,
  [\href{http://arxiv.org/abs/1407.6966}{{\tt arXiv:1407.6966}}].

\bibitem{Cabrera:2012vu}
M.~E. Cabrera, J.~A. Casas, and R.~Ruiz~de Austri, {\it {The health of SUSY
  after the Higgs discovery and the XENON100 data}},  {\em JHEP} {\bf 07}
  (2013) 182, [\href{http://arxiv.org/abs/1212.4821}{{\tt arXiv:1212.4821}}].

\bibitem{Lyons:1900zz}
L.~Lyons, {\it {Open statistical issues in Particle Physics}},  {\em Ann. Appl.
  Stat.} {\bf 2} (2008) 887--915.

\bibitem{Demortier:2007zz}
L.~Demortier, {\it {P values and nuisance parameters}},  in {\em {Statistical
  issues for LHC physics. Proceedings, Workshop, PHYSTAT-LHC, Geneva,
  Switzerland, June 27-29, 2007}} (2007) 23--33.

\bibitem{Gross:2010qma}
E.~Gross and O.~Vitells, {\it {Trial factors for the look elsewhere effect in
  high energy physics}},  {\em Eur. Phys. J.} {\bf C70} (2010) 525--530,
  [\href{http://arxiv.org/abs/1005.1891}{{\tt arXiv:1005.1891}}].

\bibitem{Aaboud:2016ejt}
ATLAS: M.~Aaboud {\em et.~al.}, {\it {Search for new phenomena in events
  containing a same-flavour opposite-sign dilepton pair, jets, and large
  missing transverse momentum in $\sqrt{s}=$ 13 $pp$ collisions with the ATLAS
  detector}},  {\em Eur. Phys. J.} {\bf C77} (2017) 144,
  [\href{http://arxiv.org/abs/1611.05791}{{\tt arXiv:1611.05791}}].

\bibitem{Hakkila:2013cra}
J.~Hakkila, T.~J. Loredo, R.~L. Wolpert, M.~E. Broadbent, and R.~D. Preece,
  {\it {A template for describing intrinsic GRB pulse shapes}},  {\em J.
  Hakkila} {\bf Symposium} (2013) [arXiv:],
  [\href{http://arxiv.org/abs/1308.5957}{{\tt arXiv:1308.5957}}].

\bibitem{10.2307/4144397}
R.~L. Wolpert and K.~L. Mengersen, {\it Adjusted likelihoods for synthesizing
  empirical evidence from studies that differ in quality and design: Effects of
  environmental tobacco smoke},  {\em Statistical Science} {\bf 19} (2004)
  450--471.

\bibitem{Bayes:1764vd}
R.~Bayes, {\it {An essay toward solving a problem in the doctrine of chances}},
   {\em Phil. Trans. Roy. Soc. Lond.} {\bf 53} (1764) 370--418.

\end{thebibliography}\endgroup

\end{document}